\documentclass[10pt,times,authoryear,doublespacing,reviewcopy]{elsarticle}

\usepackage{float}
\usepackage{color}
\usepackage{graphicx}
\usepackage{subfigure}
\usepackage{subcaption}
\usepackage[margin=2.5cm]{geometry}
\usepackage{empheq}
\usepackage{algorithm}
\usepackage{algpseudocode}
\usepackage{tabularx}
\usepackage{fancyhdr}

\usepackage{natbib}

\usepackage[colorlinks=true,
            linkcolor=black,
            citecolor=black,
            urlcolor=black]{hyperref}

\usepackage{amssymb}
\usepackage{amsmath}
\usepackage{mathtools}

\usepackage{amsthm}

\begin{document}
\let\WriteBookmarks\relax
\def\floatpagepagefraction{1}
\def\textpagefraction{.001}

\begin{frontmatter}



\title{A multiphysics model for triboelectric nanogenerator design with explicit surface roughness representation} 

\author[GCEC]{MD Tanzib Ehsan Sanglap} 
\author[MMRG]{Jack Perris}
\author[MMRG]{Rudra Mukherjee}
\author[MMRG,IITM]{Charchit Kumar}
\author[GCEC]{Lukasz Kaczmarczyk}
\author[GCEC]{Chris J. Pearce}
\author[MMRG]{Daniel M. Mulvihill}
\author[GCEC]{Andrei G. Shvarts$^*$}

\affiliation[GCEC]{organization={Glasgow Computational Engineering Centre (GCEC), James Watt School of Engineering, University of Glasgow},
            city={Glasgow},
            postcode={G12 8QQ},
            country={UK}}
            
\affiliation[MMRG]        
{organization={Material and Manufacturing Research Cluster (MMRC), James Watt School of Engineering, University of Glasgow},
            city={Glasgow},
            postcode={G12 8QQ},
            country={UK}}
\affiliation[IITM]        
{organization={Department of Mechanical Engineering, Indian Institute of Technology Madras},
            city={Chennai, Tamil Nadu},
            country={India}}

\begin{abstract}
The design of triboelectric nanogenerators (TENGs) for efficient energy harvesting requires predictive models that capture the interplay between surface roughness, real contact area, and electrostatic behaviour across diverse tribolayer materials and roughness levels. To address this demand, this paper presents a multiphysics finite element framework that couples mechanical contact analysis with electrostatic simulations, considering exact surface roughness representations rather than idealised statistical approximations. Compared with optical interference microscopy measurements, the framework predicts the real contact area ratio more accurately than analytical models. The proposed approach captures the electrostatic behaviour by scaling the TENG surface charge density with the real contact area ratio between the rough tribolayers, computed for a given mechanical load. This method improves agreement with experiments for open-circuit voltage and capacitance relative to approximate analytical models. To represent the TENG circuit, a time-dependent ordinary differential equation is integrated, enabling evaluation of electrical responses under varying load conditions and elucidating the roles of surface roughness, mechanical load, contact–separation frequency, and resistive load. The framework provides a robust, scalable tool for performance optimisation across dielectric materials, mechanical behaviours, and operating conditions and is readily extendable to other surface-dependent energy-harvesting devices.
\end{abstract}








\begin{keyword}
 TENG \sep Finite Element Method \sep Surface Roughness \sep Electrostatics \sep Contact Mechanics

\end{keyword}

\end{frontmatter}



\begin{figure}[H]
    \centering
    \includegraphics[width=0.86\textwidth]{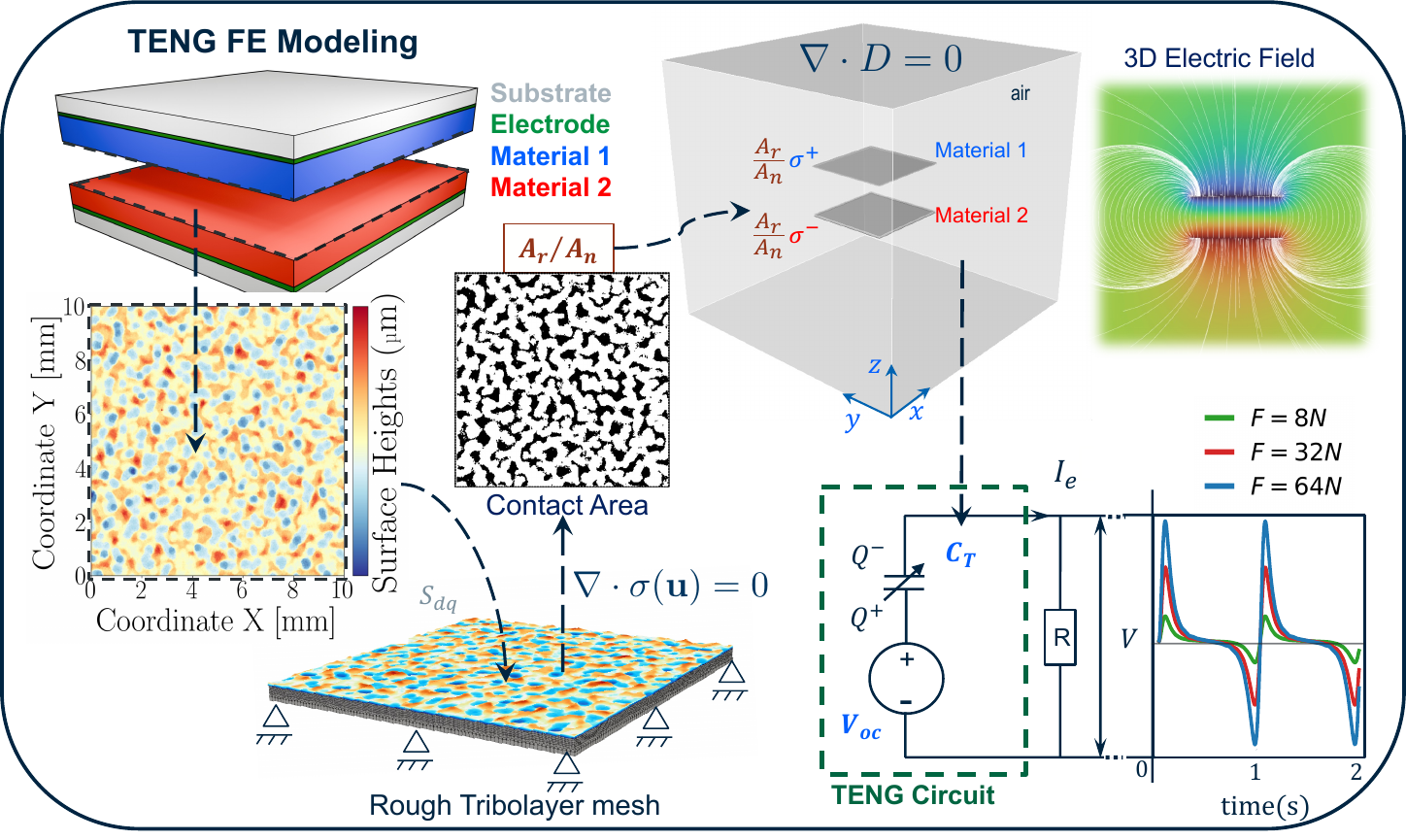}\
    \caption*{}
    \label{fig:placeholder}
\end{figure}

\thispagestyle{fancy}
\fancyhf{}
\fancyfoot[L]{%
\rule{0.5\textwidth}{0.1pt}\vspace{1pt}\\
{\footnotesize\itshape $^*$Corresponding author email: andrei.shvarts@glasgow.ac.uk}
}

\section{Introduction}
\label{sec:Introduction}
Ambient energy harvesting is a promising strategy for utilising sustainable and autonomous micro-nanoscale technologies designed to convert mechanical or thermal energy into electricity. Among these technologies, nanogenerators offer promising solutions as compact and versatile energy harvesters. Early efforts primarily focused on piezoelectric nanogenerators for efficient generation of electricity from mechanical deformation of piezoelectric materials (e.g. ZnO nanowires) as well as pyroelectric nanogenerators to enhance energy harvesting from time-dependent temperature fluctuations and thermoelectric nanogenerators for converting temperature gradients into electrical power via the Seebeck effect \citep{wang2006piezoelectric, wang2011nanogenerators, fan2012flexible, liu2021recent, mulvihill2025test}. 
\begin{figure}[!b]

    \centering
    \includegraphics[width=0.46\linewidth]{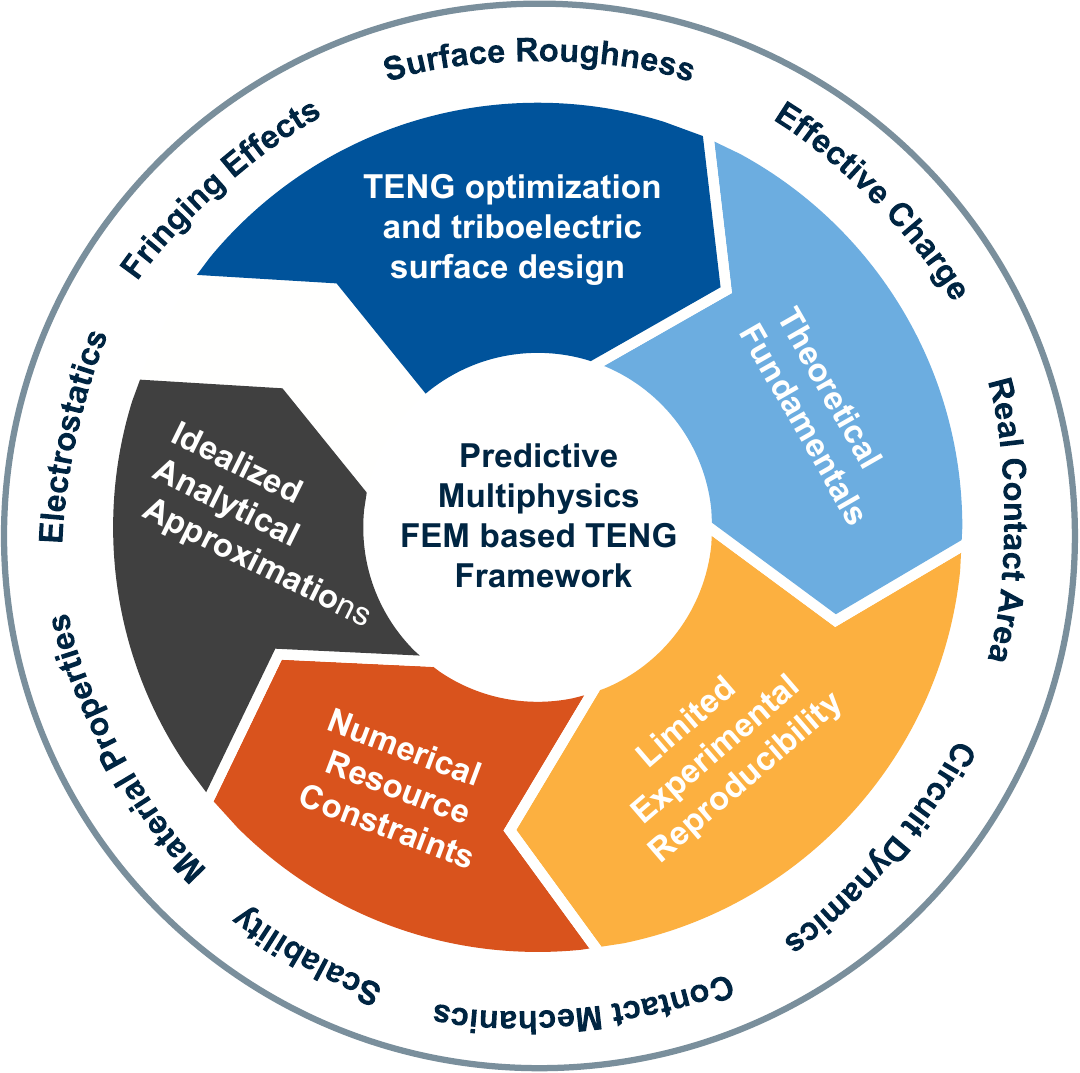}
    \caption{Scope of a predictive multiphysics numerical framework for systematic TENG design and optimisation. Addressing the existing research gaps requires a predictive tool for investigating distinct physics, considering coupling between finite surface roughness, contact mechanics, triboelectric charge transfer, and electrostatics analyses within a single computational pipeline.}
    \label{fig1:TENG_research_gap}
\end{figure}
Despite these advances, the efficiency of many conventional nanogenerators is limited when harvesting at low frequency irregular mechanical motions that dominate most ambient environments. This limitation motivated the fabrication of the Triboelectric Nanogenerators (TENGs) \citep{fan2012flexible} that utilise contact electrification and electrostatic induction. Over a short period, TENGs demonstrated superior performance compared to conventional energy harvesting technologies due to their versatility, high output potential, low cost, ease of production, and resource efficiency, making them a promising candidate for several industries \citep{wang2023handbook, gao2024spontaneously}. Recent studies demonstrate potential applications of TENGs not only in wearable and medical electronics \citep{paosangthong2019recent,pan2020ultrastretchable,khandelwal2020triboelectric, xiao2021wearable, cao2022application,walden2022opportunities, islam2025covalent}, but also in industries such as civil, textiles, robotics, automotive, and aerospace \citep{pang2024triboelectric, phan2017aerodynamic, jin2020triboelectric, yong2021auto}. Therefore, research interest is growing in optimising TENG circuit design, charge transfer mechanisms, material, electrodes, and geometry of tribo-layers~\citep{niu2014theoretical, el2021active,zhao2021universal,lv2023high}. The substantial experimental investigations are also concentrated on understanding TENG behaviour and identifying more suitable surfaces and material parameters \citep{armitage2022investigating,ibrahim2021surface, kumar2023multiscale, bairagi2023mechanical,hussain2024exploring,mulvihill2025test}. 

TENGs operate in various modes and configurations. The basic TENG modes typically consist of two or more dissimilar surfaces that exchange electrons, ions, and/or material components upon external mechanical force, enforcing contact \citep{kim2021triboelectric}. Once these tribolayers get separated, opposite charges get bound at the tribolayer interfaces, resulting in a potential drop. Further, when metallic electrodes are placed on the outer sides of the tribolayers and are electrically connected into a circuit, an alternating current is generated due to the flow of electrons through repeated cycles of mechanical contact and separation~\citep{wang2023handbook}. In addition to the conventional TENG operating modes, including contact-separation, lateral sliding, freestanding, and rolling, recent investigations have also explored hybrid operational modes \citep{el2021active,zhao2021universal, baburaj2024high, bairagi2023wearable, li2024triboiontronics} and surface configurations \citep{shi2023progress, xiong2024endowing, li2024triboelectrification, kumar2023mechanics}. However, despite their prominence, experimental investigations rely primarily on extrapolated predictions and repetition across diverse configurations, resulting in challenges such as scalability and the difficulty of isolating individual physical and geometrical parameters of innovative TENG performance strategies.

Distinct physical factors have been taken into account in various analytical models of TENGs, including dielectric properties of tribo-layers, thickness, air gap between the layers, surface roughness, charge density, charge regeneration effects, etc. \citep{shao20193d, shao2020theoretical,armitage2021investigation, li2022field,olson2024puts, mulvihill2025test}. However, most models approximate and simplify physical reality, either by assuming infinite lateral TENG length \citep{niu_Voc_linear, rathi2022tribo}, neglecting bending of the electric fields near device edges \citep{XUVoc, zhao2024theoretical, dharmashenaVoc}, approximating contact mechanics using statistical asperity-based contact models \citep{wen2021load}, or not considering the load-dependent behaviour of TENGs\citep{venugopal2022effective}. As a result, these approaches exhibit limited accuracy in predicting TENG performance. In this context, numerical methods can offer promising alternatives for resolving complex, coupled multiphysical interactions for individual physics in various applications. However, numerical tools tailored for modelling TENG accounting for roughness remain limited due to computational resource constraints and the difficulty of implementing robust coupling methods  \citep{polonsky2000fast,thicknessanddielectriceffect, Cheng2023, kim2025coupling}. Consequently, key research gaps remain unexplored, particularly concerning how electrostatic field distributions in a finite-sized device are influenced by realistic rough surface topographies. \textbf{Figure~\ref{fig1:TENG_research_gap}} conceptually summarises the current progress in TENG modelling and the scope of optimisation of surfaces with the distinct TENG physics. To address these gaps and challenges, a multiphysics numerical framework that consistently captures contact mechanics, electrostatic field equations, mechanical and dielectric material behaviour and circuit dynamics is required. Such a framework would enable simulation of TENG within a single computational pipeline with real and/or explicitly represented triboelectric rough surfaces, allow rigorous experimental validation, and predict charge generation and performance for diverse TENG surface configurations.

\begin{figure}[!b]
    \centering
    \includegraphics[width=0.55\linewidth]{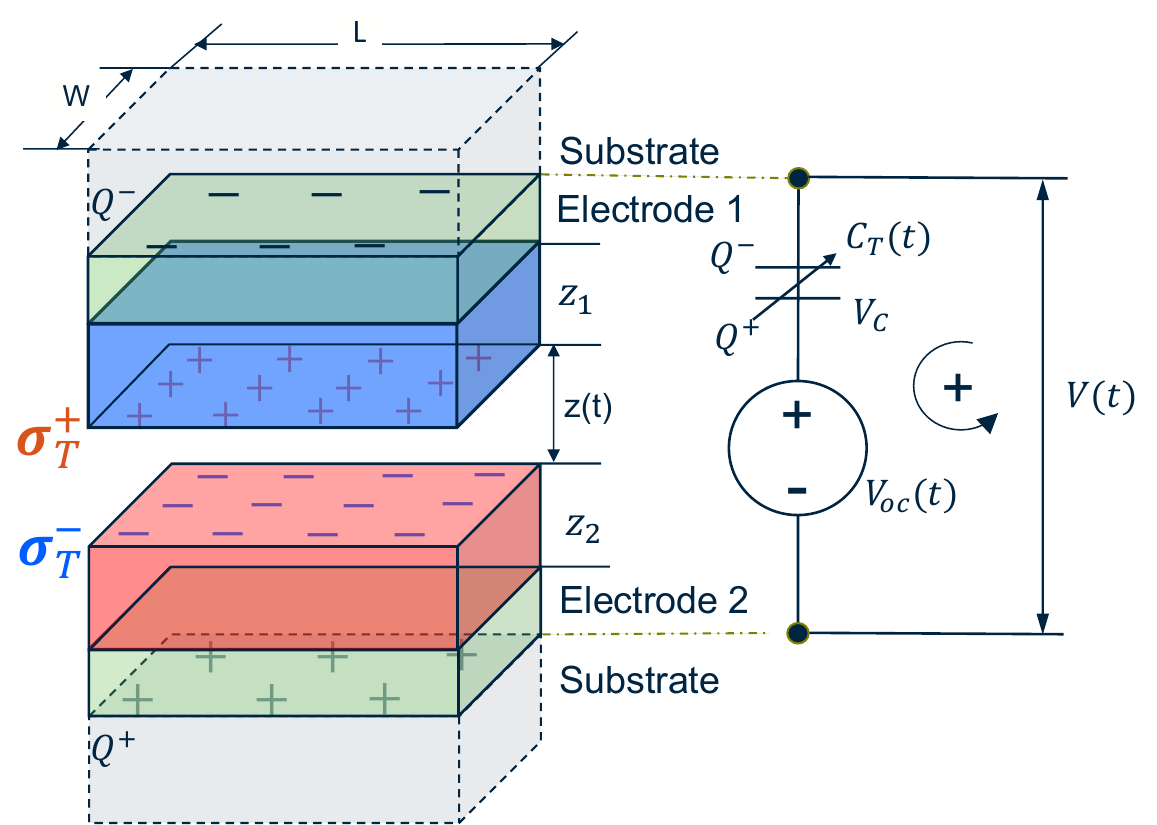}
    \caption{Schematic of a contact-separation TENG with two dielectric layers (blue, red) and electrodes (green) showing a time-dependent open-circuit voltage $V_{OC}(t)$ and capacitance $C_T(t)$ in the external circuit where mechanical contact generates equal and opposite tribo-charges $\sigma_T^{+}$ and $\sigma_T^{-}$, while charges $Q^{+}$ and $Q^{-}$ are induced in the separation stage on the electrodes.}
    \label{fig:Schematic_TENG}
\end{figure}

\subsection{Fundamental electrostatic behaviour of TENG}
In most TENG theories, the electrostatic behaviour of TENG is commonly interpreted using the voltage-charge-separation ($V-Q-z$) relation proposed by ~\cite{wang-CapacitiveBehaviorVQX}. This description captures the relationship between \textit{contact electrification} and \textit{electrostatic induction}. The former depends on the surface deformation, such that when layers are pressed against each other, the equal but opposite tribocharges, denoted by $\sigma_T$, are produced. When the layers are separated, these charges remain bound to the surfaces and produce an open-circuit potential difference $V_{OC}$ between electrodes. Typically, these electrodes are connected in a circuit, and the induced charges are transferred between the electrodes, such that TENG acts as a capacitor with time-varying charges~\citep{wang2023introduction, wang2023handbook, cheng2023triboelectric} as shown in \textbf{Figure~\ref{fig:Schematic_TENG}}.

\begin{figure}[t]
    \centering
    \begin{minipage}{0.49\textwidth}
        \centering
        \includegraphics[width=0.9\linewidth]{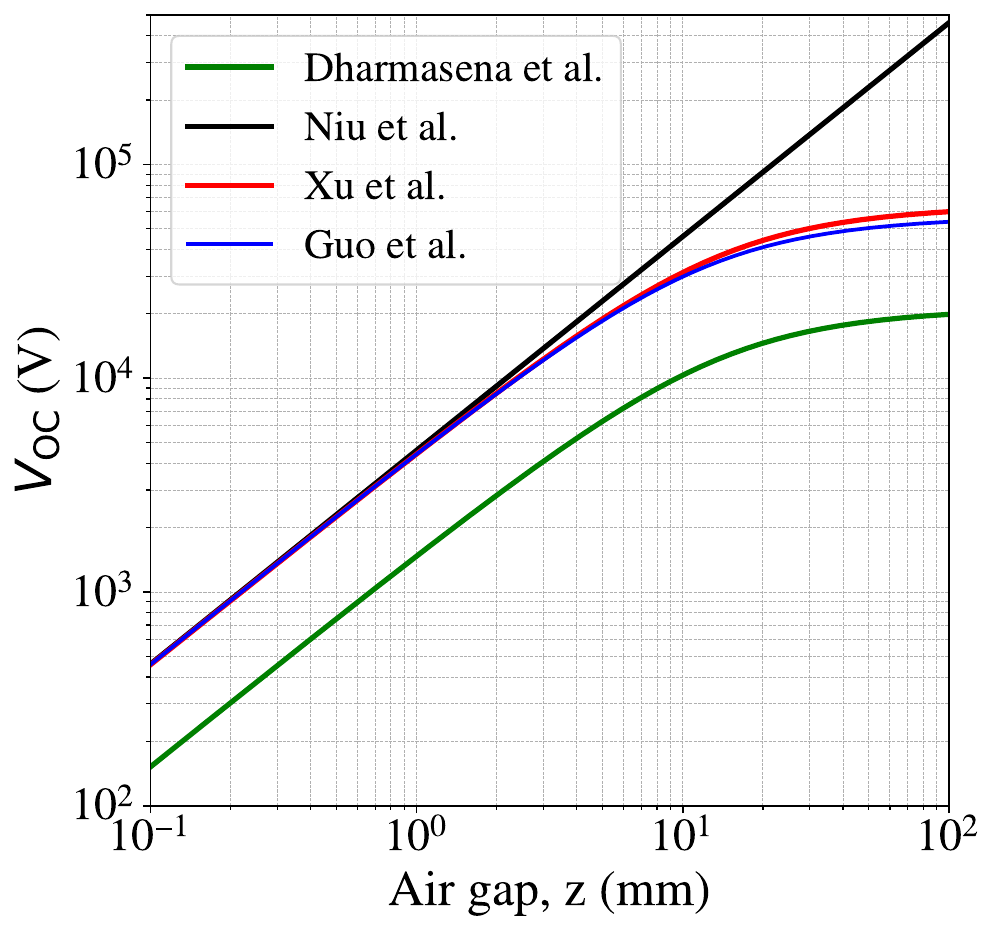}
        \\(a)
    \end{minipage}
    \hfill
    \begin{minipage}{0.5\textwidth}
        \centering
        \includegraphics[width=0.9\linewidth]{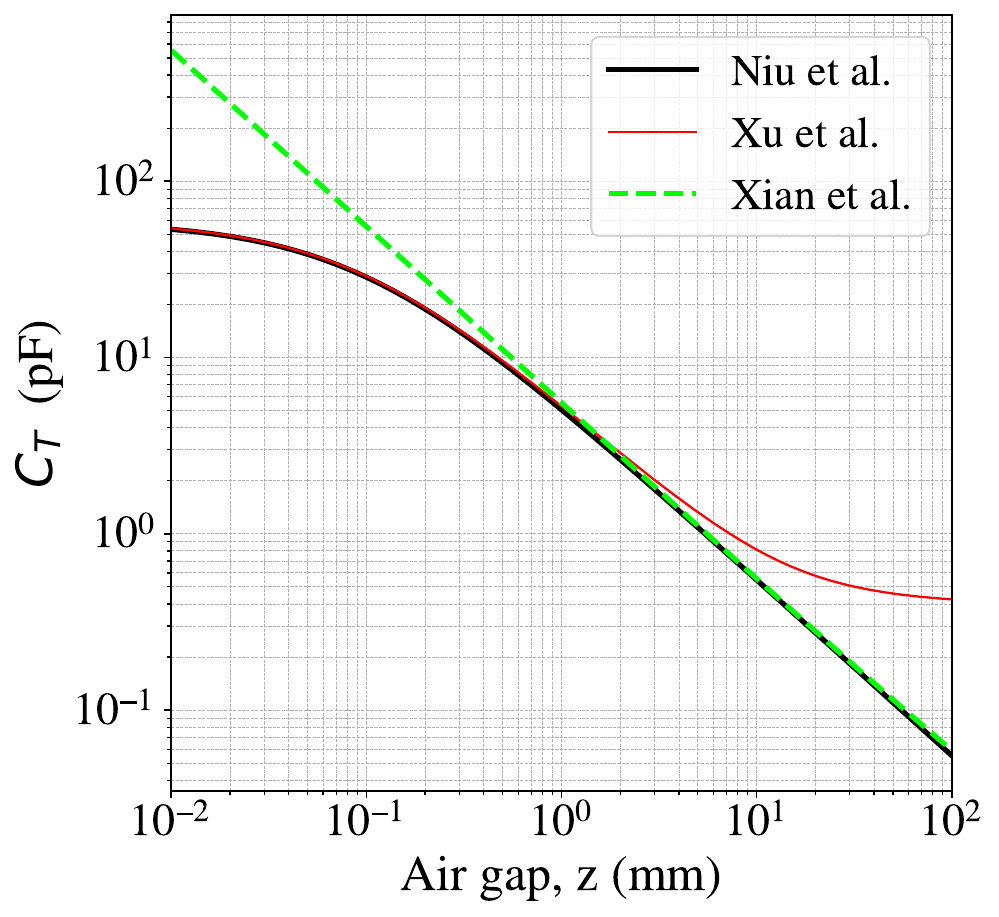}
        \\(b)
    \end{minipage}

    \caption{TENG characteristics as functions of air gap using the material parameters reported in \cite{XUVoc}, showing:(a) analytical approximations of the open-circuit voltage $V_{OC}$ \citep{XUVoc, niu_Voc_linear, guo2020derivation} (b) effective capacitance $C_T$ derived from the analytical models \citep{niu_Voc_linear, XUVoc, xiang2006electrostatic}.}
    \label{fig_0}
\end{figure}

Thus, the $V-Q-z$ relation incorporates $V_{OC}(z)$, device capacitance, $C(z)$, the air gap $z$ between the two layers and accounts for the charge in the external circuit $Q$ over the electrode area $A_n$, resulting in a net voltage $V$:
\begin{equation}
V = V_{OC}(z) -\frac{Q}{C_T(z)}, 
\label{eq:basic}
\end{equation}
see \textbf{Figure~\ref{fig_0}(a)}. In this context, Dharmashena et al. \cite{dharmashenaVoc} observed the dependency of $V_{OC}$ on the air gap, considering a distance-dependent electric field for a finite square-shaped TENG. Xu et al. \cite{XUVoc} improved this model by ensuring the asymptotic agreement with the simplified model of \cite{niu_Voc_linear} for small gaps and first described the load-dependent TENG behaviour analytically through scaling of the electric field (or charge density) with the real contact area ratio. Among the quantitative formulations reported in the literature, the most efficient analytical formula for circular equivalent TENG while calculating $V_{OC}$ is found in the derivation by  \cite{guo2020derivation}, which, however, overlooks the permittivity of the TENG layers. Furthermore, while the models depict electrodes in the setup, the mathematical electrostatic formulations have not accounted for them. Critically, most analytical models also neglect the non-uniformity of the electric field (fringing)  near the boundaries of charged surfaces \citep{shao2020theoretical,chen2020universal,chu2021theoretical}. The comparison between the aforementioned analytical approximations is presented in Figure~\ref{fig_0}(b), where a noticeable discrepancy in predicting TENG performance is evident when computing the $V_{OC}$ dependency on the air gap.

In microelectronics, electrostatic interactions are commonly analysed using the parallel-plate approximation to model the capacitive behaviour. However, simple analytical models are ineffective in accounting for complex electrophysical properties, e.g., dielectric inhomogeneity, interface state and non-uniform electric fields ~\citep{huang1992design}. For two-dimensional parallel finite plates, \citep{xiang2006electrostatic} derived the capacitance formula accounting for fringing of the electric field using conformal mapping techniques. Subsequent studies demonstrated the advantages of FEM in obtaining more accurate capacitance estimates~\citep{9015111}. Correspondingly, \cite{liu2023plate} conducted benchmark studies on 3D parallel plate capacitance using the FEA tool FEniCS, emphasising the length-to-gap dependence. The distinguishable characteristics with air gaps in different analytical approximations for the same smooth TENG surfaces are also noticeable in \textbf{Figure~\ref{fig_0}(b)}, indicating the limitations and distinct behaviour of different models. Thus, investigations are necessary to model the 3D capacitance of TENGs by accounting for finite electrode surfaces to realistically capture electrostatic interactions under dielectric heterogeneity and to predict experimental behaviour under reasonable operating conditions.

Another critical electrostatic aspect associated with efficient charge collection and transfer in TENG is related to the electrodes. In particular, TENG tribo-layers are thin (10–500 $\mu$m) and are usually coated with thinner lined electrodes (0.1–4 $\mu$m). Despite their importance to device performance, the electrode geometry and thickness have largely been neglected in existing TENG modelling approaches. In mathematical models, the electrode surface can be simulated under floating or equipotential electrode conditions~\citep{ma2007integrated, dudem2018triboelectric}. Substantially, modelling a floating electrode on TENG requires accurately computing the electric flux through the surface on the interface \citep{papaioannou2011floating, michalas2015dielectric, chen2021hybridizable, amann2014simulation}. In this regard, a limited number of numerical approaches have been investigated. Among them, the Boundary Element Method (BEM) provides improved computational efficiency compared with mesh-free methods \citep{yin2025mesh}. However, accurately handling complex materials such as anisotropic media remains challenging \citep{amann2014simulation}. Consequently, these limitations motivate assessments of analytical predictions and accurate full-scale simulations of TENG dielectric layers considering electrodes explicitly, e.g. using FEM.

\subsection{Influence of surface roughness on TENG performance}

Throughout TENG development efforts, numerous studies highlighted the significant influence of surface topography, applied forces, and loading frequencies on real contact area evolution and efficiency of tribocharge generation. Recent studies have increasingly leveraged engineered surface properties to enhance TENG performance, showing promising results through microscopic and controlled roughness surface treatment~\citep{zhao2016size, choi2016high, pan2018fully, sriphan2018facile, ameer2022influence, sun20223d, zamani2024high, kumar2023multiscale}. 
In various experimental protocols, a clear dependence between triboelectric charge and the real contact area has been demonstrated, which in turn significantly affects TENG performance \citep{li2022numerical, gomes2018influence, jin2016contact}. Thus, accurate prediction of both the charge quantity and its spatial distribution is crucial.
Notably,~\cite{sriphan2018facile} reported increased electrical output from surfaces with increased roughness, while \cite{jiang2023nanocavities} demonstrated that rough surfaces with nano-cavities stored significantly more charge, boosting power output by 10,000 times compared to smooth surfaces. In contrast, a higher voltage was observed with a lower surface roughness when a soft and rough PVS layer contacting against a rigid-flat mica surface in an experiment by~\cite{kumar2023multiscale}. These findings reflect the interconnected, often non-intuitive relationship between surface morphology and TENG performance, underscoring the need for detailed investigations of roughness configurations and accurate computation of contact area and charge for TENG.

From an engineering perspective, all real surfaces exhibit roughness at a certain scale. Early analytical studies laid the foundation for understanding the evolution of contact area increasing pressure using statistical multi-asperity models, such as the BGT model \citep{bush1975elastic} and the GW model \citep{greenwood1966contact}. Recent work on TENG has incorporated surface roughness effects using Persson’s contact model, which does not rely on assumptions on the geometry or distribution of asperities \citep{persson2002elastic,persson2006contact}, and predicted a higher output voltage as surface roughness decreases~\citep{wen2021load,XUVoc}. However, analytical contact models, including Persson's, were derived under the assumption of small slopes of the surface and a certain, e.g., Gaussian, distribution of surface heights. In cases of large slopes or tuned/patterned or otherwise engineered surfaces, these models can be inaccurate. Moreover, the particular structure of TENG, which may include multiple layers made of different materials, potentially with inclusions, requires heterogeneous models, for which analytical approaches are not designed. This limits their applicability for the accurate prediction of the contact area for innovative TENG designs.

To overcome these limitations, numerical methods have become increasingly influential for predicting the contact area between rough surfaces. The boundary element method (BEM) has demonstrated rigorous computational efficiency for computing contact area, producing results that fall between those predicted by Persson’s and BGT models under linear elasticity assumptions~\citep{frerot2020tamaas, yastrebov2017accurate, rottger2022contact}. Nevertheless, similarly to analytical models, BEM faces limitations when applied to realistic, non-periodic, and highly nonlinear contact problems. At the same time, FEM simulations, which are free from assumptions on the surface geometry or material properties, have proven effective for modelling mechanical contact between rough surfaces \citep{hyun2004finite} and have gained popularity with the growth of computational power \citep{yastrebov2011computational,carvalho2022efficient,MARULLI2025118200}. Moreover, FEM models have been applied successfully to characterise contact interactions at rough interfaces in coupled problems~\citep{shvarts2021computational}. Critically, FEM enables spatially resolved computation of deformation gradients and contact area across the tribolayers, which is an important consideration not only for contact mechanics but also for electromechanical coupling mechanisms~\cite{deng2023mixed}. 

\subsection{Objectives of this paper}
This work proposes a tool for advancing the understanding of TENGs by numerically resolving the evolution of real contact area through contact mechanics simulations, incorporating experimentally measured surface roughness and by coupling this with the electrical response. An open-source finite element framework is developed that robustly couples the mechanical contact problem with load-dependent electrostatic interactions, including a rigorous formulation of the floating electrode and an ODE-based circuit model for predicting time-dependent electrical output. The resulting numerical predictions are validated through comparison with available approximate analytical solutions and experimental measurements to demonstrate the accuracy and predictive capability of the proposed approach.

In this paper, Section \ref{sec:Introduction} introduces the existing analytical approximations and methodology for TENGs and defines the need for a TENG numerical modelling framework for surface roughness characterisation. Following this, Section \ref{sec:Unified computational framework for TENG} presents the unified modelling methodology and workflow, detailing the finite element formulation for rough-surface contact mechanics, the electrostatic field computation incorporating fringing effects, and an ODE-based circuit model used to predict transient electrical output. Section \ref{sec:Result_and_Discussion} demonstrates numerical results, including the validation of the contact mechanics model against experimental measurements and classical contact theories, highlighting the role of contact area evolution, followed by electrostatic analyses that compare FEM predictions with existing analytical models and experimental observations. Finally, after the concluding remarks, Section \ref{app1} details the description of the experimental setup designed for the numerical analysis and for the validation of the proposed framework. 

\section{Unified computational framework for TENG}
\label{sec:Unified computational framework for TENG}

\begin{figure}[!b]
    \centering
    \includegraphics[width=\linewidth]{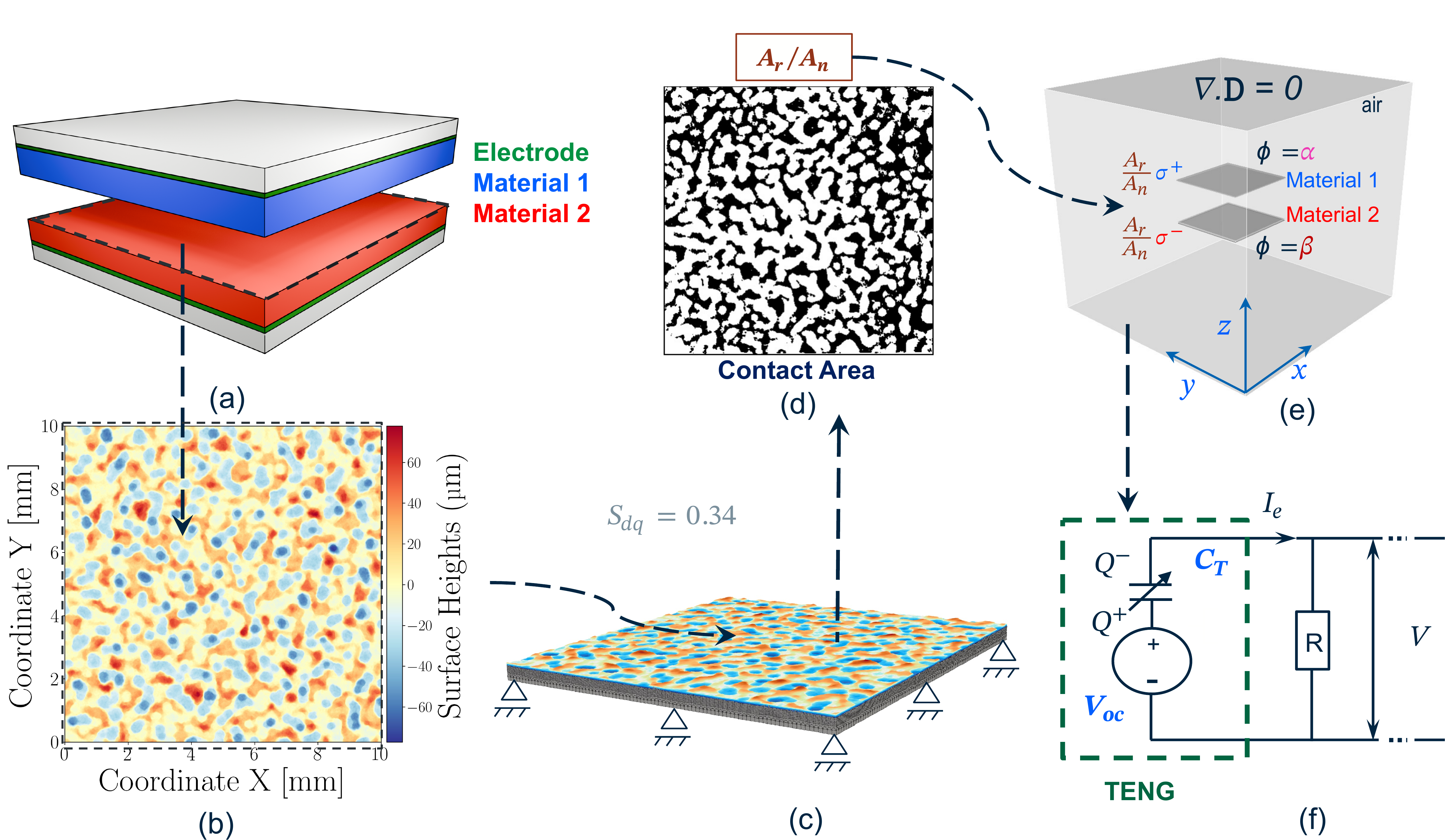}
    \caption{
Schematic of the simulation pipeline for TENG accounting for surface roughness, showing: 
(a) Layered TENG structure substrate (grey) at the top and bottom with attached electrodes (green) and triboelectric materials 1 and 2 in between. 
(b) Combined rough surface topography of both layers.
(c) 3D finite element mesh upon projecting the measured surface height, used to compute the mechanical contact response upon indentation by a rigid counter-surface. 
(d) simulation result showing contact (black) and non-contact (white) zones, permitting the computation of $A_r/A_n$.
(e) Electrostatic model with the TENG layers containing surface charge densities $\sigma^{\pm}$ scaled by the contact area fraction to evaluate the $V_{OC}$ and $C_T$. 
(f) Equivalent circuit model of the TENG, comprising a time-dependent $V_{OC}$ and $C_T$, delivering an output voltage $V$ and short-circuit current $I_{e}$ across a resistive load.
}
\label{fig:workflow}
\end{figure}

Although the TENG contact-separation mode appears well-understood, its performance is determined by complex multiscale interactions among 3D rough surfaces, electrostatic fields, and circuit dynamics. To capture this multi-physics, an accurate, robust and scalable numerical framework has been developed to analyse TENG within the open-source finite-element library MoFEM \citep{kaczmarczyk2020mofem}. The proposed framework incorporates explicit surface roughness modelling, enabling accurate representation of complex interface geometries, material heterogeneities, and non-uniform contact conditions, which are essential for realistic TENG analysis.

\subsection{Resolution of the coupled multiscale problem}
\label{sec:multiscale}
In this scheme, the workflow has been structured into three distinct subsequent steps: mechanical contact analysis, electrostatics simulation, and electric circuit integration. The complete simulation workflow is illustrated in \textbf{Figure ~\ref{fig:workflow}}.

To perform the contact area computation, the framework requires surface data in the form of surface height at the exact coordinates of the real/random or engineered surface, which can be extracted, e.g., using an optical profilometer or functionally produced. The microstructural characteristics of the contact interface are accurately captured by maintaining the unique roughness profile of the real surface in the numerical model. The contact problem is modelled by considering a deformable TENG layer pressed against a rigid plane, where the surface roughness of the deformable body represents the effective roughness of the two tribolayers and the solid’s elastic modulus captures the effective modulus of the contacting pair. Such an approach permits simplifying the contact problem without loss of accuracy for the cases of sufficiently small surface slope and linear elastic response of the material~\citep{barber2003bounds, persson2006contact, XUVoc}. These conditions hold for the experiments considered in this paper for the verification of the proposed framework and are representative of those typically encountered in TENGs. Subsequently, the height measurements are mapped for each surface node and scaled along the height of the volumetric mesh, which represents the deformable tribolayer. Afterwards, the simulation is performed with this layer pressed against a rigid flat to analyse the evolution of the real contact area ratio, $A_r/A_n$, which depends on the applied load. It is important to note that the proposed FE framework allows a range of extensions, such as modelling a rough rigid surface or incorporating inelastic material behaviour, enabling its application to innovative TENG design.

Following the mechanical analysis, a full-scale electrostatic FE model is formulated in which the tribolayers are separated by arbitrary air gaps over the full length of the nominally flat TENG surfaces possessing the specific tribocharge density. Critically, for a given mechanical load applied during contact, the surface charge density is scaled by the $A_r/A_n$ ratio computed at the contact stage. This linear scaling of $\sigma_T$ has been validated across charged patches of varying sizes, from full-scale regions down to small-scale patches with different sizes representative of the contacting rough surface. This homogenisation approach was explained in detail in section SI-3 in the supporting information (Figure S2).

After conducting the electrostatic analysis to obtain $V_{OC}$ and $C_T$ for a given air gap, defined as a time-dependent displacement $z(t)$, the transferred charges and voltages are calculated by solving the TENG circuit ordinary differential equation. This allows the framework to predict the electrical performance of the TENG with different materials, surface roughness, mechanical motion, frequency load, and circuit characteristics. To aid in reproducibility, the simulation pipeline implemented in MoFEM~\citep{kaczmarczyk2020mofem} is described in section SI-1 of the supplementary information.

\subsection{Mechanical contact modelling of TENG}
\begin{figure}[t]
    \centering
    \includegraphics[width=0.7\linewidth]{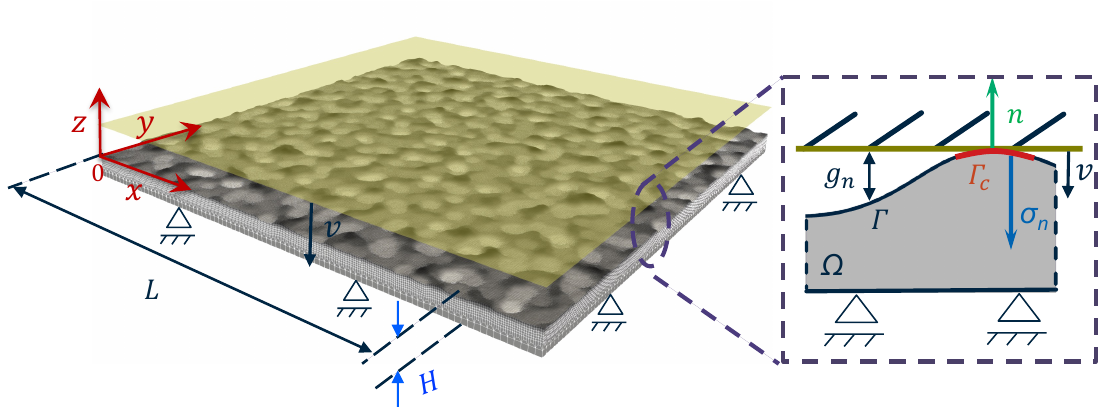}
    \caption{Schematic of the contact problem with a rigid surface indenting on a deformable body with surface roughness, where fixed boundary conditions are applied at the bottom in addition to the symmetric boundary conditions to the vertical faces. A rough surface topology is introduced at the top of the deformable layer, where the active contact zone $\Gamma_c$ is a subset of the potential contact zone $\Gamma$ such that $\Gamma_c \subseteq \Gamma$.}
    \label{fig:contact_and_area}
\end{figure}

Under the assumptions of a frictionless, non-adhesive TENG contact between a deformable layer and a rigid plane (see \textbf{Figure~\ref{fig:contact_and_area}}), the problem is described as follows, where fixed boundary conditions are applied at the bottom in addition to the symmetric boundary conditions on the vertical faces of TENG, such that:
\begin{subequations}\label{eq:balance_momentum}
\begin{empheq}[left=\empheqlbrace]{align}
&\nabla \cdot \boldsymbol{\sigma} (\mathbf{u}) = 0, 
  && \text{in } \Omega := (0,L)\times(0,L)\times(-H,0), \label{eq:balance_of_linear_momentum}\\
&g_n (\mathbf{u}) \geq 0,\;\; p_n \leq 0,\;\; p_n g_n (\mathbf{u}) = 0, 
  && \text{on } \Gamma, \label{eq:KKT_conditions} \\
&u_x = u_y = u_z = 0, 
  && \text{on } z=-H, \\
&u_x = 0, 
  && \text{on } x=0,\,x=L, \label{eq:bc_bottom} \\
&u_y = 0, 
  && \text{on } y=0,\,y=L. \label{eq:bc_xsym}
\end{empheq}
\end{subequations}
where the Cauchy stress tensor $\mathbf{\sigma}$ in equation \eqref{eq:balance_of_linear_momentum} depends on the displacement field $\mathbf{u}$ via a linear elastic constitutive law, such that $\mathbf{\sigma} = \mathbf{C}: \nabla^s \mathbf{u}$, where $\mathbf{C}$ is the fourth-order elasticity tensor and $\nabla^s \mathbf{u}$ denotes the symmetric gradient of $\mathbf{u}$. Notably, the framework is equally applicable for nonlinear materials, e.g., hyperelasticity, plasticity, or viscoelasticity.

The contact interactions on the potential contact boundary, $\Gamma$ are formulated using the Karush–Kuhn–Tucker (KKT) conditions as in equation \eqref{eq:KKT_conditions}, which impose complementarity constraints between the normal gap function $g_n$ and the contact pressure $p_n$~\citep{yastrebov2013numerical}. Here, $g_n(u)$ denotes the gap that is defined as the signed normal distance between the potential contact surface of the deformable body and the rigid body, such that: 
\begin{equation}
    g_n(\mathbf{u}) = g_0 + \mathbf{u\cdot v} - \mathbf{c}(t)\cdot \mathbf{v},
\end{equation}
where $g_0$ is defined as the initial gap, $\mathbf{v}$ is the outward normal to the rigid flat and $\mathbf{c}(t)$ is the translation vector of the rigid body. Thus, a positive $g_n$ indicates an open gap and  $g_n = 0$ means contact. These cases are briefly tabulated in \textbf{Table \ref{tab:KKT}}.

\begin{table}[h]
\centering
\caption{Physical complementarity of the contact condition}
\begin{tabular}[htbp]{@{}l l l}
\hline\hline
\textbf{Situation} & \textbf{Gap} & \textbf{Pressure} \\
\hline

No contact on $\Gamma \setminus \Gamma_c$ & $g_n > 0$ & $p_n = 0$ \\
Contact on $\Gamma_c$ & $g_n = 0$ & $p_n < 0$ \\
\hline
\hline
\end{tabular}
\label{tab:KKT}
\end{table}


Subsequently, the contact KKT conditions \eqref{eq:KKT_conditions} have been enforced by the complementarity function~\citep{hueber2005primal}: 
\begin{equation}
\label{eq:c_fun}
C\bigl(g_n(\mathbf{u}), p_n\bigr) = \frac{1}{2}\left(c_n g_n - p_n - |p_n+c_n g_n|\right) = 
\begin{cases}
  c_n g_n, & \text{if}\; p_n \leq -c_n g_n, \\
  -p_n, & \text{if}\; p_n > -c_n g_n.
\end{cases}
\end{equation}

The KKT conditions are fulfilled when $C\bigl(g_n, p_n\bigr) = 0$ in equation \eqref{eq:c_fun}, where $c_n$ is a regularisation (augmentation) parameter which does not affect the result and is only used to improve the convergence of the nonlinear (e.g. Newton-Raphson) finite element solver required for the contact problem \eqref{eq:balance_momentum}~\citep{yastrebov2013numerical}. In most FE solvers, the contact pressure $p_n$ appearing in~\eqref{eq:c_fun} is typically approximated using scalar~\citep{hueber2005primal} or vectorial~\citep{gitterle2010finite} Lagrange Multipliers. In the proposed framework, a novel approach is employed, where the contact pressure $p_n$ is considered as the normal component of the normal trace of a tensorial field of Lagrange multipliers, which is approximated using a Raviart-Thomas $H(\text{div})$ FE approximation space constructed on elements adjacent to the contact boundary~\citep{shvarts2026_arbitrary_order}. Such an approach ensures stability and accuracy of the solution, particularly in the computation of the extent of the real contact area using FEM, discussed in subsection \ref{sec:Contact area calculation}. Moreover, this approach is highly scalable and suited to high-performance computing (HPC) environments, enabling the simulation of the large-scale problems required for the explicit representation of surface roughness.

\subsubsection{Contact area calculation}
\label{sec:Contact area calculation}
\begin{figure}[t]
        \centering
        \includegraphics[width=0.34\linewidth]{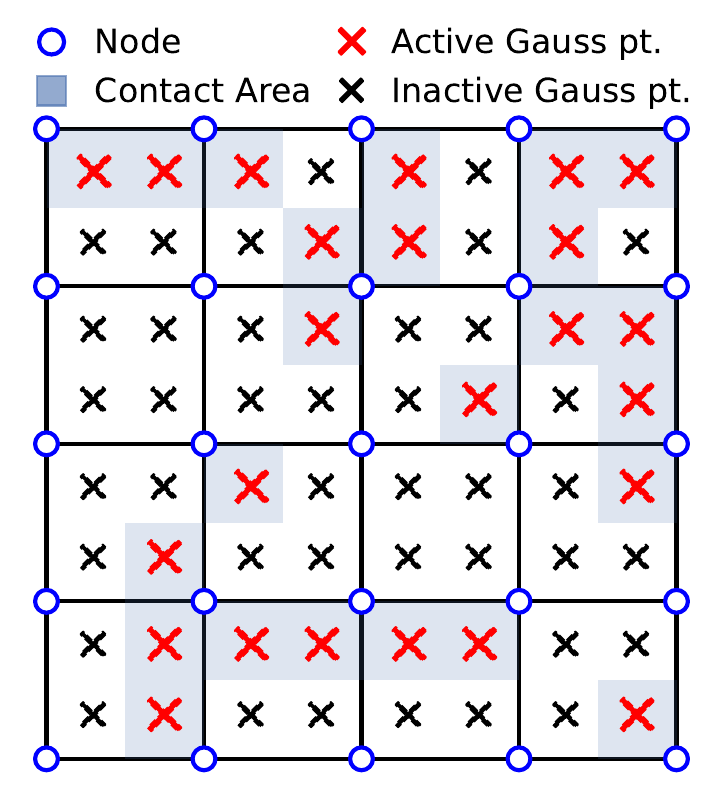}
    \caption{ Contact area computation using FE discretisation of the contact interface, showing quadrilateral contact elements. Numerical evaluation of the real contact area (shaded in grey) is obtained by summing area contributions associated with active contact Gauss points (highlighted in red).}
    \label{fig:contact_and_area_computation}
\end{figure}

The real contact area depends on the gap distribution, the applied load, and the geometry of the TENG layers. Its accurate computation is challenging in the discretised setting~\citep{yastrebov2017accurate}. Following the numerical strategy initially proposed by~\cite{oliveira2008algorithms} and later adopted in \cite{SHVARTS2021113738}, the real contact area is computed by summing up the areas corresponding to all active integration (Gauss) points across the elements in contact (see \textbf{Figure ~\ref{fig:contact_and_area_computation}}).
However, in contrast to the previous works, the proposed formulation considers the status of the Gauss points, which is determined based on the values of the fields (gap, contact pressure) at these points. Specifically, in agreement with the definition of the complementarity function~\eqref{eq:c_fun}, where the Gauss points are classified as \emph{active} if the approximated value of $p_n + c_n g_n \leq 0$ and \emph{inactive} otherwise, the total active real contact area is defined as follows:
\begin{equation}
A_{\mathrm{real}} =
\sum_{e \in \mathcal{C}}
\sum_{i \in \mathcal{G}_e}
\chi_i (g_{ni}, p_{ni})\, w_i \, J_i, \quad \chi_i = 
\begin{cases}
  1, & \text{if}\; p_{ni} \leq -c_n g_{ni}, \\
  0, & \text{if}\; p_{ni} > -c_n g_{ni};
\end{cases}
\end{equation}
where $\mathcal{C}$ denotes the set of contact elements, $\mathcal{G}_e$ is the set of Gauss points associated with element $e$, $w_i$ is the quadrature weight corresponding to $i$-th Gauss point of the element, $J_i$ is the Jacobian determinant evaluated at this point, and $\chi_i$ is an indicator function depending on the values of gap $g_{ni}$ and contact pressure $p_{ni}$ computed at the same point using the FE approximation. Moreover, in the current work, to increase the accuracy of the real contact area computation, the calculation of the value of Jacobian $J_i$ at the $i$-th Gauss point of the element is performed in the current (deformed) configuration.

\subsection{Electrostatics modelling of TENG}
\label{Electrostatics modelling of TENG}
\begin{figure}[t]
    \centering
    \includegraphics[width=0.37\linewidth]{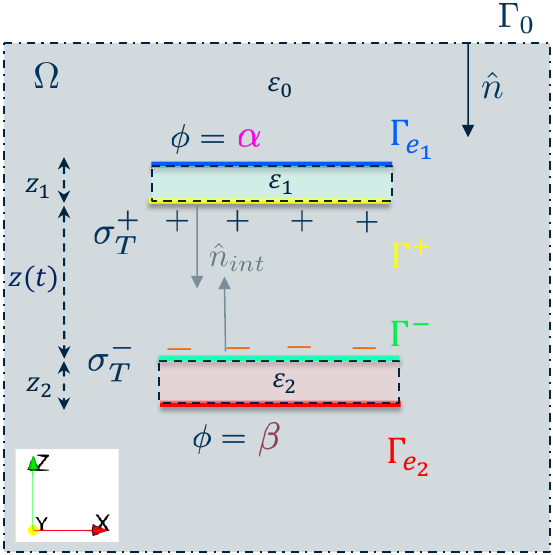}
    \caption{Schematic of TENG electrostatic problem setup comprising two dielectric tribolayers 
    ($\epsilon_1$, $\epsilon_2$) separated by air ($\epsilon_0$). 
    The charge densities $\sigma_T^+$ and $\sigma_T^-$ 
    reside on the interfaces $\Gamma^+$ and $\Gamma^-$, respectively. 
    Thin electrodes are modelled as boundaries $\Gamma_{e1}$ and $\Gamma_{e2}$ respectively with constant potentials $\alpha$ and $\beta$. 
    The external boundary $\Gamma_0$ represents the far field and is treated 
    as a grounded surface.}
    \label{fig:electrostatic_model}
\end{figure}

The fundamental subset of Maxwell's equations \cite{huray2009maxwell} governs the behaviour of stationary or slow-moving electric charges. Importantly, TENG layers are typically thin, and the charges appear only on the surfaces. Consequently, the electrostatic problem has been formulated under the quasi-static approximation, with no volumetric charge or magnetic effects. Primarily, the objective of the electrostatics analysis of TENG with interface conditions is to find the accurate potential difference between the electrodes when $\sigma_T^+$ and $\sigma_T^-$ are present on the interfaces, considering the air gap between, as shown in \textbf{Figure \ref{fig:electrostatic_model}}. Therefore, the electrostatic problem is formulated for the potential $\phi$ in domain $\Omega$, including unknown constant potentials $\alpha$ and $\beta$ on the electrodes, such that:
\begin{subequations}\label{eq:electrostatics_bvp}
\begin{empheq}[left=\empheqlbrace]{align}
&\nabla \cdot \mathbf{D} = 0 
  && \text{in } \Omega, \label{eq:poisson_equation} \\
&\phi = 0 
  && \text{on } \Gamma_0, \label{eq:Dirichlet}\\
&\llbracket \mathbf{D} \rrbracket \cdot \hat{\mathbf{n}}_{\text{int}} 
  = \pm\,\sigma_\text{T} \tfrac{A_r}{A_n} 
  && \text{on } \Gamma^{\pm}, \label{eq:inteface_condition}\\[6pt]
&\phi = \alpha 
  && \text{on } \Gamma_{e1},\label{eq:contstant_pot} \\
&\phi = \beta 
  && \text{on } \Gamma_{e2}, \\
&\int_{\Gamma_{e1}} \llbracket \mathbf{D} \rrbracket \cdot \hat{\mathbf{n}} \, d\Gamma_{e1}
  = -Q
  && \text{on } \Gamma_{e1}, \label{eq:floating_electrode_cond_1}\\
&\int_{\Gamma_{e2}} \llbracket \mathbf{D} \rrbracket \cdot \hat{\mathbf{n}} \, d\Gamma_{e2} 
  = +Q
  && \text{on } \Gamma_{e2}.\label{eq:floating_electrode_cond_2}
\end{empheq}
\end{subequations}

 \begin{figure}[t]
        \centering
        \includegraphics[width=0.28\linewidth]{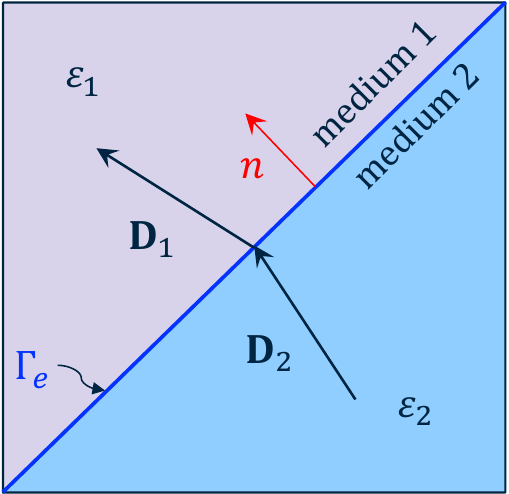}
        \caption{ Electric displacement jump across an interface between two materials with different permittivity}
        \label{fig:EDisplacemnt}
\end{figure}
In equation \eqref{eq:electrostatics_bvp}, $\hat{\mathbf{n}}$ and $\hat{\mathbf{n}}_{int}$ denote the unit normal vectors to the $\Gamma_0$ and $\Gamma^{\pm}$ respectively. The constitutive relation for the Poisson equation~\eqref{eq:poisson_equation} connects the electric field, $\mathbf{E} = - \nabla \phi$ and the electric displacement, $\mathbf{D} = \varepsilon_0 \varepsilon_r  \mathbf{E}$, where $\varepsilon_0$ and $\varepsilon_r$ represent absolute and relative dielectric permittivities, respectively. The interface conditions~\eqref{eq:inteface_condition} follow from Gauss’s law, and require computing $\llbracket\mathbf{D}\rrbracket$, which represents the jump of the electric displacement field across an interface between two materials, as shown in \textbf{Figure \ref{fig:EDisplacemnt}}. Numerically evaluating this quantity is non-trivial since the gradient of the potential needs to be computed from FE shape functions of volume elements, as shape functions defined on the interface do not directly provide gradients. To address this, the approach proposed in \cite{ATHANASIADIS2023116129} is adopted, where the potential gradients at the interface element's integration points are reconstructed using the field approximations from the adjacent volume elements. Importantly, the interface conditions in equation~\eqref{eq:inteface_condition} for tribo-layers $\Gamma^{\pm}$ account for positive and negative surface charge density $\sigma_T$ and are scaled by the real contact area ratio. 

Furthermore, the condition~\eqref{eq:contstant_pot} represents the floating electrode behaviour \citep{amann2014simulation}, where the potential remains constant over the surfaces, but the constants (suppose $\alpha$ and $\beta$) need to be calculated.
In addition, the electrode boundary conditions are applied according to equations~\eqref{eq:contstant_pot}--\eqref{eq:floating_electrode_cond_2} for the top and bottom surfaces of the TENG, where the total surface charge induced on each electrode is equal to an integral over the normal component of $\llbracket\mathbf{D}\rrbracket$, which follows from Gauss’s law applied to an interface containing surface charge \citep{griffiths_2017, jackson1999classical}. In this regard, when a non-zero charge $Q$ is present on the electrodes, the flux through the electrode boundaries is defined as in Equations~\eqref{eq:floating_electrode_cond_1} and \eqref{eq:floating_electrode_cond_2}. In contrast, when no charge is transferred in the external circuit, i.e. $Q = 0$, total electric flux through the electrode vanishes. 

\subsubsection{Calculation of open-circuit voltage}

When the triboelectric layers are separated under the open-circuit conditions, they produce $V_{OC}$ due to the presence of $\pm\sigma_T$ on tribosurfaces. At this stage, since the circuit is not closed, no charge transfers through the circuit. Thus, to calculate the values of the constant potential due to $\pm\sigma_T$ while $Q=0$ on the electrodes, given the linearity of the problem, the superposition principle can be used. Three auxiliary potential solutions  $\phi_a(x)$, $\phi_b(x)$, and $\phi_c(x)$ are introduced, where each solution satisfies Equation \eqref{eq:poisson_equation}, boundary conditions in Equations \eqref{eq:poisson_equation}-\eqref{eq:Dirichlet}  and distinct boundary conditions on $\Gamma_{e_1}$ and $\Gamma_{e_2}$, namely:

\begin{subequations}
\label{eq:solset}
\begin{equation}
\phi_a(x) =
\begin{cases}
  0 & \text{on }\Gamma_{e_1}, \\
  0 & \text{on }\Gamma_{e_2},
\end{cases}
\end{equation}

\begin{equation}
\phi_b(x) =
\begin{cases}
  1 & \text{on }\Gamma_{e_1}, \\
  0 & \text{on }\Gamma_{e_2},
\end{cases}
\end{equation}

\begin{equation}
\phi_c(x) =
\begin{cases}
  0 & \text{on }\Gamma_{e_1}, \\
  1 & \text{on }\Gamma_{e_2}.
\end{cases}
\end{equation}
\end{subequations}

A linear combination of the above solutions with unknown scalar coefficients $\alpha$ and $\beta$ is ultimately the solution of the problem~\eqref{eq:electrostatics_bvp} when:
\begin{equation}
\label{eq:linear_combination}
    \phi(x) = \phi_a(x) + \alpha \left[\phi_b(x) - \phi_a(x)\right] + \beta \left[\phi_c(x) - \phi_a(x)\right],
\end{equation}
where $\alpha$ and $\beta$ respectively provide the constant potential values at the top and bottom electrode. Thus, once the auxiliary solutions $\phi_a(x)$, $\phi_b(x)$, and $\phi_c(x)$ are obtained, substitution of the linear combination from Equation~\eqref{eq:linear_combination} into Conditions~\eqref{eq:floating_electrode_cond_1} and~\eqref{eq:floating_electrode_cond_2} leads to the following system of two linear equations:
\begin{equation}\label{eqn:EDequations1}
\begin{bmatrix}
B_1 - A_1 & C_1 - A_1 \\
B_2 - A_2 & C_2 - A_2
\end{bmatrix}
\begin{bmatrix}
\alpha \\
\beta
\end{bmatrix}
=
-
\begin{bmatrix}
A_1 \\
A_2
\end{bmatrix},
\end{equation}
If $i = 1, 2$ is considered as the electrode number, then the terms in the above system can be considered as $A_i$, $B_i$, and $C_i$ where,
\begin{equation}
\label{eq:system_defs}
A_i := \int_{\Gamma_{e_i}} \llbracket \mathbf{D}_a \rrbracket \cdot \hat{\mathbf n}\, d\Gamma,
\qquad
B_i := \int_{\Gamma_{e_i}} \llbracket \mathbf{D}_b \rrbracket \cdot \hat{\mathbf n}\, d\Gamma,
\qquad
C_i := \int_{\Gamma_{e_i}} \llbracket \mathbf{D}_c \rrbracket \cdot \hat{\mathbf n}\, d\Gamma,
\end{equation}
In the above,  $\llbracket\mathbf{D_a}\rrbracket$,  $\llbracket\mathbf{D_b}\rrbracket$, and $ \llbracket\mathbf{D_c}\rrbracket$ represent the jumps of the electric displacement at the respective electrode boundary $\Gamma_{e_i}$ which are calculated from the solutions $\phi_a(x)$, $\phi_b(x)$ and $\phi_c(x)$. Once the system~\eqref{eqn:EDequations1} is solved for two unknowns $\alpha$ and $\beta$, the difference between them provides $V_{OC}$ such that $V_{OC} = \alpha - \beta$.

\subsubsection{Calculation of capacitance}
\label{sec:capacitance_computation}
The device capacitance $C_T$ can readily be calculated using the charge conservation relation, such that:

\begin{equation}
C_T = \frac{Q}{V_{\text{out}}},
\end{equation}

In the above, the induced potential difference $V_{\text{out}}$ is obtained from the governing equation and boundary conditions given in \eqref{eq:electrostatics_bvp}. The $V_{\text{out}}$ calculation follows an approach analogous to that used for $V_{OC}$; however, at this stage, the charges are transferred to the electrodes $(Q \neq 0)$ while the tribocharges are not considered, i.e.
$\sigma_T = 0$. Consequently, the solution $\phi_a$ = 0 in Equation~\eqref{eq:linear_combination}, and therefore $A_i =0$, see Equation~\eqref{eq:system_defs}. This leads to the following system of equations for recalculating $\alpha$ and $\beta$:
 
\begin{equation}\label{eqn:EDequations_capa}
\begin{bmatrix}
B_1 & C_1  \\
B_2 & C_2 
\end{bmatrix}
\begin{bmatrix}
\alpha \\
\beta
\end{bmatrix}
=
\begin{bmatrix}
-Q \\
\,\,\,Q
\end{bmatrix},
\end{equation}
such that $V_{\text{out}} = \alpha - \beta$.

\subsection{Electric circuit ODE and time dependency}
\label{subsection:circuit_ODE}
In most cases, TENGs are investigated as dynamic systems with periodic time-dependent mechanical excitation. Thus, the gap $z(t)$ is introduced as a periodic function that represents the loading and unloading stages. Throughout the remaining work, the function $z(t)$ has been defined as a periodic function expressed as: 
\begin{equation}
z(t) =  z_{max}(1 - \cos \omega t)
\end{equation}

where $\omega = 2\pi/T$, and $T$ denote the time taken to complete one full cycle, i.e. the inverse of the tapping frequency $f$. The $z_{max}$ is the maximum separation gap over which the device operates.

In the proposed modelling workflow, FEM simulations are performed to compute $V_{OC}$ and $C_T$ for a discrete range of air gap values from $[0, z_{max}]$. These quantities are then interpolated to obtain their continuous representation for all values of $z(t)$ using a 1D smoothing spline based on the Univariate Spline approach \cite{dierckx1981improved}. 

The fundamental V-Q-z relation in equation \eqref{eq:basic} for TENG consists of two main components: (i) the $V_{oc}$ generated due to $\sigma_T$ and (ii) internal capacitance due to the induced charges. To analyse arbitrary external circuit configurations, it is required to compute net voltage $V(t) = I_e R_L$ over time from the TENG internal circuit as shown in \textbf{Figure~\ref{fig:circuiteqv}}. Here, $I_e$ and $R_L$ are the current and resistance on the external circuit of TENG. Applying Kirchhoff’s law to the equivalent TENG circuit yields the following first-order ordinary differential equation:
\begin{figure}[t]
    \centering
    \includegraphics[width=0.35\linewidth]{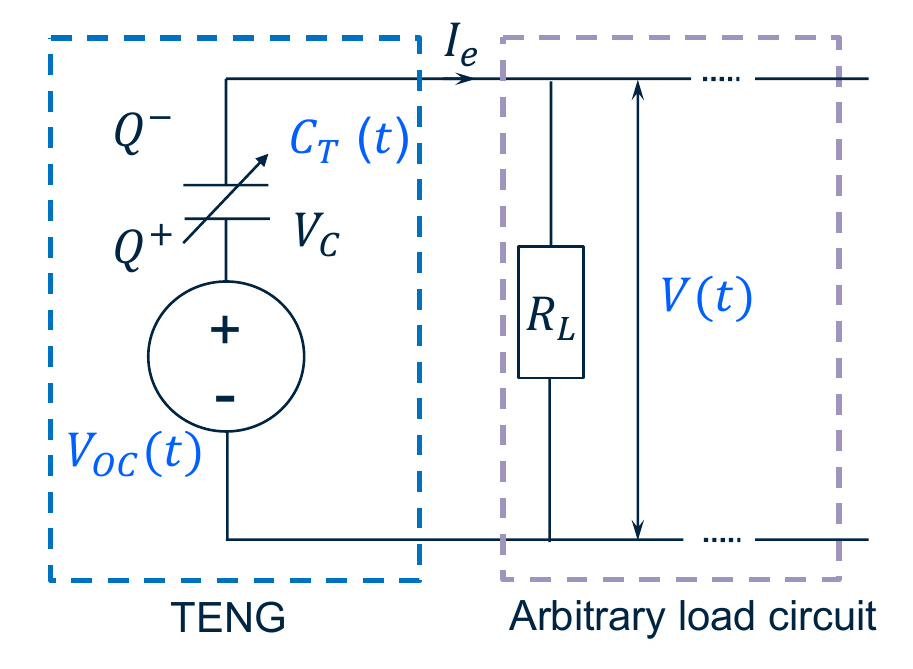}
    \caption{Equivalent TENG circuit.}
    \label{fig:circuiteqv}
\end{figure}
\begin{equation}
    \frac{dQ(t)}{dt} R_L = -\frac{Q(t)}{C_T(t)} + V_{OC}(t),
    \label{circuit_ODE}
\end{equation}
The ODE in Equation \eqref{circuit_ODE} is solved numerically using the implicit Runge-Kutta method implemented in the function \textit{solve\_ivp} of the SciPy library (specifically, employing the `Radau' solver). The result governs the time evolution of the transferred charge $Q(t)$ in the external load circuit. 

\section{Results and Discussion}
\label{sec:Result_and_Discussion}
\begin{figure}[!b]
    \centering
     \begin{minipage}{0.57\linewidth}
        \centering
        \includegraphics[width=\linewidth]{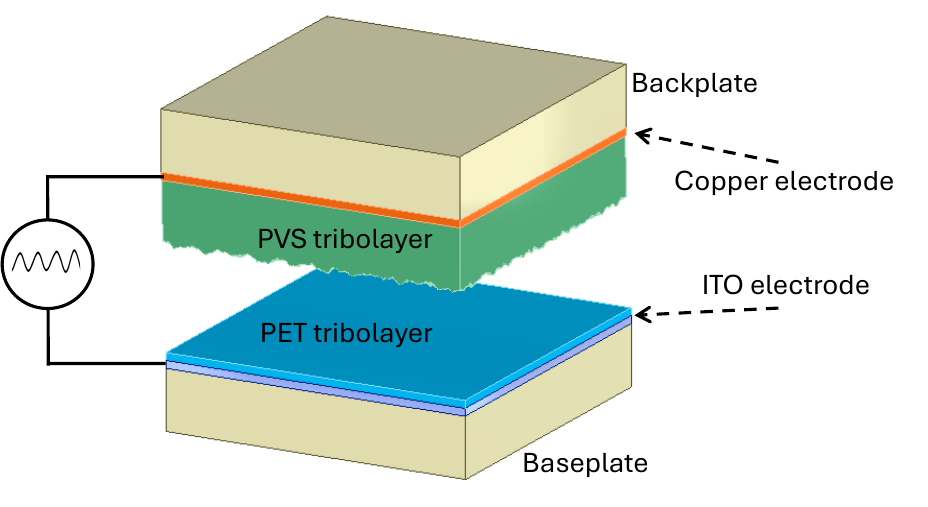}
        \caption*{(a)}
    \end{minipage}
    \begin{minipage}{0.34\linewidth}
    \hfill
    \includegraphics[width=\linewidth]{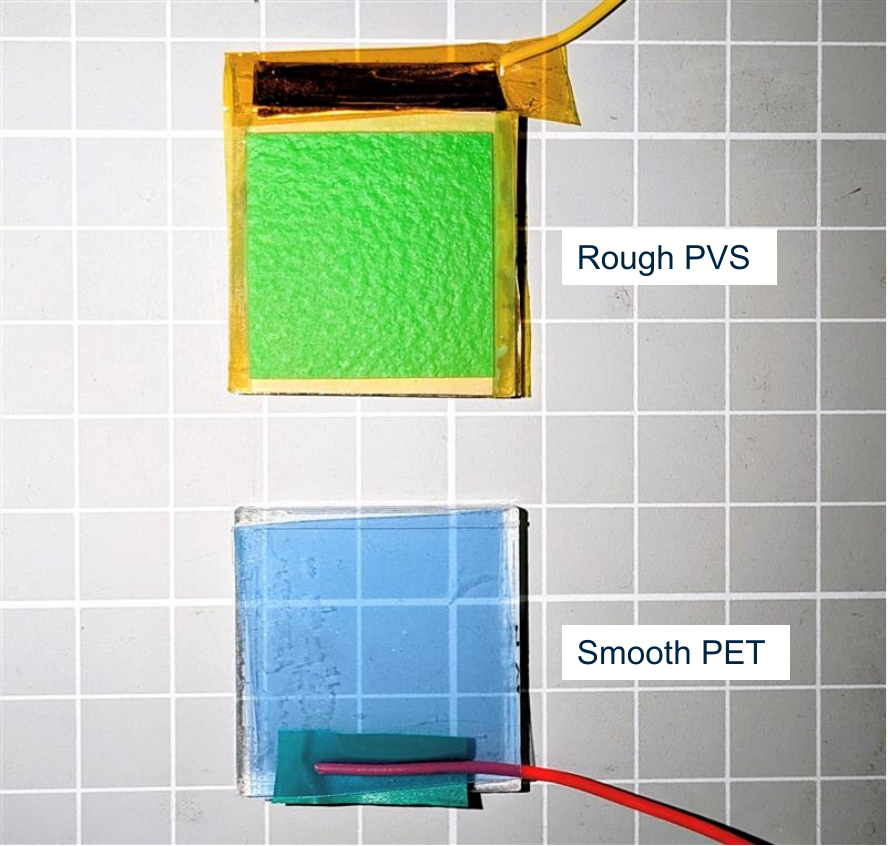}
    \caption*{(b)}
    \end{minipage}
    \caption{Device architecture of the contact-separation PVS-PET TENG showing (a) a schematic diagram of the TENG layers connected to the circuit and (b) experimental TENG tribo-layers, PVS and PET, with the attached copper electrode and ITO coating, respectively (top view).}
  \label{fig:TENG_Setup}
\end{figure}

This section presents the application of the proposed numerical framework to simulate the performance of the TENG setup shown in \textbf{Figure~\ref{fig:TENG_Setup}}. This contact-separation TENG consists of two dielectric layers: a rough PVS layer on the top and a smooth PET layer at the bottom, separated by an air gap. Each tribolayer is supported by a rigid substrate at the back, referred to as the ``Backplate'' and ``Baseplate'', respectively. A copper electrode is embedded into the interface between the PVS and its substrate, while the PET is coated with indium tin oxide (ITO) on the side facing its substrate. During operation, the air gap varies dynamically according to $z_t \in [0, z_{\max}]$, where $z_{\max}$ denotes the maximum separation distance between the tribolayers. 

Notably, during FEM simulation, the substrates have been treated as rigid, since their Young's modulus, $E$ is several orders of magnitude higher than the effective elastic modulus of the contact tribolayer pair. Furthermore, the electrode layers are orders of magnitude thinner than the dielectric layers and also possess significantly higher elastic stiffness compared to the compliant tribo-layer. Therefore, the electrodes have been modelled as infinitesimally thin.

To facilitate the experimental validation, simulations are carried out for a model of a real TENG device fabricated specifically for this study (see Section~\ref{app1}). To demonstrate the capability of the framework in capturing the effects of surface roughness and applied mechanical load on the device output, several distinct experimental configurations were considered. The corresponding experimental protocols are summarised in \textbf{Table \ref{tab:Exp_catagories}}.

\begin{table}[t]
\centering
\caption{List of experimental configurations and measured quantities}
\renewcommand{\arraystretch}{1.5}
\begin{tabular}{p{0.05\linewidth} p{0.13\linewidth} p{0.45\linewidth} p{0.22\linewidth}}
\hline\hline
\textbf{Setup} & \textbf{Experiment} & \textbf{Configuration / Method} & \textbf{Purpose/Use} \\
\hline

A 
& Interference Reflection Microscopy (IRM) 
& A $10\times10\,\mathrm{mm}^2$ rough PVS pressed against PET under controlled load. The contact interface has been captured using an IRM-based CMOS camera (see Section ~\ref{app1.2}). The material properties are listed in \textbf{Table~\ref{tab:Mech_parameters}}.
& Validation of the contact area computed by the FEM contact model under different loads.\\

B 
& Surface charge measurement
& Charge transfer was measured between the same samples as in setup A over the real contact area (see Figure~\ref{fig:TENG_Setup}). The measurement procedure follows~\cite{kumar2026_truetribocharge} and properties are noted in \textbf{Table~\ref{tab:Elecmaterial_properties}} and in Section~\ref{app1.3}.
& Calculation of $\sigma_T$ for FEM input in electrostatic simulations. \\

C 
& Capacitance measurement
& Capacitance has been measured between two 25$\times$25\,$\mathrm{mm}^2$ ITO electrodes forming a capacitor structure. 
& Validation of the electrostatic capacitance model. \\

D 
& Load-dependent $V_{\mathrm{OC}}$ measurement
&  Data from Xu et al. \cite{XUVoc} for 25$\times$25 $\mathrm{mm}^2$ PET-PDMS with composite $S_{dq} = 0.22$, $\sigma_T^\pm = \pm 40.7 \mu C/m^2$. 
&  Validation of load-dependent responses from FEM outcomes. \\

E 
&  TENG output voltage measurement
& The $10\times10\,\mathrm{mm}^2$ Cu–PVS and ITO–PET TENG as in \textbf{Figure~\ref{fig:TENG_Setup}}, were tested under load variation considering parameters and procedures followed from \cite{kumar2023mechanics} and summarised in \ref{app1.1}. 
& Validation of ODE-based TENG circuit simulations at different loading frequencies. \\
\hline\hline
\end{tabular}
\label{tab:Exp_catagories}
\end{table}
\begin{table}[H]
    \centering
    \caption{Mechanical properties, geometry and measured surface parameters of tribolayers in the TENG setup `A' and `E' for the validation of the framework~\cite{dogru2018poisson, johnston2014mechanical}.}
    \renewcommand{\arraystretch}{1.1}
    \begin{tabular}{@{}lll@{}}
        \hline\hline
        \textbf{Parameter} & \multicolumn{2}{c}{\textbf{Material}} \\
        \cline{2-3}
        & \textbf{PET} & \textbf{PVS} \\
        \hline
        \hline
        Surface area, $A_n$ & \multicolumn{1}{c}{\quad\quad\quad$10$ mm $\times$ $10$ mm} \\
        Thickness, $z$ & $0.18$ mm & $0.25$ mm \\
        Young’s modulus, $E$ & $1.4$ GPa & $3.5$ MPa \\
        Poisson’s ratio, $\nu$ & $0.48$ & $0.40$ \\
        Surface RMS height, $S_q$ & $12.6\,$nm & $25\,\mu$m \\
        Surface RMS gradient, $S_{dq}$ & - & $0.34$ \\
        \hline\hline
    \end{tabular}
    \label{tab:Mech_parameters}
\end{table}
\begin{table}[H]
    \centering
    \caption{Dielectric material properties of tribo-layers where $\sigma_T$ has been measured in the configuration of setup `B' and `E'. The air permittivity is $\varepsilon_0 = 8.85 \times 10^{-9} \, \mu\mathrm{F}/\mathrm{mm}$.}
    \renewcommand{\arraystretch}{1.2}
    \begin{tabular}{@{}lll@{}}
        \hline\hline
        \textbf{Material} 
        & \textbf{Relative permittivity, $\varepsilon_r$} 
        & \textbf{Tribo-charge density, $\sigma_T$} \\
        \hline
        PET & $3.3$ & $+80~\mu$C/m$^2$ \\
        PVS & $2.72$ & $-80~\mu$C/m$^2$ \\
        \hline\hline
    \end{tabular}
    \label{tab:Elecmaterial_properties}
\end{table}

The mechanical properties of the tribolayers are presented in \textbf{Table~\ref{tab:Mech_parameters}}, where the root-mean-square surface height ($S_q$) of PET measured via atomic force microscopy was found to be negligible ($\approx 500$ times lower) than that of PVS. The dielectric parameters of the tribolayers are presented in \textbf{Table~\ref{tab:Elecmaterial_properties}}. In particular, the method for the tribocharge density measurement has been detailed in experimental Section~\ref{app1.3}.

\subsection{Experimental validation of the FEM contact mechanics model}
\begin{figure}[!b]
    \centering
    \includegraphics[width=\linewidth]{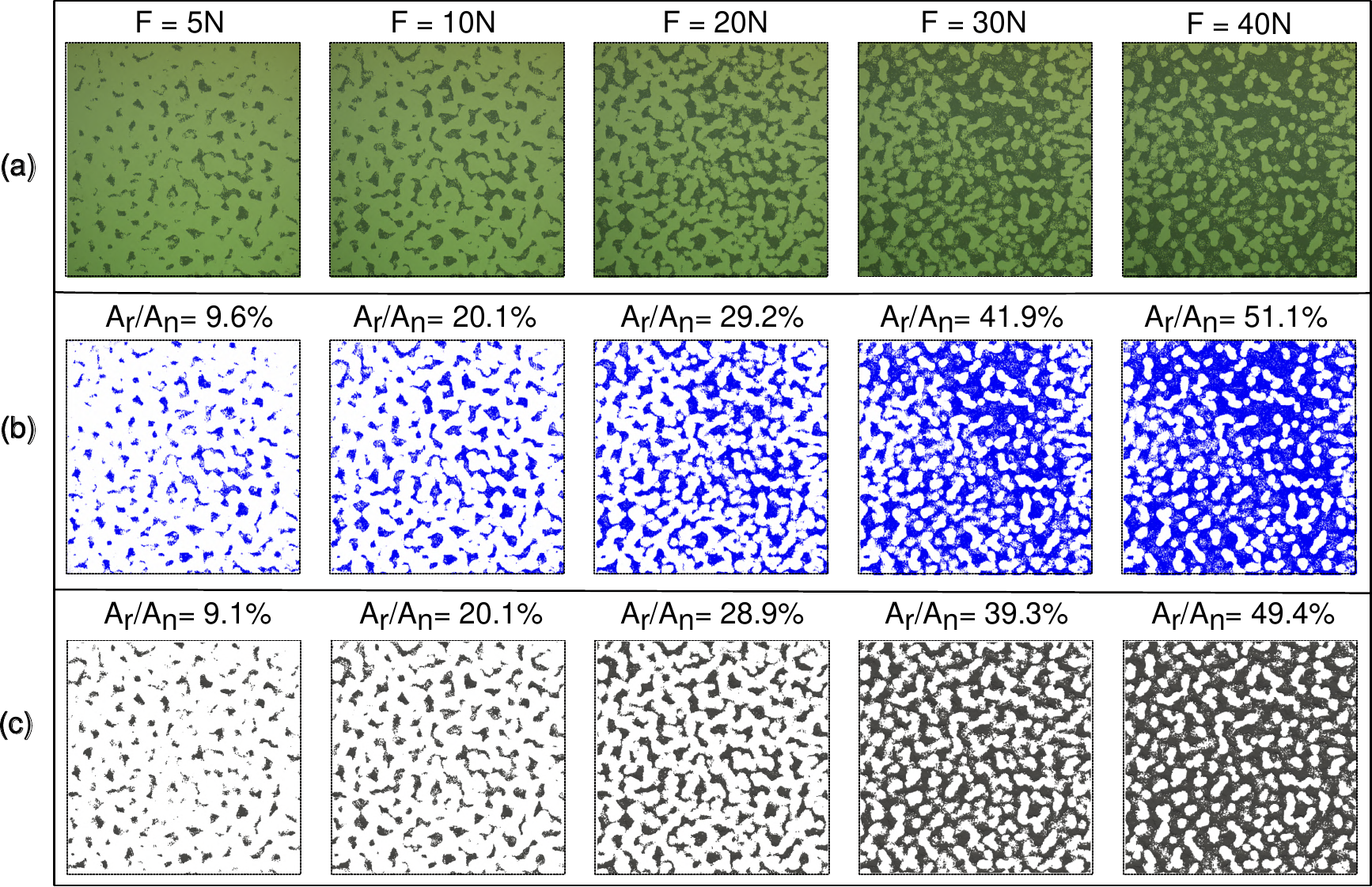}
    \caption{Evolution of real contact area with increasing normal load. Row (a) shows raw experimental optical images (RGB) of the contacting interface corresponding to a range of mechanical loads (5N, 10N, 20N, 30N and 40N). Row (b) presents the corresponding binarised contact pixel maps (blue colour represents contact zones and the white colour is for out-of-contact zones). Row (c) shows finite element simulation results predicting the contact area under the same mechanical loads (black colour represents the real contact area).}
    \label{fig:Binfig}
\end{figure}

\begin{figure}[t]
    \centering
    \includegraphics[width=0.6\linewidth]{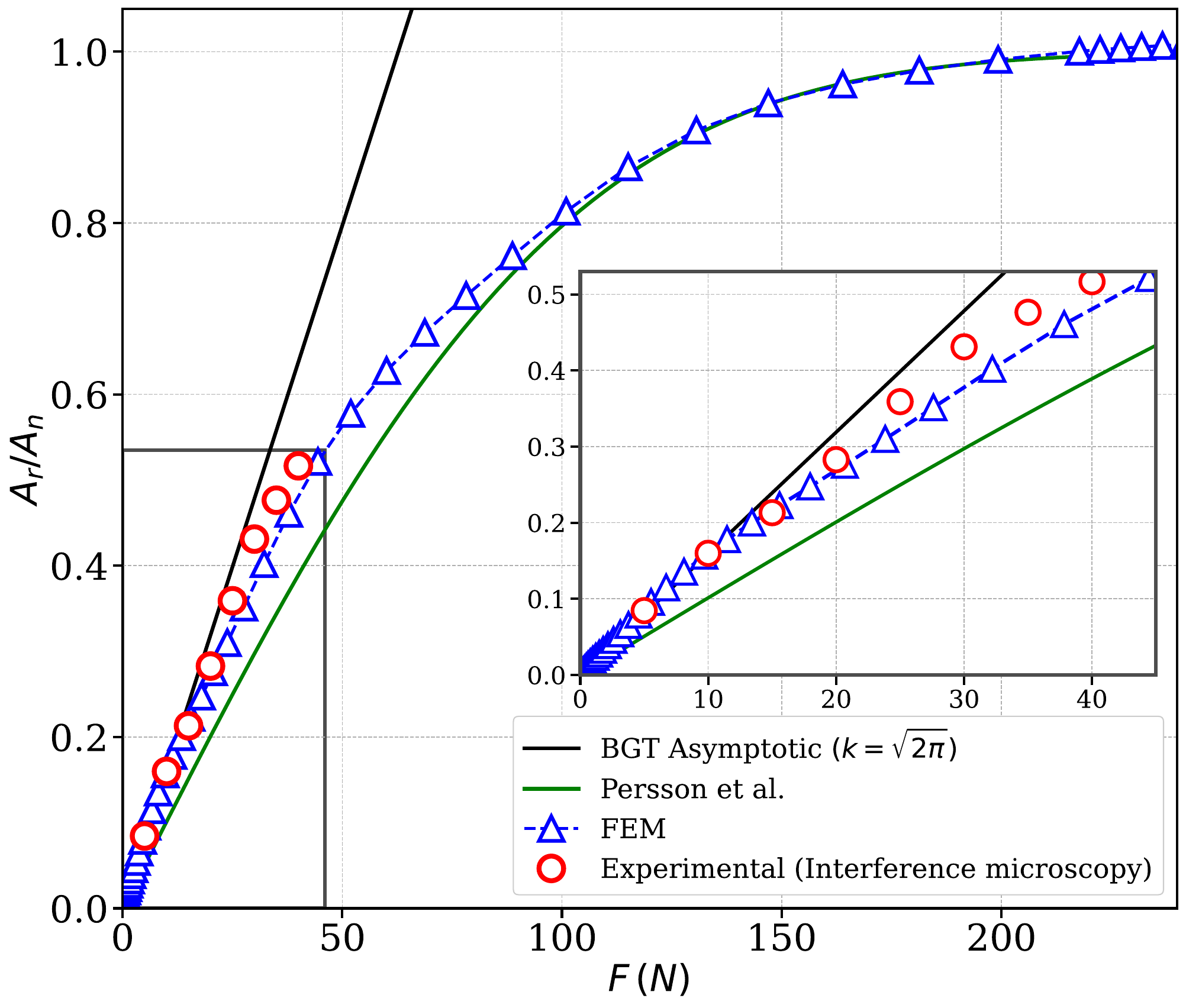}
    \caption{Comparison of experimental results (based on the setup `A' in Table \ref{tab:Exp_catagories}) and MoFEM contact simulations with Persson's contact theory \citep{persson2006contact} for the evolution of real contact area with the increasing mechanical load. The inset shows a zoom-in on the part of the plot corresponding to the range of mechanical loads used in the experiment.} 
    \label{fig:contact_expvsFEM}
\end{figure}

Since the real contact area critically influences TENG performance, validation is first performed to assess the capability of the contact mechanics numerical model to predict the real contact area accurately. An experiment was conducted for different forces, focusing on the contact between a newly fabricated rough PVS tribolayer and a PET counter-surface, as referred to in the setup `A' in Table~\ref{tab:Elecmaterial_properties}. Notably, the setup excluded electrodes and substrates to enable direct optical observation and measurement of the real contact area using interference microscopy; further details are provided in Section~\ref{app1.2}. In this context, experimental contact areas were obtained by binarising the interference images using the \texttt{OpenCV} Python library~\citep{bradski2000opencv}, allowing for pixel-wise contact mapping, as illustrated in \textbf{Figure~\ref{fig:Binfig}(a--b)}.

The surface roughness of the entire  $10\,\text{mm} \times 10\,\text{mm}$ PVS layer was measured before the TENG experiment, as discussed in subsection~\ref{app1.1}. The processed surface grid data were subsequently projected onto the finite-element mesh consisting of $1.9$ millions of hexahedral elements and $631\times631$ nodes on the top surface.  This level of refinement permits preserving the actual surface representation, maintaining the probability density function (PDF) correctly, as shown in Figure~\ref{fig:test_rig}(c-d). The contact simulation was performed in a distributed-memory parallel environment utilising 128 compute cores with 3.8 GB of RAM per core and required $\sim$20 minutes per mechanical load case. Importantly, the ability to represent roughness across the full specimen within practical simulation wall-clock times underscores the scalability of the framework and its suitability for virtual design and optimisation.

The resulting FEM predictions of the evolution of the real contact area with increasing mechanical load exhibit excellent qualitative and quantitative agreement with the experimental observations (compare \textbf{Figure~\ref{fig:Binfig}(b-c)} and see also \textbf{Figure~\ref{fig:contact_expvsFEM}}. The experiment was limited to a mechanical load range of $5$--$40\,\text{N}$ due to the experimental setup constraints. Within this range, the FEM results show absolute deviation from the experiment of only $0.5$--$2.6\%$ of contact area fraction, indicating that the model reliably captures both the growth and spatial distribution of the real contact area.

Furthermore, Figure~\ref{fig:contact_expvsFEM} demonstrates that the FEM results exhibit a transitional behaviour between two classical analytical contact models. At low normal loads (e.g., 8N) where contact gets initiated, and both the experimental data and FEM predictions closely follow the BGT model~\cite{bush1975elastic}, which shows up to $54\%$ higher real contact area ratio than the analytical model in \cite{persson2006contact}. With increasing load, the experimental and numerical outcomes progressively deviate from the BGT model, while maintaining the same mutual trend. At higher loads, the FEM results gradually converge towards Persson’s contact model~\citep{persson2006contact} at 220 N when the interface approaches full contact. 

Overall, Persson's theory was previously used to predict the contact area in TENG applications~\cite{XUVoc} where a pronounced difference was observed between the experimental results and the statistical model of \cite{persson2002elastic}, e.g., reaching $30\%$ more for the loads in the range $30-40$ N. The proposed FEM framework accurately predicted the actual contact area even for realistic engineered surface topography and deformation under a wide range of mechanical loading conditions. These capabilities are particularly important for linking mechanical contact behaviour to electrostatic performance predictions in TENG systems, discussed in the following subsections. 

\subsection{Validation of the electrostatics model}
This section validates the implementation of the electrostatic model for the TENG  experimental setup `D' in Table \ref{tab:Exp_catagories}. The TENG device presented in Figure~\ref{fig:TENG_Setup} was modelled using the numerical scheme described in Section \ref{Electrostatics modelling of TENG}, considering air around the TENG (see Figure~\ref{fig:electrostatic_model}).   

\begin{figure}[t]
    \centering
    \includegraphics[width=0.5\linewidth]{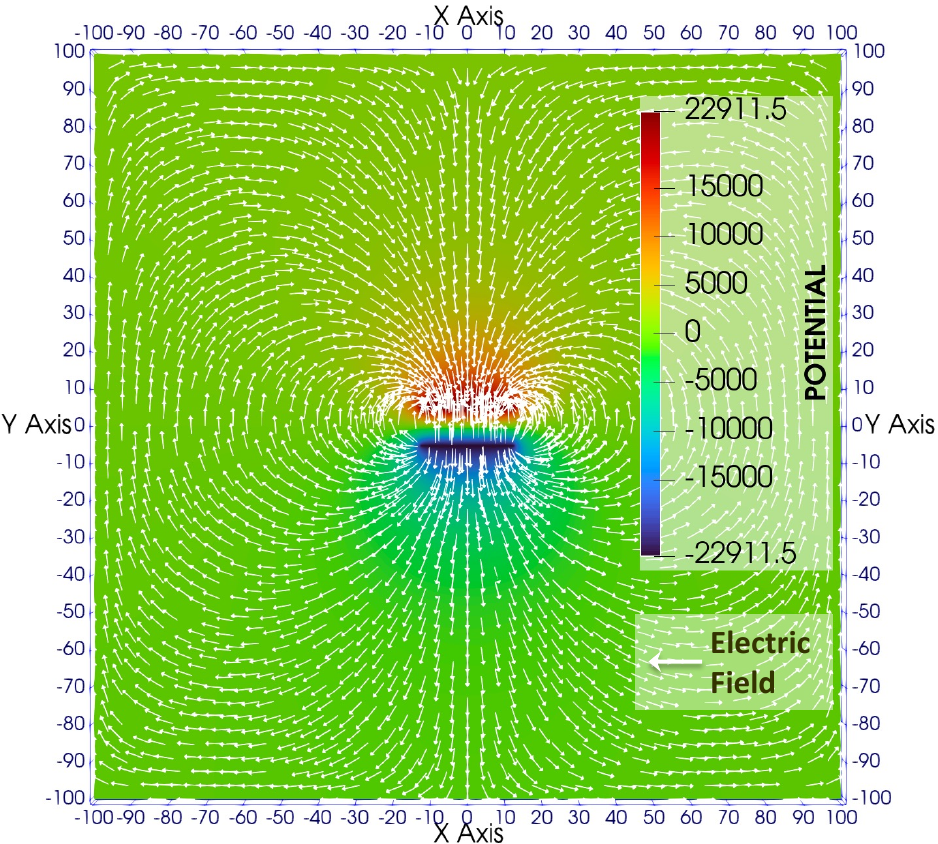}
    \caption{A cross-section of TENG for setup `D' showing 3D potential (colour bar) and electric field (highlighted with the white arrows) distribution for a $10$mm air gap between the dielectric layers.}
    \label{fig:Efield}
\end{figure}

\subsubsection{Open Circuit Voltage}
\begin{figure}[!b]
    \centering
    \includegraphics[width=0.5\linewidth]{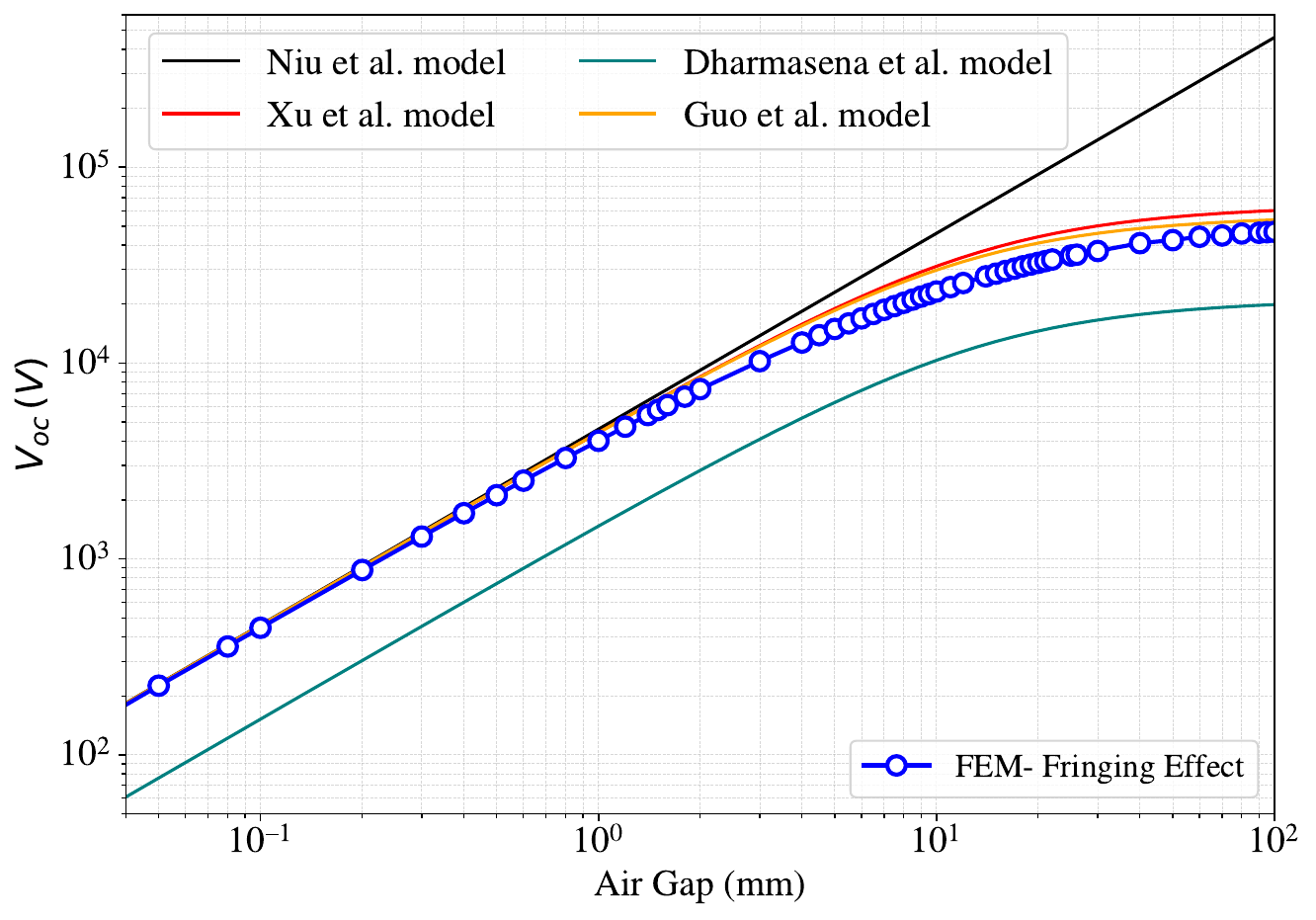}
    \caption{Open-circuit voltage ($V_{OC}$) comparison between FEM results and available analytical TENG models for setup `D' \citep{dharmasena2017triboelectric, niu_Voc_linear, guo2020derivation}}
    \label{fig:voc_compana}
\end{figure}

The first electrostatic study focused on the dependence of the $V_{OC}$ on the air gap between the charged tribo-layers. A detailed representation of the FEM electric field for the considered TENG model while computing $V_{OC}$ is visualised in \textbf{Figure~\ref{fig:Efield}}. Further, \textbf{Figure~\ref{fig:voc_compana}} highlights that FEM-computed  $V_{OC}$ with different air gaps closely matches the analytical approximations of~\cite{XUVoc} and \cite{ guo2020derivation}, including the characteristic saturation of $V_{OC}$ at larger air gaps. For higher gaps, FEM predicts $17.5\%$ lower $V_{OC}$ along the air gaps compared to the Xu et al. model, reflecting the growing influence of 3D fringing effects. Unlike Xu’s model, which assumes a uniform electric field in the horizontal ($x$) direction across the dielectric, the FEM results reveal noticeable bending of the electric field near the plate edges. This highlights the substantial impact of fringing and field non-uniformity on TENG performance, illustrating the importance of capturing full 3D electrostatics effects for device predictions.

\subsubsection{Capacitance}
\begin{figure}[t]
    \centering
    \includegraphics[width=0.5\linewidth]{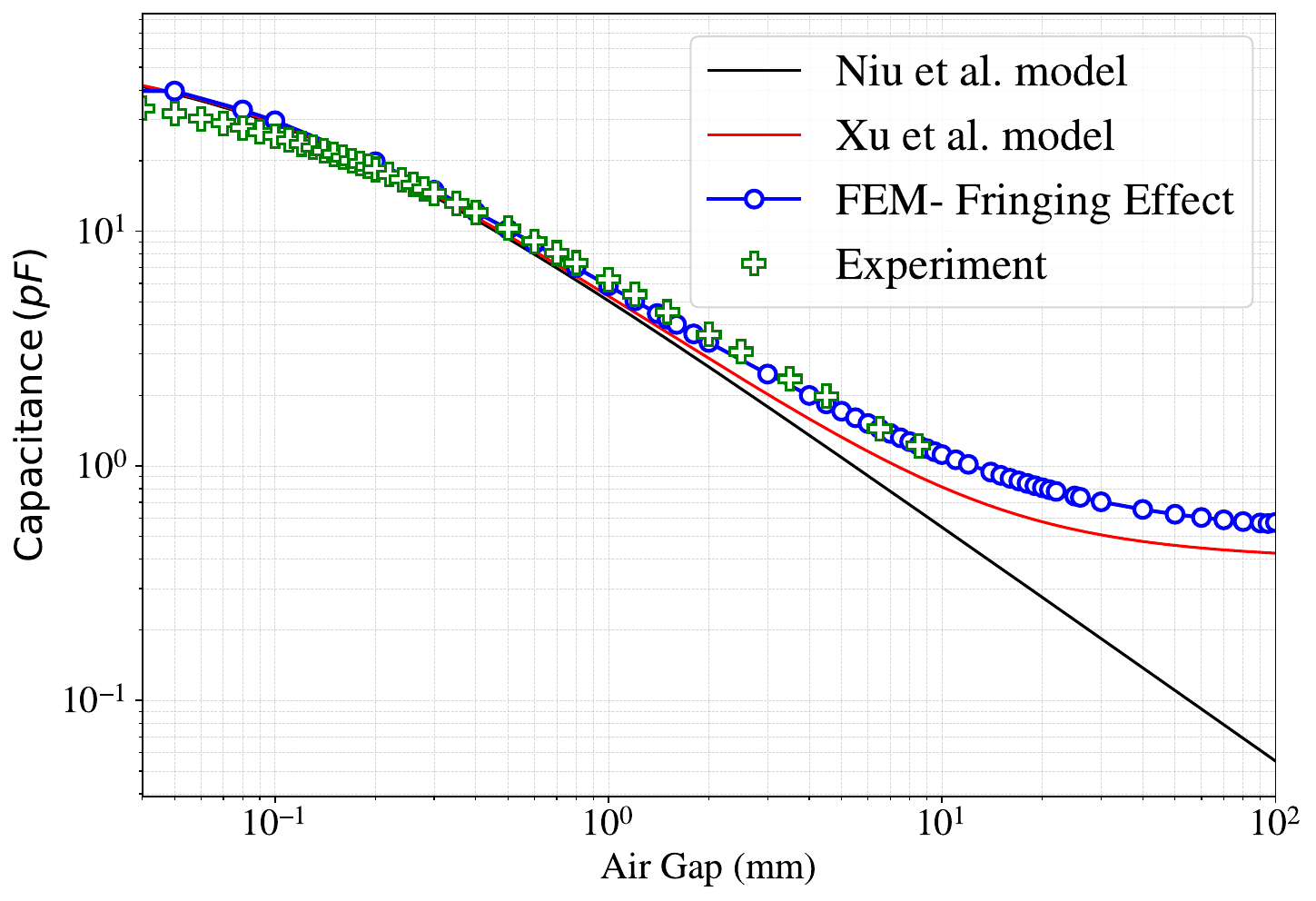}
    \caption{Validation of the FEM capacitance model with experiment setup `C' and comparison with analytical models derived from \citep{XUVoc, niu_Voc_linear}.}
    \label{fig:capa_result}
\end{figure}

Further validation of the electrostatic FEM model has been performed by computing the finite capacitance for the experimental setup `C'. In the experiment, two parallel ITO-coated glass electrodes were used, with the electrical connections established using conductive clips attached to the ITO edges. The air gap between electrodes was controlled using an electromechanical testing system, and capacitance values were recorded at each gap using an LCR meter. The experimental details are further elaborated in Section \ref{app:exp_capa}. In \textbf{Figure~\ref{fig:capa_result}}, the outcomes of the FEM simulation using the methodology explained in Section \ref{sec:capacitance_computation} show significantly better agreement with the experiment than the analytic forms at the higher gaps, where 3D fringing electric field effects become dominant.
 
\subsubsection{Time dependency}  
\begin{figure}[http]
    \centering
    \includegraphics[width=0.5\linewidth]{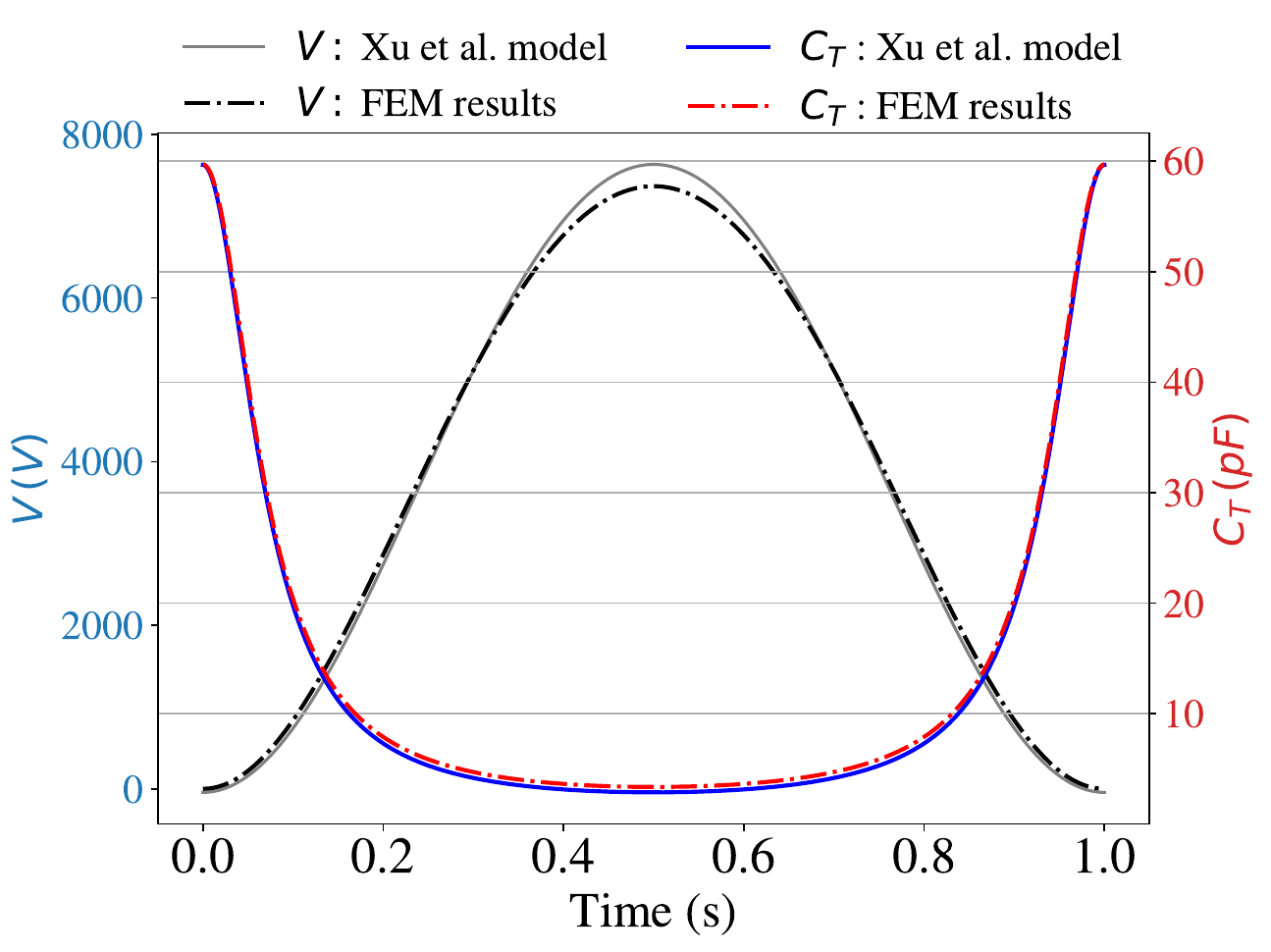}
    \caption{Comparison of the net voltage and capacitance response over time under a loading frequency of $1$ Hz with displacement from the range $[0, 2]$ mm. The results from the Xu et al. model \citep{XUVoc} (solid lines) and FEM simulation (dashed lines) are shown, considering the case when $\sigma_U = 0.9 \sigma_{T}$.}
    \label{fig:capacitance_comp}
\end{figure}

The output voltage visualised in \textbf{Figure~\ref {fig:capacitance_comp}}  is computed from combining the FEM output of $V_{OC}(t)$ and $C_T(t)$. Upon considering 3D fringe-field effects, FEM results (dashed lines) for a 2 mm maximum separation gap and $1~Hz$ frequency closely follow Xu et al.’s analytical model (solid lines) with time and are slightly higher quantitatively at intermediate gaps than the \cite{XUVoc} analytical model.

Additional investigation confirmed that $V_{OC}$ and the capacitance increased with lateral size (10$\times$10 $mm^2$, 25$\times$25 $mm^2$, and 50$\times$50 $mm^2$). These apparent details have been explored, showing the effect of TENG lateral sizes in section SI-2a of the supplementary material (see Figure S2). 

\subsection{Application of the coupled model to study mechanical load effect}

\begin{figure} [httb]
    \centering
    \includegraphics[width=0.5\linewidth]{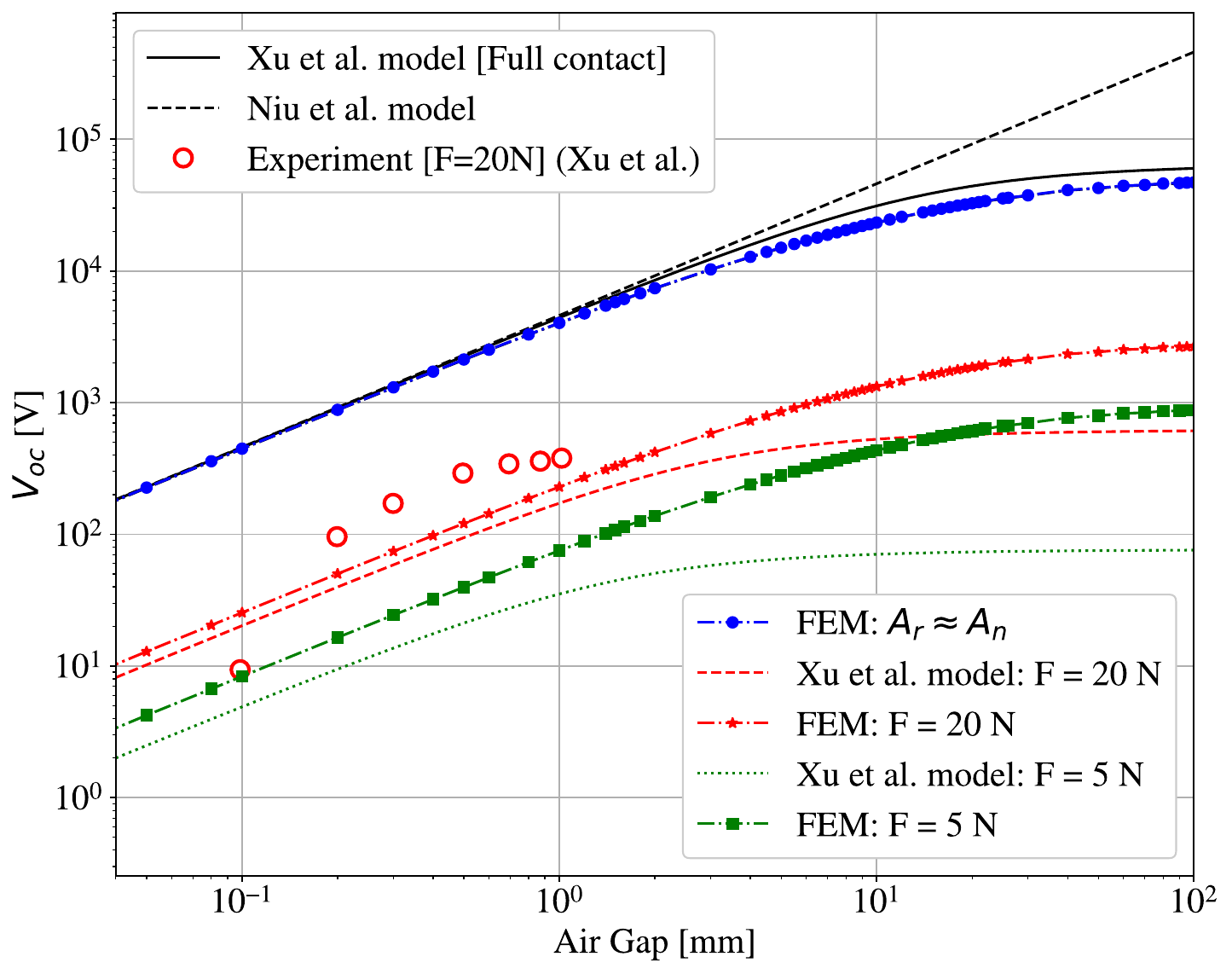}
    \caption{The dependency of $V_{OC}$ on air gap for different mechanical loads, comparing FEM results with analytical approximations and experimental setup `D', where 20 N of mechanical force was applied.}
    \label{fig:voc_vs_contact}
\end{figure}

Following the outcomes of the previous subsection, $V_{OC}$ has been computed for a range of mechanical loads for the corresponding contact area ratio. The results are compared against the experimental parameters in setup `D' and the analytical predictions \cite{XUVoc}.  The FEM contact area simulation for a PDMS surface was fabricated with $S_{dq} = 0.22$. The surface representation for this and the comparison of contact area analyses are detailed in section SI-5 of the supplementary information (see Figure S7). At lower loads in \textbf{Figure~\ref{fig:voc_vs_contact}}, the FEM results show a lower value $V_{OC}$ than the full load due to the small contact area. For the case of 20 N load, FEM shows a closer prediction to the experiment. At the same time, the FEM curves for any load reach saturation at a higher air gap than the analytical load-dependent model by~\cite{XUVoc}. 
\begin{figure}[!b]
    \centering
    \includegraphics[width=0.6\linewidth]{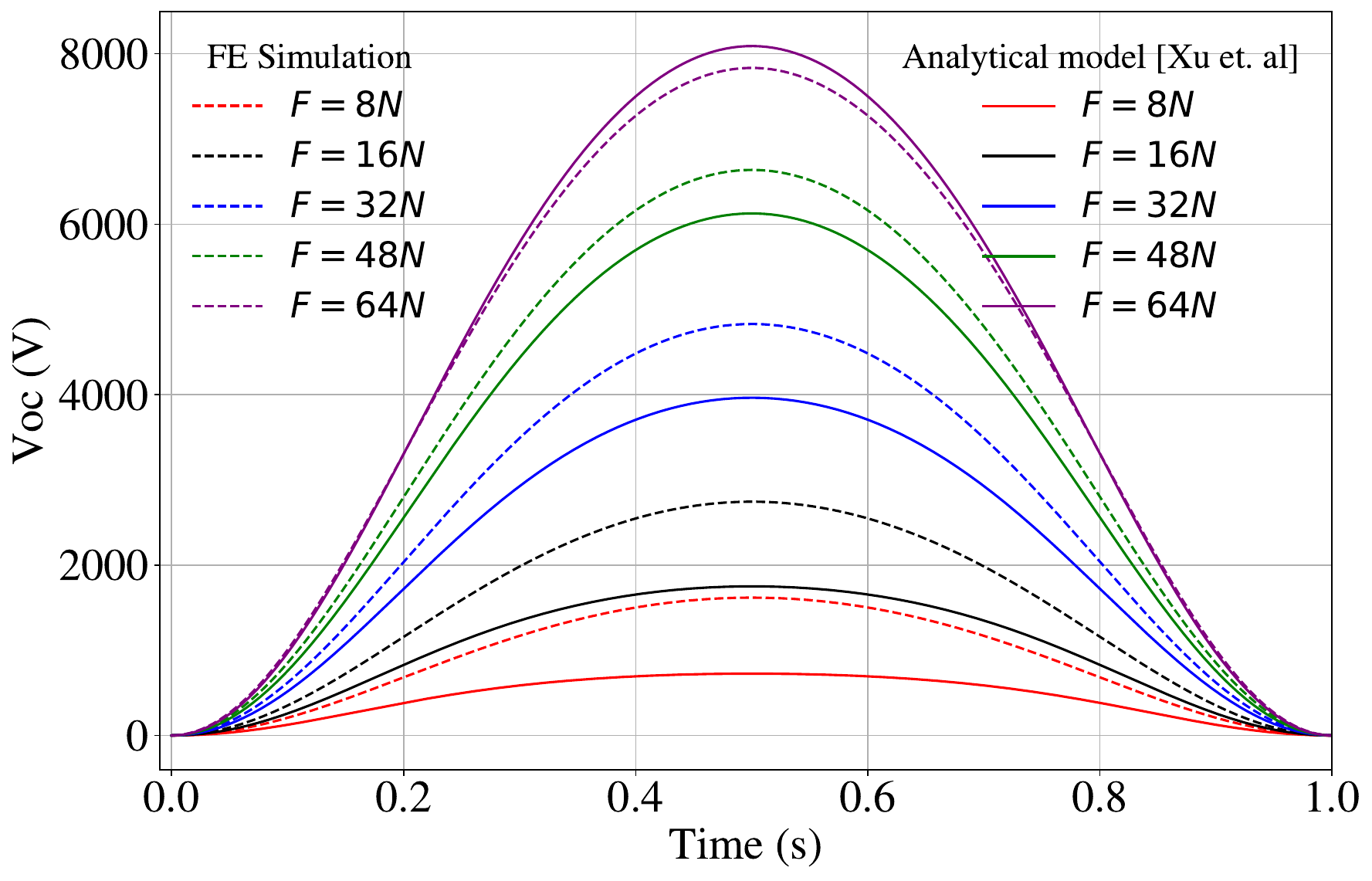}
    \caption{The transient response of the $V_{OC}$ when layers gets separated by $z_{max} \in [0, 2]$ at $1$ Hz frequency, for  different force values ($F$), comparing the FEM simulation (dashed lines) and the analytical model \cite{XUVoc} (solid lines) for experimental setup 'E'.}
    \label{fig:voc_comp}
\end{figure}
This discrepancy arises from the underlying assumptions in the analytical model. In reality, the tribo-charge spots generated by rough surface contact are spatially distributed as irregularly shaped asperity contact patches. This complex distribution is reduced in the analytical model to a single equivalent circular disc of area $A_r$, which does not fully capture the heterogeneous nature of real contact and becomes substantially inaccurate for small contact areas, resulting in the underestimation of the $V_{OC}$. In contrast, the FEM framework explicitly resolves the actual contact area, while the consideration of scaled tribo-charge distributed over the surface provides accurate results, which has also been discussed in section SI-3 of the supplementary material.

Following the verification of the load-dependent behaviour, further investigation continued with setup `E' as shown in Figure~\ref{fig:TENG_Setup}. The electrostatic properties for this simulation are presented in Table \ref{tab:Elecmaterial_properties}.

\textbf{Figure~\ref{fig:voc_comp}} illustrates the time and load-dependent response, where each curve corresponds to a distinct applied force and corresponding contact area fraction. In all cases, the FEM results (dashed lines) predict an increase of $V_{OC}$ with increasing load. However, the difference between the simulation and analytical prediction changes from positive to negative. At higher loads (e.g., $64\, N$), fringing field effects shown in the electrostatic FEM simulation reduce the predicted $V_{OC}$ by $3.1\%$ relative to the analytical model (see Figure~\ref{fig:voc_compana}). In contrast, at lower applied loads (e.g., $8N$), fringing has less effect compared to the effect of merged contact area, resulting in underestimation of the $V_{OC}$, such that FEM predicts a higher (e.g., $56\%$) $V_{OC}$ compared to the analytical model~\cite{XUVoc}. 

\begin{figure}[H]
    \centering
    \includegraphics[width=0.5\linewidth]{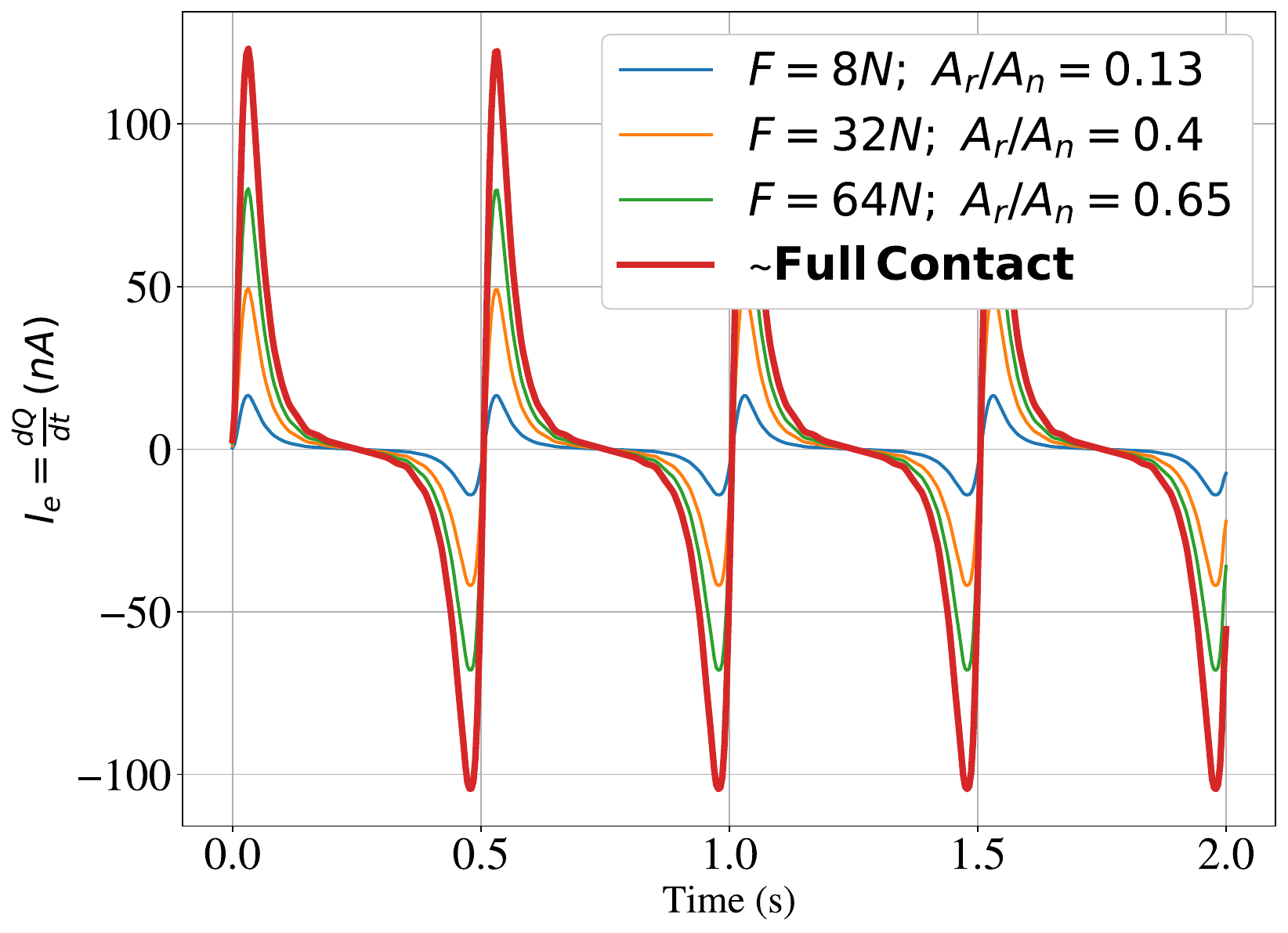}
    \caption{Short-circuit current response showing current increase with increasing mechanical load (\textit{F}) that affects the real contact area ratio (an external resistance of $R_L=100 M \Omega$ has been considered).}
    \label{fig:current_voltage_sim}
\end{figure}

Finally, once $V_{OC}(t)$ and $C_T(t)$ have been computed, the transient response of transferred charges to the outer circuit is obtained upon solving the ODE for the circuit shown in Figure~\ref{fig:circuiteqv} in Section \ref{subsection:circuit_ODE}. The derivative of the charge $Q$ over time provides the output current, as shown for a range of forces in \textbf{Figure~\ref{fig:current_voltage_sim}}. In this regard, the absolute values of positive and negative current peaks increase with increasing mechanical load. These results also confirm that the magnitude of charge transfer per cycle increases with force. At the same time, the frequency of the mechanical motion primarily governs the timing of charge exchange, which is elaborated further in section SI-4 of the supplementary document (see Figure S5).

\subsection{Effect of operating frequency on TENG output} 
\begin{figure}[H]
    \centering
    \includegraphics[width=\linewidth]{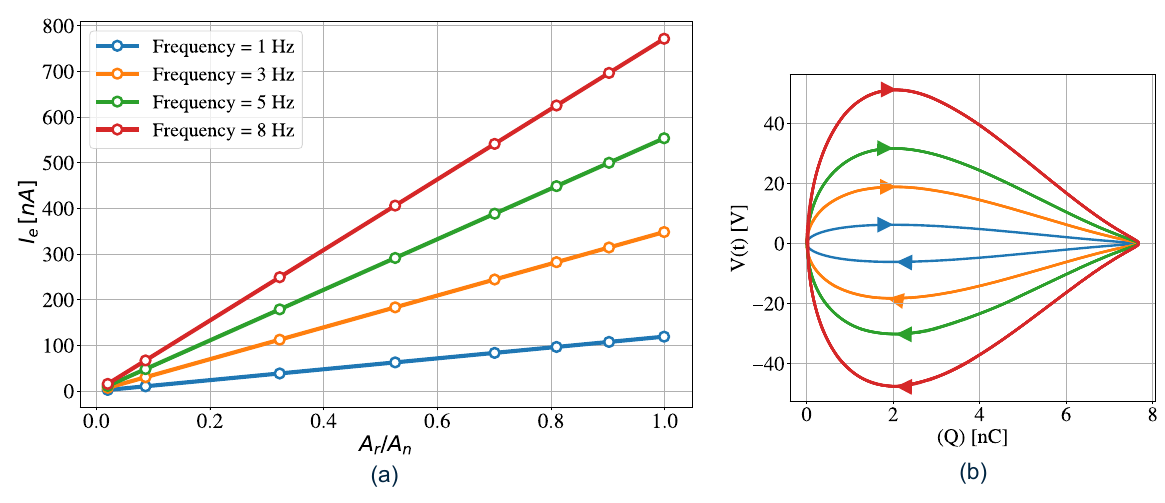}
    \label{fig:Isc_area_ratio}
    \caption{TENG output dependency on frequency: (a) Variation of maximum current ($I_{e}$) with contact area ratio for different frequencies, showing that higher frequencies result in greater current output. (b) V--Q hysteresis loops at different excitation frequencies. The loop width increases with frequency, indicating enhanced charge transfer and higher voltage output compared to the lower frequencies.}
  \label{fig:Freq_vs_Ar_an}
\end{figure}

The accumulated transferred charges and their dependence on the loading and unloading frequency are also crucial in governing the TENG characteristics. The effect of these parameters has been examined at 100~M$\Omega$ resistance to verify the framework using the experimental setup `E'. \textbf{Figure~\ref{fig:Freq_vs_Ar_an}(a)} shows the variation of maximum external current ($I_{e}$) as a function of $A_{r}/A_{n}$. This result demonstrates that the current response increases linearly with the contact area ratio and nonlinearly with the tapping frequency. The curves corresponding to higher frequencies exhibit steeper slopes and larger current outputs. For example, at full contact ($A_r/A_n \approx 1$), $I_e$ rises from approximately 120 nA at 1 Hz to about 780 nA at 8 Hz. However, the relative current gain diminishes at higher frequencies, such that $I_e$ increases approximately 3 times when frequency increases from 1 Hz to 3 Hz, while $I_e$ at frequency 8 Hz is only 6.5 times greater than the current at 1 Hz. This demonstrates that TENG output behaviour is limited by charge-transfer per cycle, suggesting that identifying an optimal operating frequency is advantageous for maximising TENG performance. The additional relevant investigations with forces are presented in section SI-4 of the supplementary documents (see Figure S6). 

Moreover, \textbf{Figure~\ref{fig:Freq_vs_Ar_an}(b)} shows the V–Q hysteresis loops for frequencies between 1.0 and 8.0 Hz. Within this range, the behaviour highlights the strong influence of operating frequency on the charge transport. It demonstrates that the transferred charge changes dynamically: first, increases and then decreases over repeated contact–separation cycles, varying broadly with frequency. At lower frequencies, the loops are found to be narrower and have less charge transfer and lower voltage levels. As the frequency increases, the voltage output increases, indicating more effective charge accumulation.

\subsection{Effect of the resistance on TENG output}
TENGs inherently have high internal impedance, which leads to output voltages in the hundreds to thousands of volts, while the current is limited to micro-nano amperes in many configurations. Hence, TENGs require power management circuits to convert and match their output to practical loads \citep{min2022optimisation, gao2024achieving, abid2024output}, and there is a need to quantify current–voltage trade-offs, power transfer characteristics, and their sensitivity to excitation frequency and applied force.

\begin{figure}[H]
    \centering
        \includegraphics[width=\linewidth]{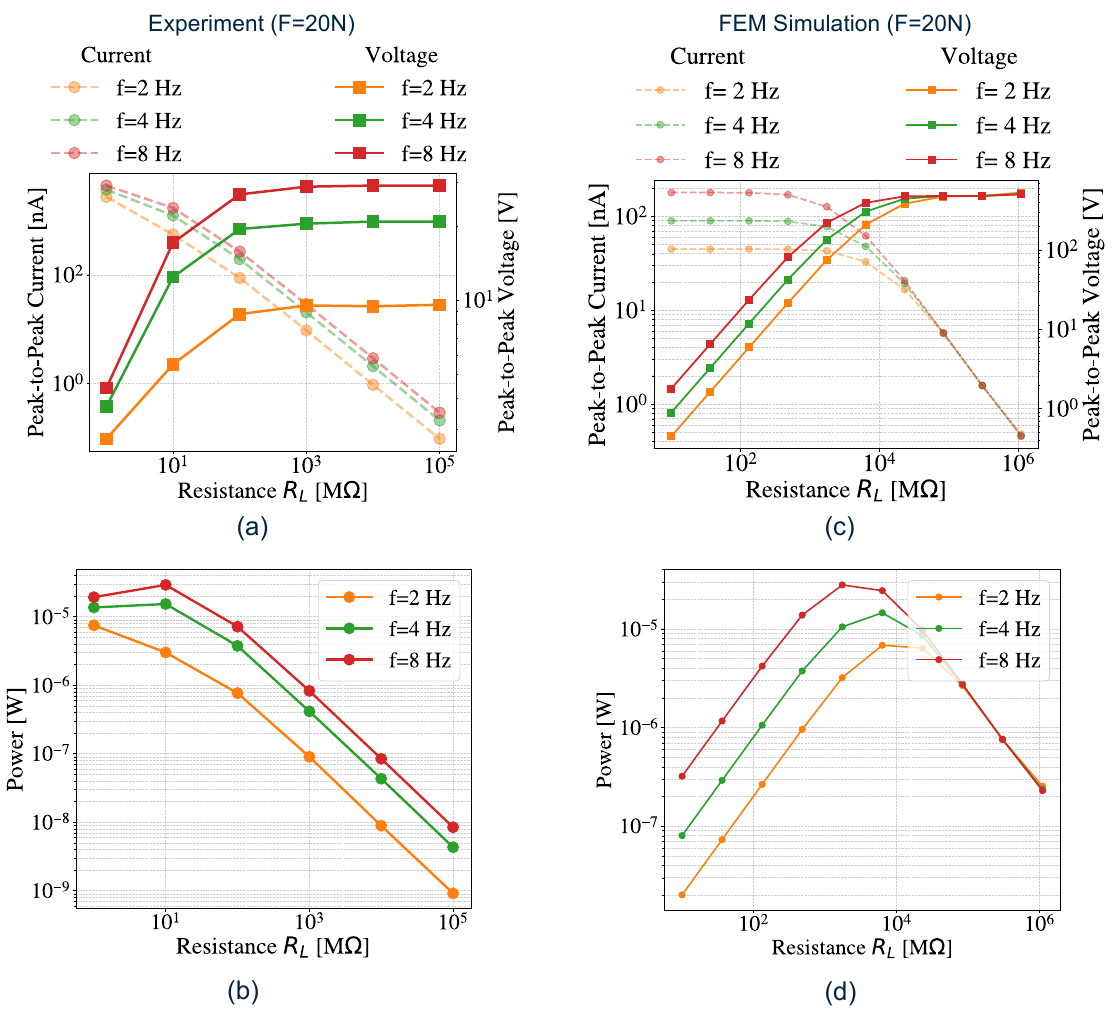}
    \caption{Comparison between experimental and simulated performance of TENG under different operational conditions showing, (a,b): experimental measurements at a mechanical load of 20 N, (c,d): simulations at the same mechanical load,  (a,c): peak-to-peak current (dashed lines) and voltage (solid lines) as a function of $R_L$ under different excitation frequencies (2 Hz, 4 Hz, and 8 Hz); (b, d): changes of TENG power outputs $(P = VI)$ with $R_L$.
    }
  \label{fig:SIM_VSEXP_area} 
\end{figure}

To account for this, electrical output has been investigated experimentally and numerically as a function of the resistive load. For a fixed normal force of $20N$, the experimental investigation shown in \textbf{Figure~\ref{fig:SIM_VSEXP_area}(a)} with setup `E', demonstrates that the peak-to-peak current decreases with increasing load resistance, while the voltage increases and distinctly saturates at high resistance. In this context, FEM simulation results at the same applied load show qualitatively similar trends in \textbf{Figure~\ref{fig:SIM_VSEXP_area}(c)} and \textbf{Figure~\ref{fig:SIM_VSEXP_area}(d)}, but quantitatively different outcomes compared to the experiments. In particular, the FEM simulations show a consistent physical trend in which the saturation of the output voltage occurs at nearly the same voltage regardless of the excitation frequency, reflecting the transition from short-circuit to open-circuit operation, as predicted by the equivalent circuit model in Section~\ref{subsection:circuit_ODE}. Remarkably, a similar saturation characteristic has also been reported in the analytical outcomes presented by Dharmasena et al.~\citep{dharmashenaVoc}, which supports the physical validity of the simulation framework.

In figure\ref{fig:SIM_VSEXP_area}, at low load resistances, output power is limited by current saturation, while at high resistances, it is constrained by voltage saturation. Both experimental measurements and simulation outcomes in Figure~\ref{fig:SIM_VSEXP_area} permit obtaining the optimal load resistance at which the output power is maximum. From the available experimental results, the optimal $R_L \approx 10 \,\text{M}\Omega$, while the FEM simulation in Figure~\ref{fig:SIM_VSEXP_area}(d) shows that as the excitation frequency increases from 2 to 8~Hz, the maximum power rises by nearly an order of magnitude and shifts toward lower resistance values, indicating an increased charge transfer rate at higher operating frequencies. However, this effect reduces the higher is the frequency. Unlike the experiments, simulations also indicate that once the voltage reaches the open-circuit value at high resistance, the output should be frequency-independent.

\begin{figure}[t]
    \centering
    \includegraphics[width=\linewidth]{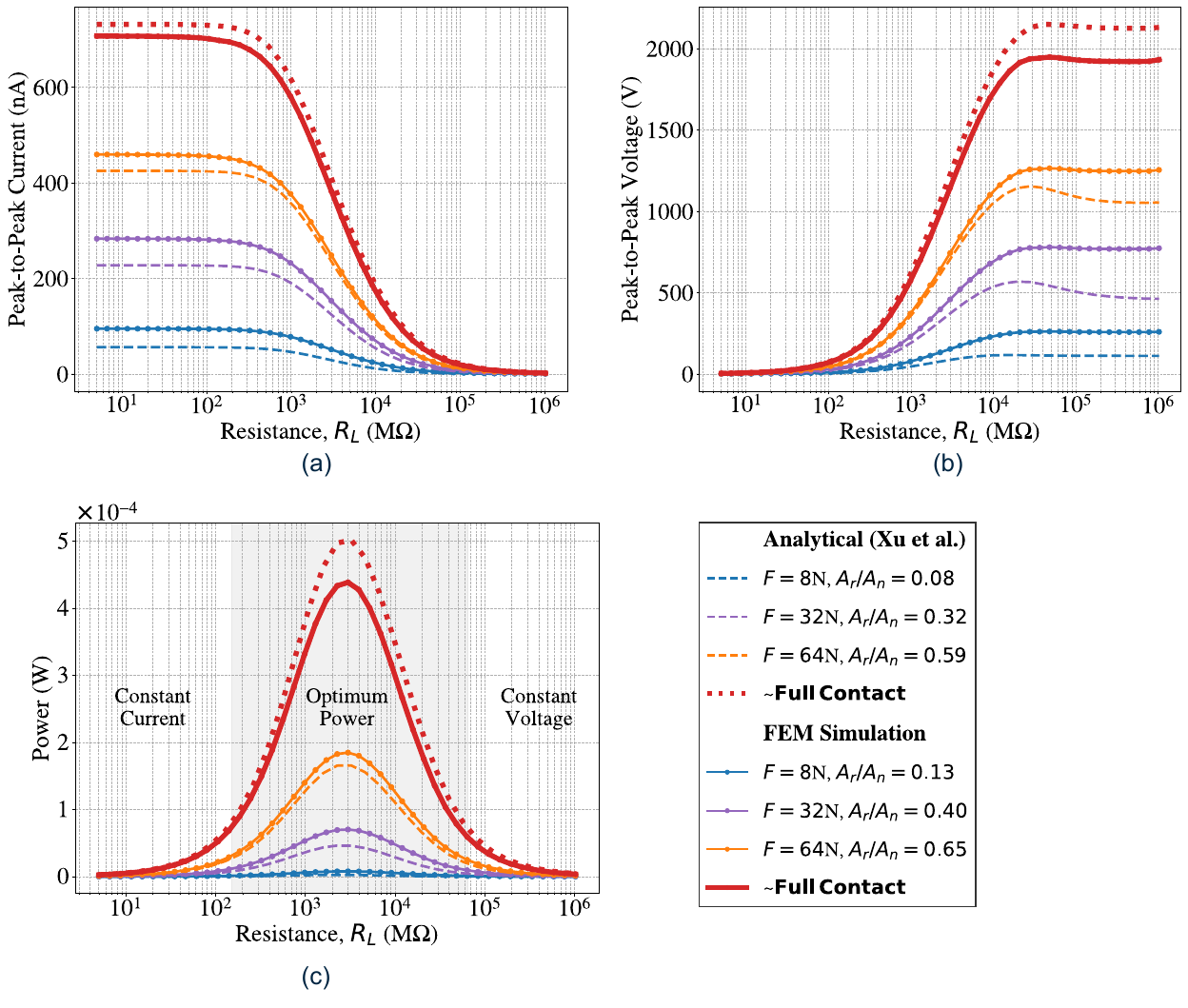}
    \caption{Electrical response of the TENG with external resistance $R_L$ at an excitation frequency of $6~Hz$ for different applied normal loads ($F=8$, $32$, and $64$). Results derived from the analytical approximation (dashed lines) are compared with FEM simulations (solid lines with markers), showing: (a) Peak-to-peak current, (b) peak-to-peak voltage, and (c) output power versus load resistance, illustrating the load-dependent transition from current-dominated to voltage-dominated regimes and the corresponding optimum power region.}
    
  \label{fig:XU_FEM_R}
\end{figure}

Furthermore, in \textbf{Figure~\ref{fig:XU_FEM_R}}, the simulation results obtained at steady state and fixed excitation frequency of 6~Hz show three phases of TENG I-V characteristics curves between resistive loading from 10 to $10^6 M\Omega$: (i) constant current at low resistance, (ii) constant voltage at high resistance and (iii) optimum power in between, confirming the regime transitions. This demonstrates the strong dependence of the electrical output on the applied normal force. Increasing the force from 8N to 220~N (capturing the evolution of TENG layers from partial to near full contact), the corresponding increase in peak power and optimal load resistance ($10^3$–$10^4$ M$\Omega$) is attributed to the force-induced increase in the effective $A_r/A_n$. The relative electrical output deviations follow the same trend, suggesting that differences in contact-area estimation dominate the quantitative discrepancy between the two approaches.  In addition, the optimum resistance has been observed as force independent, but it changes with frequency, device size and capacitance. 

Although quantitative differences between experimental and simulated values are observed in Figure~\ref{fig:SIM_VSEXP_area}, particularly at high load resistances, these may arise primarily from high internal impedance, contacting surfaces alignment, charge leakage, viscoelastic deformation of the dielectric layers, and/or parasitic electrical capacitance effect. The FEM predictions show consistently higher output performances than the analytical approximation when in partial contact and lower when layers are tending to full contact. The FEM framework resolves three-dimensional electrostatic field distributions, including fringing-field effects and spatially nonuniform charge and capacitance distributions, which are simplified or neglected in the analytical formulation. Despite these analytical assumptions, the strong agreement in scaling trends, saturation behaviour, and optimal loading conditions confirms the predictive capability of the FEM-based model. The results demonstrate that the model reliably captures the correct trends and the dominant physics governing resistive load optimisation and contact area computation for contact-separation TENGs and provides a framework for device design under varying mechanical and excitation conditions, considering surface roughness.

\section{Conclusions}
This paper presented a multiphysics finite element framework that serves as a predictive tool for TENG design, incorporating mechanical contact analysis, three-dimensional electrostatic field computation, and ODE-based circuit model within a single computational pipeline, implemented in the open-source library MoFEM~\citep{kaczmarczyk2020mofem}. The key novelty of this framework is that it captures the coupled physics governing contact-separation TENG performance, taking into account the exact measured surface roughness rather than using idealised statistical approximations.

The framework was validated against experimental measurements of the real contact area using optical interference microscopy between a rough PVS tribolayer and a PET counter-surface. The outcomes deviate only  0.5–2.6 percentage points across the entire experimental testing range. FEM predictions also exhibit transitional behaviour between the classical analytical models for small and full contact. Importantly, the ability to model roughness over the full real specimen within practical wall-clock times underscores the scalability of the framework.

The electrostatic module, which includes floating electrode boundary conditions and full three-dimensional field resolution, was validated through comparison with available analytical models and measurements. When in full contact, the FEM computed $V_{OC}$ agrees closely with the analytical approximations of Xu et al.\cite{XUVoc} and Guo et al. model\cite{guo2020derivation}. It also reflects the influence of fringing fields, resulting in approximately $17.5\%$ lower $V_{OC}$ responses for finite TENG, which had been partially disregarded by most existing analytical models. 

By coupling these results with the ODE model, the transient electrical response of the TENG was predicted under varying mechanical loads, operating frequencies, and resistive loads. The model showed qualitative agreement with both the analytical model of Xu et al. and experimental measurements, confirming that real contact area is the dominant parameter governing charge generation and electrical output. For resistive load optimisation, TENG I–V characteristics identified the optimal load resistance, which was compared with the experimental outcomes. This result demonstrates the robustness and suitability of the proposed predictive framework for investigating the effect of the real contact area in rough-surface TENG systems.

The present study assumed elastic material behaviour for tribo-layers (PVS and PET) to match the experimental configuration for the available loading range.  Notably, the framework is compatible with extended material models available in MoFEM and can be adapted to incorporate viscoelasticity, plastic deformation and surface wear for further research. In this implementation, the triboelectric charge density is experimentally calibrated and prescribed as an input parameter in the electrostatic simulations. Consequently, the model does not account for the dynamic evolution of charge with contact history or environmental factors. This limitation is common to many existing analytical and numerical TENG models and represents an important direction for future research aimed at achieving more comprehensive predictive capabilities.  In addition, charge leakage and interfacial recombination are also not explicitly modelled, which may explain remaining discrepancies between simulated and measured outputs, particularly at high load resistances in TENG circuit design. Further developments will focus on incorporating complex material behaviour, such as viscoelasticity,  large-deformation contact for large slopes of roughness and modelling dynamic charge evolution and environmental effects. 

The contact FEM approach enables direct evaluation of deformation gradients within the tribolayer, which can aid in resolving stress distributions and material response, i.e., flexoelectric material behaviour \citep{marks2025flexoelectricity}. From computational perspectives, multiscale strategies of combining the detailed contact resolution with electrostatic models can enable the simulation of other TENG modes (sliding, free-standing, rolling, etc.). Moreover, the methodology developed in this work as an open-source package is also readily transferable to other contact-driven advanced energy harvesting, virtual design and optimisation,  material optimisation, functional surface characterisation and tribological interfaces. 


\section{Definition of the experimental setup used for validation of the proposed framework}
\label{app1}

\subsection{Rough surface sample fabrication and characterisation}
\label{app1.1}

\subsubsection{Tribolayer preparation} To obtain a PVS (Polyvinyl Siloxane) sample with characteristic surface roughness, a random rough topography with a predefined areal RMS surface roughness (Sq) was designed and fabricated. The surface was designed numerically using an algorithm that considers a predefined power spectrum density. This numerical surface design tool was reported in \citep{perris20233d}. The surface design aimed to simulate naturally occurring multiscale topographies found in real engineering surfaces. The roughness of the designed surface had an $S_q$ value of $25 \mu m$.
\begin{figure}[!b]
    \centering
     \begin{minipage}{0.38\linewidth}
        \centering
        \includegraphics[width=\linewidth]{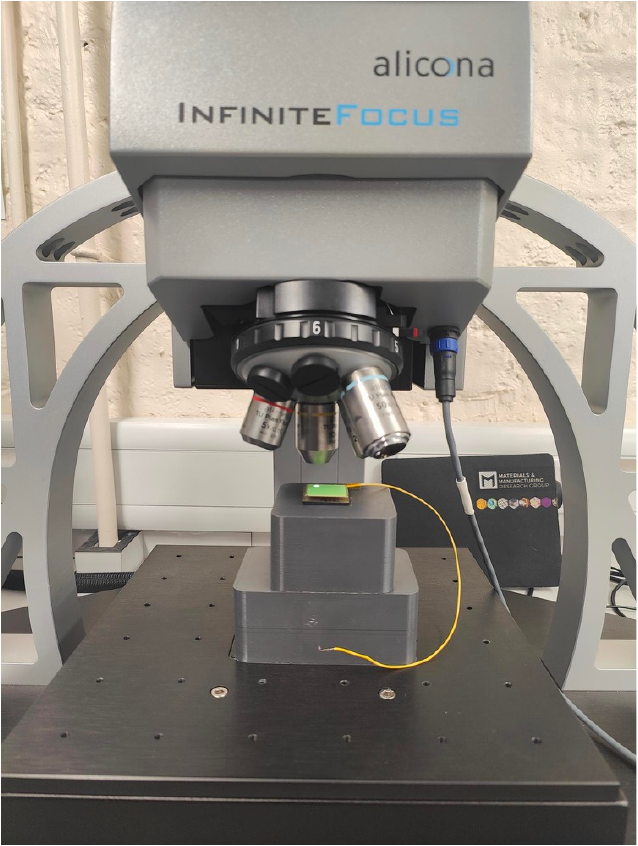}
        \caption*{(a)}
        \label{fig:alicona}
    \end{minipage}
    \hfill
    \begin{minipage}{0.48\linewidth}
        \centering
        \includegraphics[width=\linewidth]{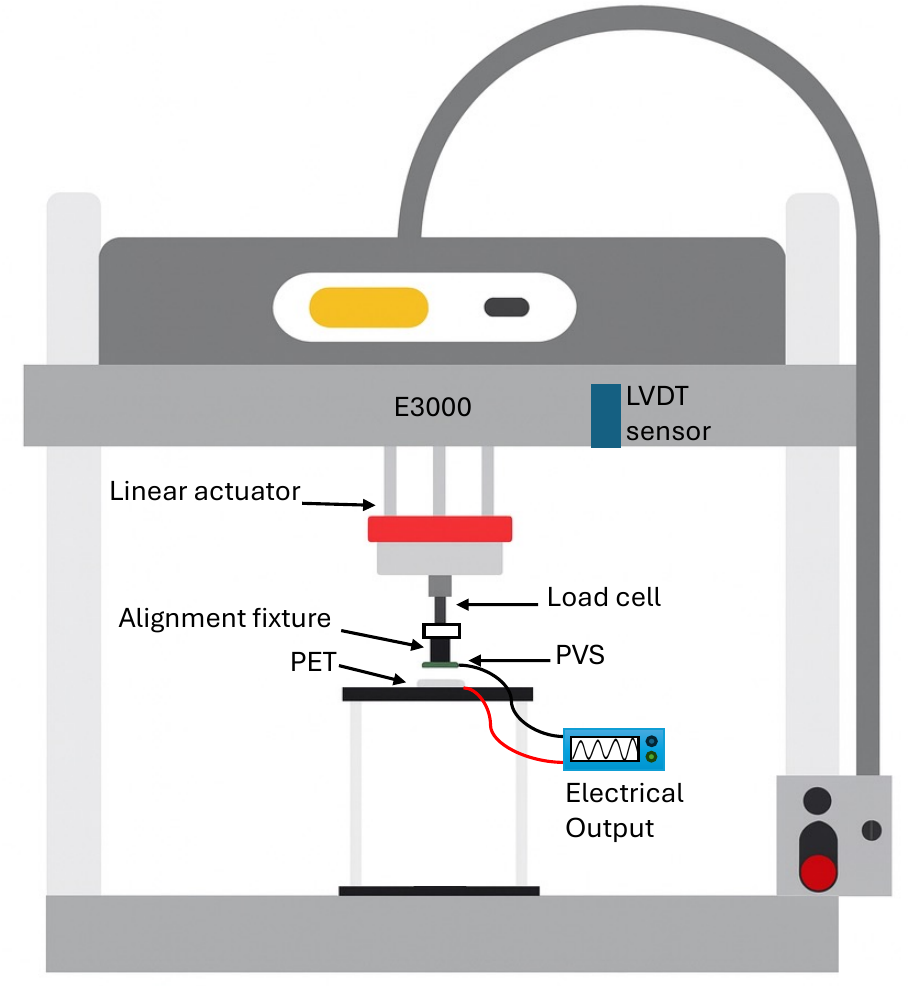}
        \caption*{(b)}
        \label{fig:test_rig_b}
    \end{minipage}
    \begin{minipage}{0.46\linewidth}
        \centering
        \includegraphics[width=\linewidth]{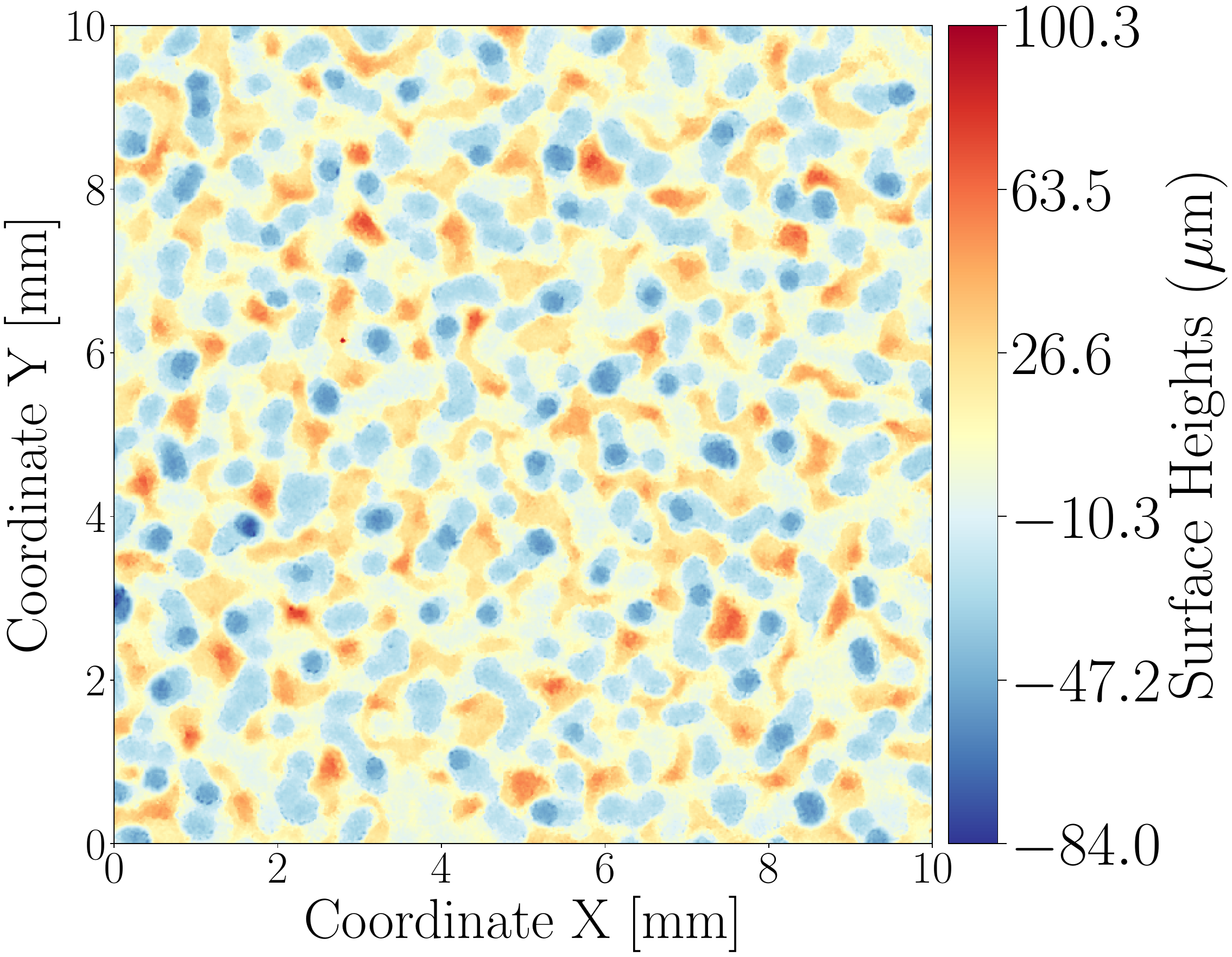}
        \caption*{(c)}
    \end{minipage}
    \hfill
    \begin{minipage}{0.51\linewidth}
        \centering
        \includegraphics[width=\linewidth]{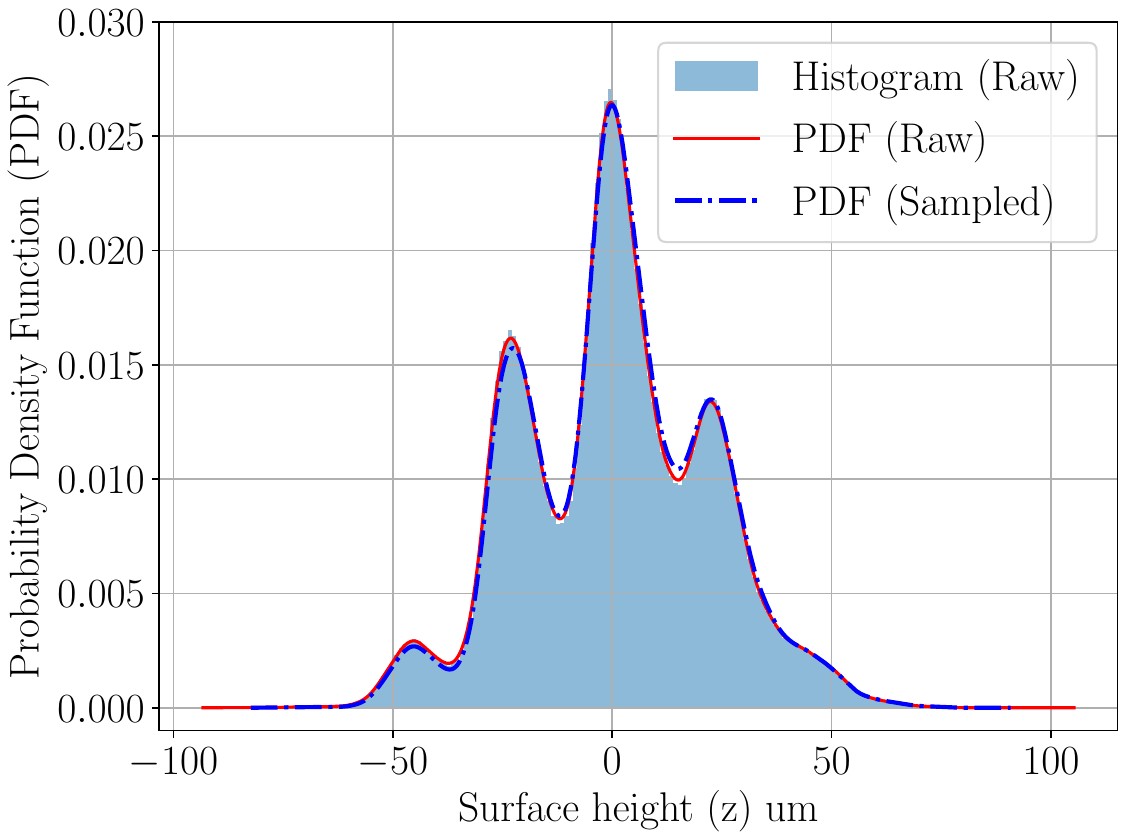}
        \caption*{(d)}
        \label{fig:suraface}
    \end{minipage}
    \caption{(a) Experimental setup for the Alicona profilometer used to acquire 3D topography of TENG PVS layers through vertical scanning optical interferometry to provide surface data for simulations and (b) schematic of the Instron test rig used for the contact separation PVS-PET TENG measurements. (c) Measured surface state for setup `A' showing experimental raw grid data showing the crests and troughs and the surface heights, and (d) raw height distribution and probabilistic surface density function (blue dashed curve) after removing outliers using local median filtering.}
  \label{fig:test_rig}
\end{figure}
A stereolithography (SLA) polymer resin technique (Formlabs Form 4 printer, USA) was used to manufacture the moulds containing the textured surfaces. Four variations of the textured surfaces were printed in the surrounding moulds. The surface topography was identical in each case. However, the depths of the mould cavities were varied to allow for optimisation of the tribolayer thickness for the mechanical testing. The polymer polyvinyl siloxane (PVS) (President - The Original, Coltene, Switzerland) was used to make replicas of the 3D printed surfaces. The two-part silicone resin was carefully applied to the 3D printed surface to avoid air bubbles and defects occurring during moulding. The uncured polymer resin is then compressed into the mould using a glass plate. The PVS resin is left to cure for 10 mins at room temperature and then separated from the mould. The imprinted textured surfaces are gently removed from the mould, and any excess polymer material is removed from the textured samples. This technique is fully outlined in our previous study on the moulding of rough surface topographies \citep{perris20233d} and in a study looking at the influence of surface roughness on TENG electrical output \citep{kumar2023multiscale}. 

\subsubsection{Surface Topography Measurement}
The surface topographies of the PVS samples were then scanned and characterised using a 3D optical profilometer (InfiniteFocus, Alicona-Bruker, Austria) as shown in \textbf{Figure~\ref{fig:test_rig}(a)}. All measurements were performed using a 5x objective lens. The optical scans confirmed that the designed topographies were 3D-printed and subsequently replicated via polymer casting, with the required surface properties for the triboelectric experiments. The optical data also ensured the topographies had been manufactured with the required properties and tolerances to allow accurate representation in the FE model. The scans were post-processed before being exported as an x-y-z point cloud dataset. 

\subsubsection{Surface data preprocessing}
The experimental surface grid data exhibited slight inconsistencies in the roughness parameter $S_{\textrm dq}$, varying between 0.33 and 0.36 with the number of grid points. Surface data also contain microscale noise and sharp peaks and valleys, which are commonly observed in experimental measurements~\citep{pradhan2025surface}. To address this, a minimal number of outliers was removed while preserving the probabilistic density distributions of the original and processed surface heights, as shown in~\textbf{Figure~\ref{fig:test_rig}(c-d)}. This allowed for mapping of the rough surface topography from the PVS sample onto the FE model.

\subsection{Experimental measurement of real contact area under varying load}
\label{app1.2}
During the experiment, the Interference Reflection Microscopy (IRM) was utilised, which operates on the principle of interference of reflected light from two surfaces: the glass–medium interface with PET and the specimen–medium of PVS interface. When the sample is illuminated with monochromatic light, part of the beam reflects from the coverslip while another portion reflects from the bottom surface of the specimen. The recombination of these beams produces constructive or destructive interference, generating contrast that depends on the distance between the specimen and the glass. Regions where the specimen is very close to the surface appear dark due to destructive interference, whereas regions farther away appear brighter. CMOS sensor camera (Pixelink C-Mount USB 3.0 Camera). 

\begin{figure}[!b]
    \centering
        \includegraphics[width=0.8\linewidth]{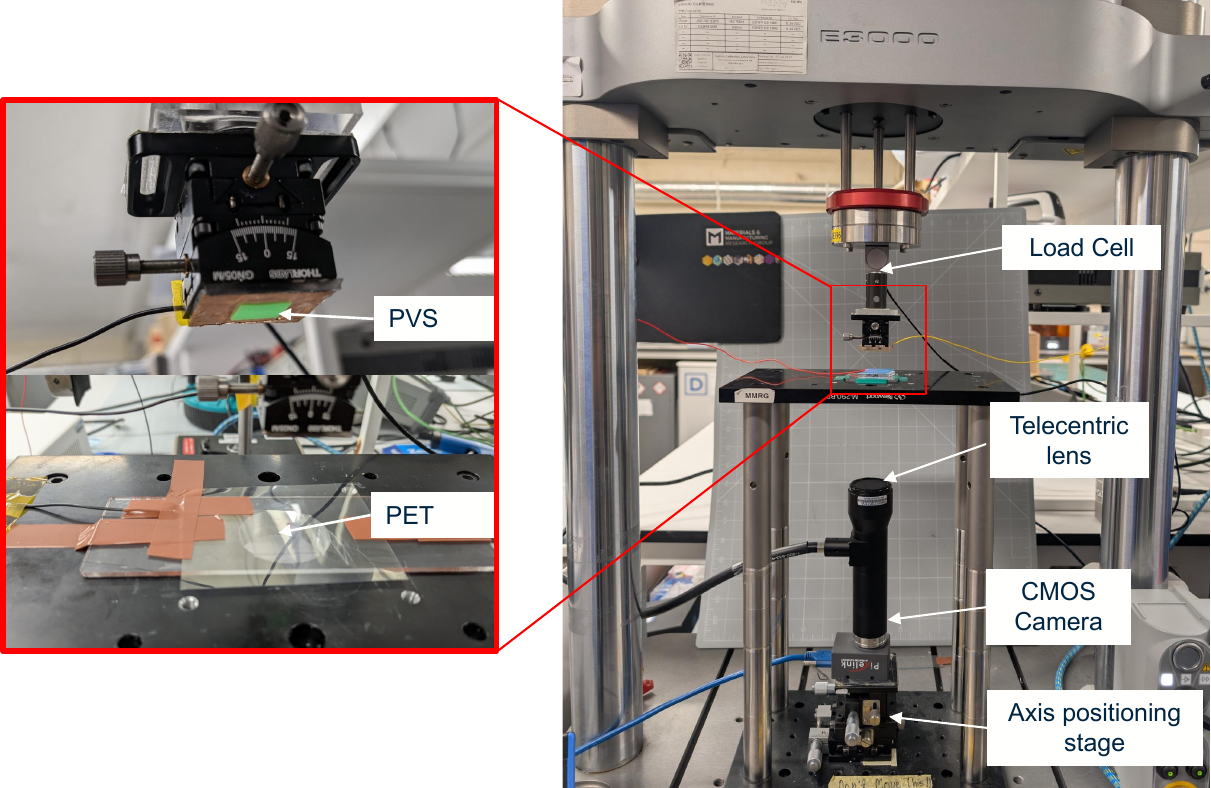}
    \caption{Optical setup for real contact area measurement between two tribo-layers}
    \label{fig:Contac_area_setup}
\end{figure}

This enables IRM to visualise cell–substrate adhesion and thin film thickness with high sensitivity.
In this work, the electrodynamic mechanical test rig was customised by~\cite{kumar2023multiscale} with a surface self-aligning facility with a 360° alignment bed and an IRM-based CMOS (Pixelink C-Mount USB 3.0 Camera) optical setup to perform highly controlled electromechanical measurements. This configuration allowed precise recording of the distribution of the real contact area under different loads. The probe was kept steady at a probe-to-surface separation distance of $2$mm. For comparison, the direct contact area illustrations were post-processed using ImageJ software and subsequently binarised with the OpenCV (Open Computer Vision) Python library.

\subsection{Charge density measurement and TENG fabrication and testing}
\label{app1.3}
Another critical triboelectric quantity, the true charge density at the PVS–PET interface, was measured using the method developed by \cite{kumar2026_truetribocharge}.
This approach enables direct and non-contact quantification of the charge generated at the interface while simultaneously capturing the true contact area during the triboelectrification event. The true charge density was calculated by dividing the total measured charge by the corresponding true contact area such that $\sigma_T = Q/A_r$. The value of true charge density was then used in the TENG FE model. 

Besides, a simple TENG device was fabricated and evaluated under a range of mechanical operating conditions to enable comparison with the TENG FE model. The device operated in vertical contact–separation mode with a dielectric–dielectric interface configuration. PVS and PET served as the active tribolayers in the true charge density measurements. Each surface had a conductive electrode bonded to its reverse side: a copper electrode for the PVS layer and a conductive indium tin oxide (ITO) coating on the PET layer. The complete device architecture is shown in \textbf{Figure~\ref{fig:test_rig}(b)}. Both electrodes were connected to an oscilloscope (InfiniiVision, Keysight Technologies, USA) to measure the open-circuit voltage generated during mechanical testing.

Mechanical loading was applied using cyclic excitation at various frequencies and normal loads to simulate operational conditions in which a TENG device experiences continuous vibration. These cyclic contact–separation events enabled continuous electrical power generation through repeated contact electrification. The mechanical testing was conducted using a high-precision linear electrodynamic fatigue testing machine (Electropuls E3000, Instron, UK). A bespoke test rig and the charge-measurement setup, detailed in \cite{kumar2023multiscale, kumar2026_truetribocharge}, were installed within the machine.

The TENG device was tested across a range of normal loads, oscillation frequencies, and external resistive loads, as outlined in Section  \ref{sec:Result_and_Discussion}. The purpose of this comparative study was to assess how variations in mechanical and electrical operating parameters influence the voltage and current output of the TENG. These experimental results were subsequently compared with the predictions from the FE model.

\subsection{Experimental setup for parallel plate capacitance measurement}
\label{app:exp_capa}
To validate the electrostatic FEM model, the capacitance of a parallel-plate configuration was measured using two ITO-coated glass substrates (sheet resistance $\leq 20 \Omega$/sq) as electrodes. Sample preparation: The ITO-coated glass substrates were cleaned using isopropanol and dried with compressed nitrogen to remove surface contaminants that could affect the measurement. Electrical connections were established using conductive clips attached to exposed ITO edges. The air gap between electrodes was precisely controlled using an Instron ElectroPulsE3000 electromechanical testing system equipped with a linear LVDT displacement sensor (accuracy $\pm$1 $\mu$m), allowing systematic variation of the gap in 10 $\mu$m increments. The electrodes were aligned using optical verification and load cell feedback ($\leq 0.5$N contact force) and then held to alignment with precision by the E3000 throughout the measurement range. Capacitance values were measured at each gap position using an LCR meter operating at 2 kHz with a 1 V AC test signal in parallel capacitance mode.  A settling time of 30 seconds was allowed before each measurement to ensure mechanical stabilisation. Three repeated measurements were taken at each gap position and averaged to improve accuracy, with environmental conditions maintained at $18^{\circ}C$ and $50\%$ relative humidity. Open-circuit compensation was performed before measurements to null parasitic capacitance from test leads and connections. The configuration is also summarised in \textbf{Table \ref{tab:capacitance_exp}}.

\begin{table}[H]
\centering
\caption{Experimental setup and measurement parameters for capacitance validation}
\label{tab:capacitance_exp}
\begin{tabular}{ll}
\hline
\hline
\textbf{Category} & \textbf{Specification} \\
\hline
Electrode material & ITO-coated glass substrates \\
Electrode size & $25 \times 25$ mm$^2$ \\
Contact force & $\leq 0.5$ N \\
Measuring Device & RS PRO LCR-6002 meter (model 117-6718)  \\
    AC test voltage & 1 V \\
Environment & 18$^\circ$C, 50\% RH \\

\hline
\hline
\end{tabular}

\end{table}
\medskip
 
\textbf{Acknowledgements} \par 
The authors acknowledge support from the UK Engineering and Physical Sciences Research Council (EPSRC) funding for the doctoral studentship of the first author at the University of Glasgow, Ref. 2812574: (‘Transient simulation of triboelectric nanogenerators considering surface roughness'). Authors also acknowledge  EPSRC part-funding of the experimental work through grant Ref. EP/V003380/1 (‘Next Generation Energy Autonomous Textile Fabrics based on Triboelectric Nanogenerators’). 

\medskip

\newpage
\medskip
\setcounter{page}{1}
\markboth{SUPPORTING INFORMATION}{SUPPORTING INFORMATION}
\renewcommand{\theequation}{E\arabic{equation}}
\setcounter{equation}{0}
\renewcommand{\thefigure}{S\arabic{figure}}
\setcounter{figure}{0}
\section*{Supporting Information} 

Supporting Information is available from the author.
\section*{SI-1: Multiphysics TENG Simulation Pipeline}
\label{app:Simulation Pipeline}
The TENG responses are computed through a sequential multiphysics mapping ($\mathcal{R} \;\rightarrow\; \mathcal{C} \;\rightarrow\;E \;\rightarrow\; O),
$ that define roughness projection ($\mathcal{R}$), contact mechanics $(\mathcal{C})$, electrostatics $({E})$, and solving circuit dynamics ODE $({O})$ are carried out in series. The detailed steps are noted below:

\begin{algorithm}
\begin{algorithmic}[1]
\caption*{\textbf{TENG modelling pipeline in MoFEM}}
\Statex
\Require Surface height heights $h(x,y)$; TENG geometry  and thickness (CAD model); mechanical parameters $(E,\nu)$; dielectric permittivity $\varepsilon$; tribocharge density $\sigma_T$; excitation kinematics $z(t,f)\in(0,z_{\max})$; MoFEM installed$^*$.
\Statex
\Statex \textbf{Step 1:  Rough surface projection on mesh; $H(x,y,z)$}
\State Read/generate surface roughness data (e.g., $h(x,y)$, $S_{dq}$, $S_{q}$)
\State Generate surface grid.
\State Construct a finite element volumetric mesh of the deformable TENG layer.
\State  Project the surface heights onto the mesh using the roughness projection tool in MoFEM.
\Statex
\Statex \textbf{Step 2: Contact problem with KKT conditions solve; $\nabla\!\cdot\!\boldsymbol{\sigma}(\mathbf{u}) = 0$}
\State Apply load increments and time step.
\State Run contact simulation employing  MoFEM contact executable.
  \State Get the contact area ratio $\alpha = A_r /A_n$ with applied pressure.

\Statex
\Statex \textbf{Step 3: Electrostatics problem with interface conditions solve; $\nabla\!\cdot(\varepsilon \nabla \phi) = 0$ }

  \State Build a multilayer dielectric electrostatic mesh with air around it. 
   \State Scale surface charge density homogenising with the contact area ratio: $\sigma_{eff} = \alpha \sigma_T$.
  \State Run consecutive electrostatic simulations utilising the MoFEM Electrostatic executable.
  \State Compute $V_{OC}$ and $C$ for different air gaps $z(t)\in(0,z_{\max})$
\Statex
\Statex \textbf{Step 4: Circuit ODE solve; $dQ/dt = \frac{1}{R} \left(V_{OC} - \frac{Q}{C}\right)$}
  \State Map the air gap with distance to time using excitation frequency.
  \State Solve TENG circuit ODE for transferred charge $Q(t)$ using solve\_ivp library in Python.
  \State Post-process quantitative load and/or frequency-dependent response, etc. 
\Statex
\end{algorithmic}
$^*$ \href{https://mofem.eng.gla.ac.uk}{https://mofem.eng.gla.ac.uk}
\end{algorithm}

\section*{SI-2: Modelling finite TENG}
\label{app:Modelling finite TENG}
\begin{figure}
    \centering
    \includegraphics[width=\linewidth]{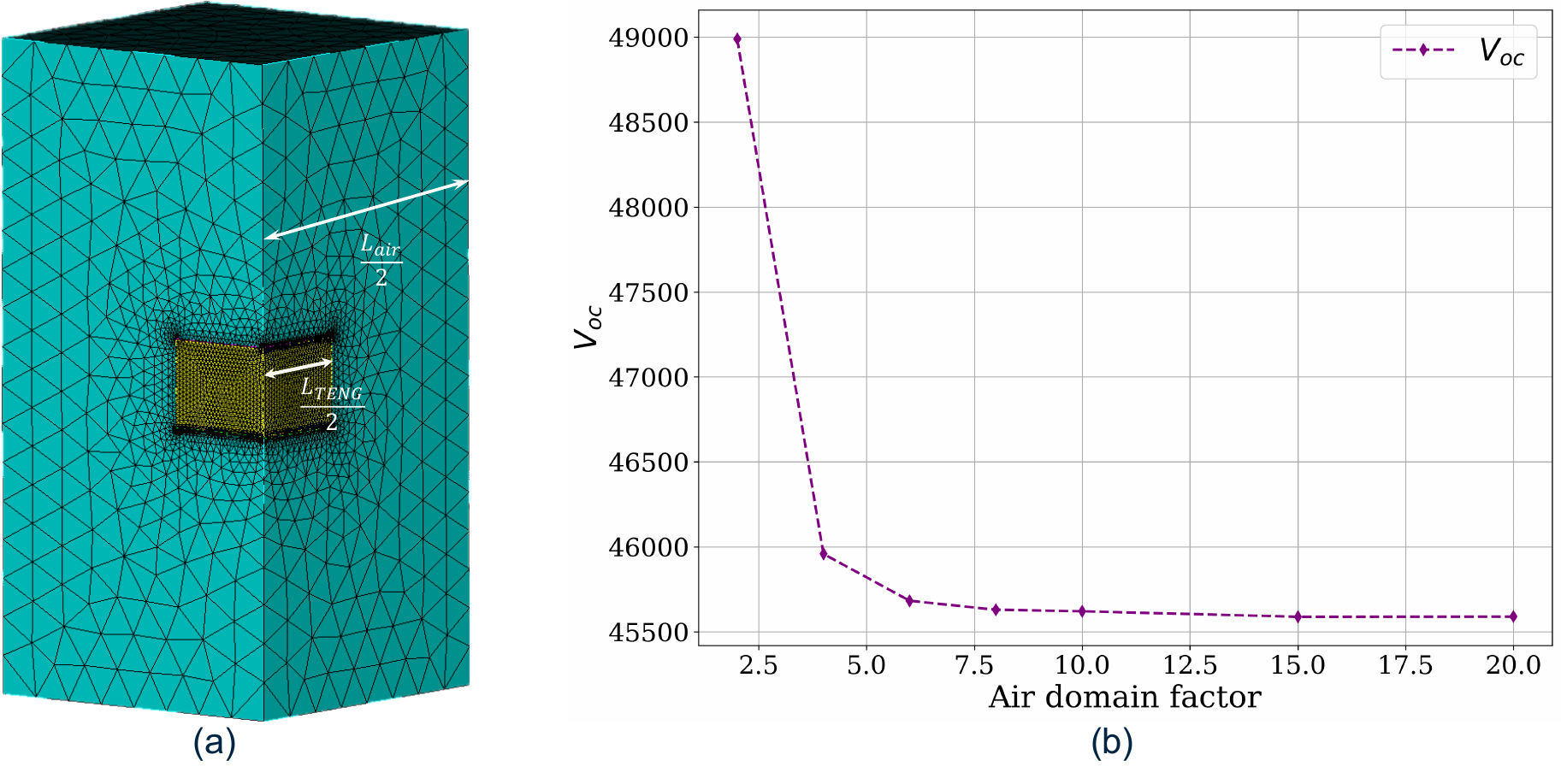}
    \caption{(a) Finite element mesh of a TENG device embedded in a surrounding air domain used to approximate the far-field electrostatic boundary. (b) Convergence study of the $V_{oc}$ with respect to the air domain factor.}
    \label{fig:Mesh_air}
\end{figure}

Modelling a finite TENG device surrounded by air requires careful consideration of the far-field boundary to avoid artificial boundary effects. Therefore, the size of the surrounding air domain was optimised to ensure that the far-field electric field distribution does not influence the computational boundary. Therefore, the mesh and computational domain were generated following the criterion that the air domain size is defined relative to the characteristic dimension of the TENG device, such that:

\begin{equation}
\text{Air domain factor} = \frac{L_{\text{air}}}{L_{\text{TENG}}},
\end{equation}

where $L_{\text{air}}$ represents the characteristic length of the surrounding air region and $L_{\text{TENG}}$ denotes the characteristic length of the TENG device. The influence of the air domain size has been studied on a $25\times25$ with $\sigma_T$ = $ \pm \,80\mu C/m^2$ TENG by varying the air domain factor, as shown in \textbf{Figure~\ref{fig:Mesh_air}}. For sufficiently large air domain factors, the results converge, indicating that the computational boundary is far enough from the device to accurately approximate the behaviour of a finite TENG. Based on this convergence study, an air domain factor of 10 was used in all subsequent simulations.

\subsection*{SI-2a: Effect of TENG Lateral sizes}
\label{app:Effect_of_TENG_Lateral_size}

\begin{figure}[H]
    \centering
      \centering
        \includegraphics[width=0.93\linewidth]{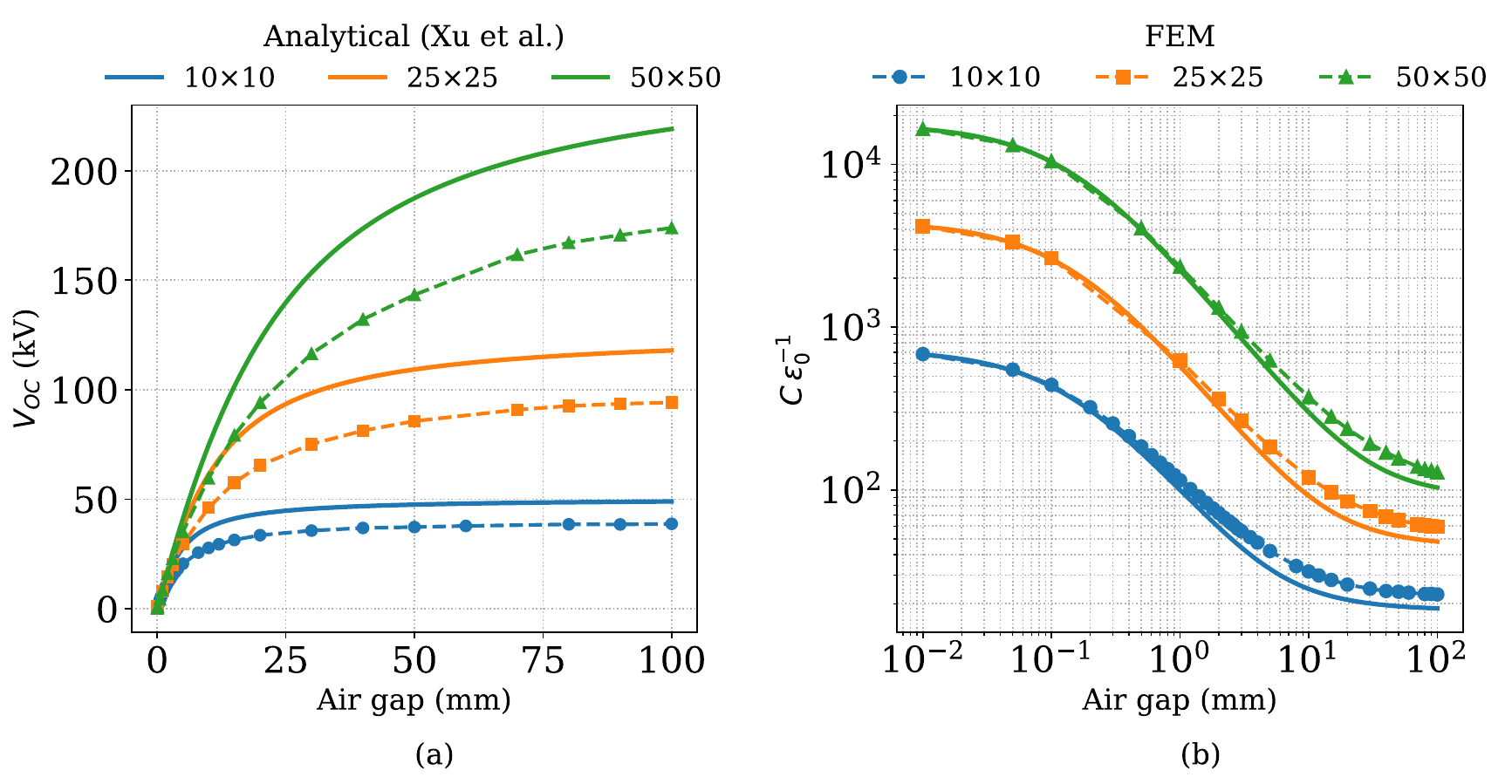}
    \caption{Effect of TENG lateral sizes ($10\times10, 25\times25,\, \text{and} \, 50\times50 mm$) highlight the influence of TENG surface area on both voltage generation and capacitance with air gap (a) $V_{OC}$ and 
    (b) corresponding capacitance variation.}
  \label{fig:Voc_capa_LW}
\end{figure}

The influence of the TENG surface when in full contact on the electrostatic response under varying air-gap separation has been presented in \textbf{Figure~\ref{fig:Voc_capa_LW}}. It shows that the open-circuit voltage increases minimally for the smaller air gaps when the device size increases. However, the increase is evident and saturates at the larger air gaps.
This enhancement arises from the increased total tribocharge associated with a larger effective surface area, which strengthens the electric field between the electrodes and results in higher potential differences. In this case, the breakdown voltage effect has been disregarded since $V_{OC}$ within the air gaps range from 0 to 100 mm, which is still lower than the air breakdown voltage at 1 atm calculated by Paschen's law [SR1, SR2].

The corresponding capacitance variation shown in Figure~\ref{fig:Voc_capa_LW}(b) also illustrates the strong dependence on both air gap and device size. For all configurations, the capacitance decreases rapidly with increasing separation due to the reduced electrostatic energy. On the flip side,
the larger devices exhibit higher capacitance values across the entire air-gap range, reflecting the direct proportionality between capacitance and TENF surface area. Notably, the rate of capacitance decay with air gap remains similar for all sizes, indicating that the geometric scaling primarily affects the magnitude rather than the functional form of the electrostatic response.

\section*{SI-3: Homogenisation of tribocharge quantities}
\label{app2}

\begin{figure}[!b]
    \centering
    \begin{minipage}[b]{0.06\linewidth}
        \centering
        \includegraphics[width=\linewidth]{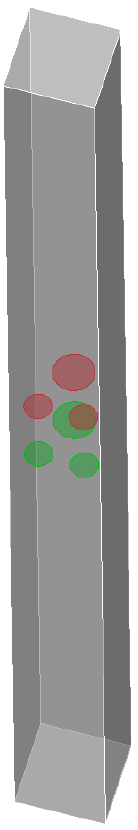}
        \caption*{(a)}
        \label{fig:patch1_setup}
    \end{minipage}
    \hfill
    \begin{minipage}[b]{0.33\linewidth}
        \centering
        \includegraphics[width=\linewidth]{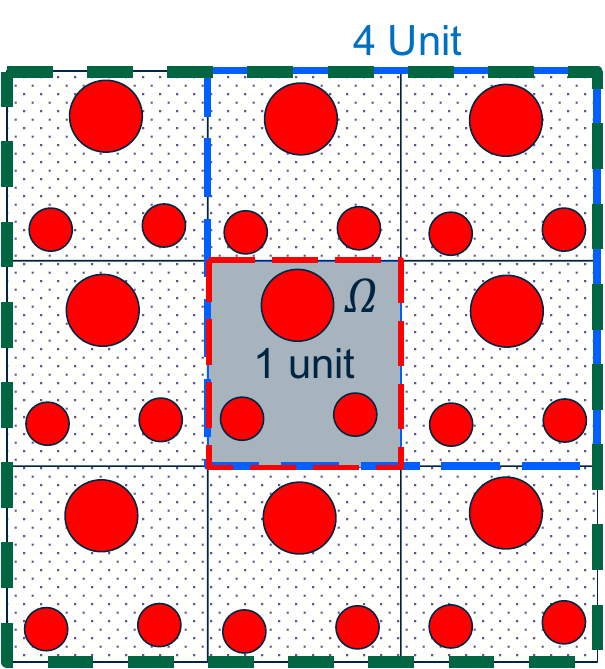}
        \caption*{(b)}
        \label{fig:patch1}
    \end{minipage}
    \hfill
    \begin{minipage}[b]{0.45\linewidth}
        \centering
        \includegraphics[width=\linewidth]{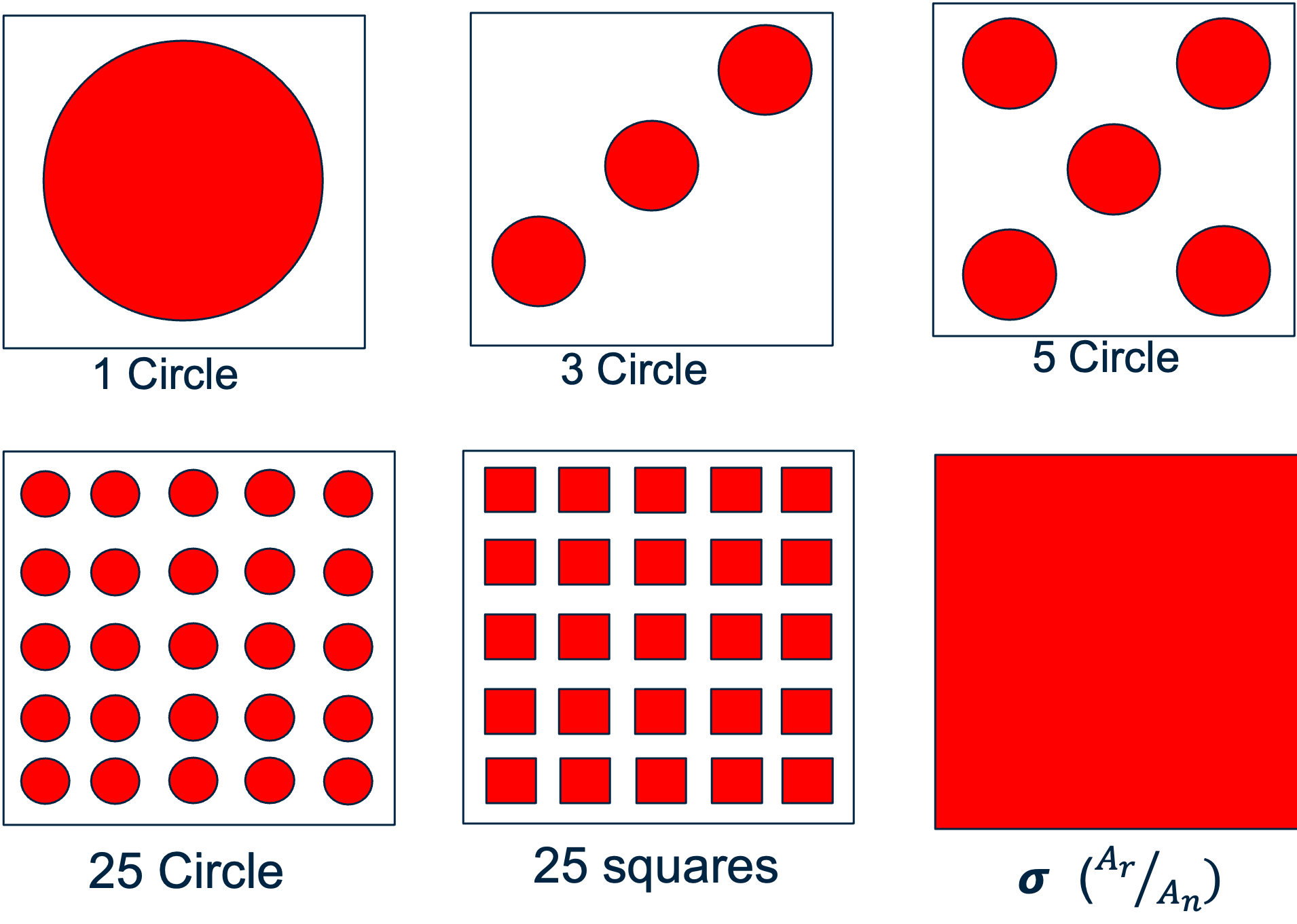}
        \caption*{(c)}
        \label{fig:patch2}
    \end{minipage}
       \caption{Setup for homogenisation analysis with different distributions of charged patches. (a) Representative unit cell with negative (red)  and positive (green) charged patches with air gaps between them. 
    (b) $\Omega$ as the unit cell and subsequent periodic arrangement for testing with 4 and 9 unit cells 
    (c) Different patch configurations are used for homogenisation, including 1, 3, and 5 circular patches, 25 square patches, and a surface with scaled charges.}   
    \label{fig:charged_patch_homogenisation}
\end{figure}

TENG voltage and current are influenced by the applied load \cite{XUVoc}, which is consistent with the evolution of the real contact area under load. This indicates that the charges primarily stay within the actual contacting regions, from which the electric fields are generated. Therefore, identifying how the electric field varies with the contact area may provide insight into the total electric field distribution on the contact surface where charges accumulate. Although triboelectric charge transfer may exhibit nonlinear dependence on pressure and contact time, several experimental studies have demonstrated an approximately proportional relationship between generated charge and real contact area for similar dielectric pairs.
For this homogenisation test, \textbf{Figure~\ref{fig:charged_patch_homogenisation}} illustrates the nine different test cases conveyed by varying the charge distribution, size, and shape of the charged patches. In the first setup, shown in Figure~\ref{fig:charged_patch_homogenisation}(a), 
a column-like representative unit cell $\Omega$ was considered, consisting of circular positive and negative charges separated by a gap of 10~cm. Multiple copies of this unit cell were then merged into a periodic arrangement and tested with $4$, $9$, and $25$ unit cells, each surrounded by air at both the top and bottom. The radius of the charged patches within each cell was systematically increased so 
that the total charged area within the unit cell can be represented by $A_r$. Then, the norm of 
the electric field was evaluated only over the central unit cell to understand how the nonuniformity introduced by neighbouring charged patches affects 
the internal field distribution. In this regard, we observed a clear linear relationship that remained consistent within the unit cell, irrespective of whether it was surrounded by $4$, $9$, or $25$ cells with alternating positive and negative 
charges at the same gap, as shown in Figure~\ref{fig:charged_patch_homogenisation}(b). In this case, the effects of the neighbouring 
cells cancel each other out, resulting in a consistent and stable linear relationship within the central unit cell.

In the second setup, shown in Figure~\ref{fig:charged_patch_homogenisation}(c), we considered different distributions of charged patches within a $10 \times 10$ surface area  $A_n$, including $1$, 
$3$, and $5$ circular patches, arrays of $25$ circular  (or $25$ square) 
patches. For each case, the radius (or side length) of the patches was increased 
to represent $A_r$.  
Additionally, we considered a uniformly charged area $A_n$, where the 
charge density $\sigma$ was scaled according to different values of $A_r/A_n$ ratios. 

\begin{figure}[H]
    \centering
     \begin{minipage}{0.45\linewidth}
        \centering
        \includegraphics[width=\linewidth]{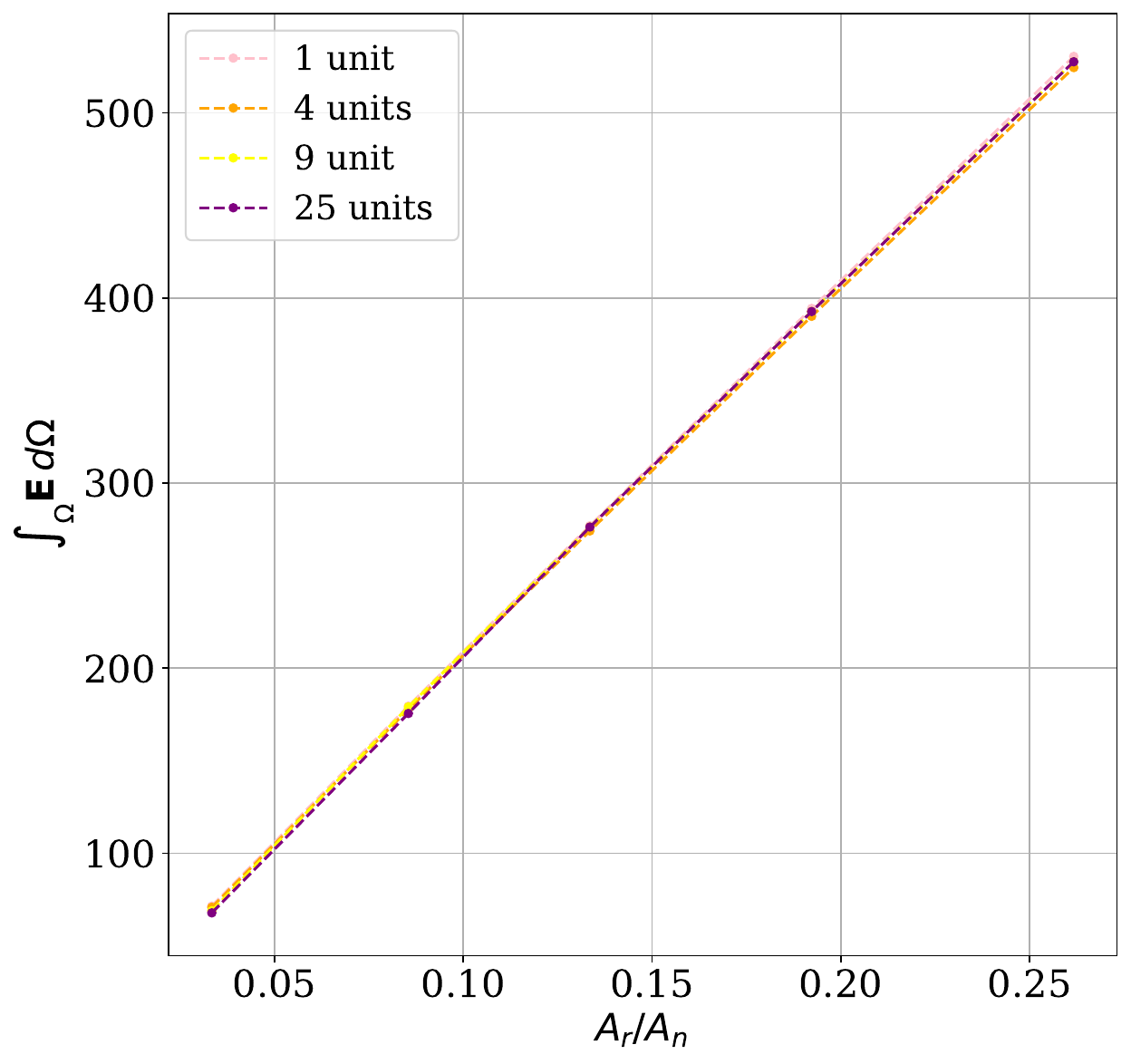}
        \caption*{(a)}
        \label{fig:sup2}
    \end{minipage}
    \hfill
    \begin{minipage}{0.46\linewidth}
        \centering
        \includegraphics[width=\linewidth]{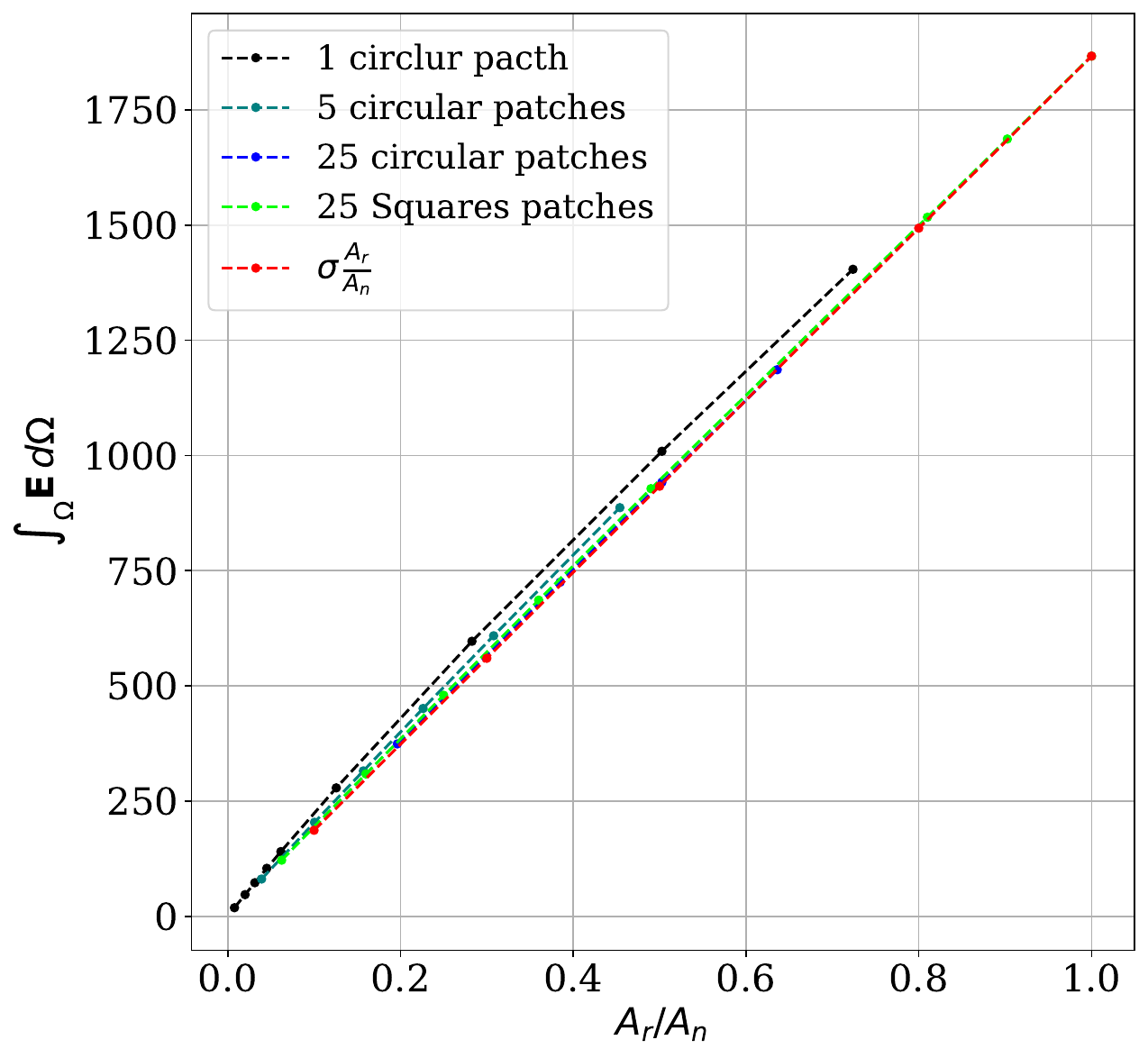}
        \caption*{(b)}
        \label{fig:sub3}
    \end{minipage}
    \caption{Homogenisation analysis of electrically charged patches.
  (a) The integral response $\displaystyle \int_{\Omega} \mathbf{E} \,\mathrm{d}\Omega$ is shown for different numbers of unit cells (1, 4, 9, and 25). 
  The outcomes overlap, confirming that even a single unit cell provides a reliable representation of the homogenised response. 
  (b) Comparison across different charged-patch configurations with different radius considering 1, 3, 5, and 25 circular patches and a surface with scaled charge distribution. 
  Despite their distinct geometrical layouts, all cases collapse onto the linear trend, highlighting that the effective electrostatic response linearly changes with contact area.}
  \label{fig:charged_patch_homogenisation_result}
\end{figure}
From these tests, first we observed the linear relationship between the electric field quantity and the charged area when the charge density was scaled as shown in \textbf{Figure~\ref{fig:charged_patch_homogenisation_result}}. The closest outcomes of it are found when densely distributed 25 circular and 25 square patches are considered, which actually can represent a rough surface closely. On the contrary, the most deviation from linearity was observed for the single circular patch case. Interestingly, it suggests that the most irregular or random patch distributions may bring it closer to perfect linearity, as seen in the case of 25 patches. This observation aligns with the fact that when the surface peaks are mostly random, their electric field effects can therefore be effectively represented by charge scaling.

\section*{SI-4: Computation of transferred charge and output voltage}
\label{app3}
\begin{figure}[t]
    \centering
    
     \begin{minipage}{0.45\linewidth}
        \centering
        \includegraphics[width=\linewidth]{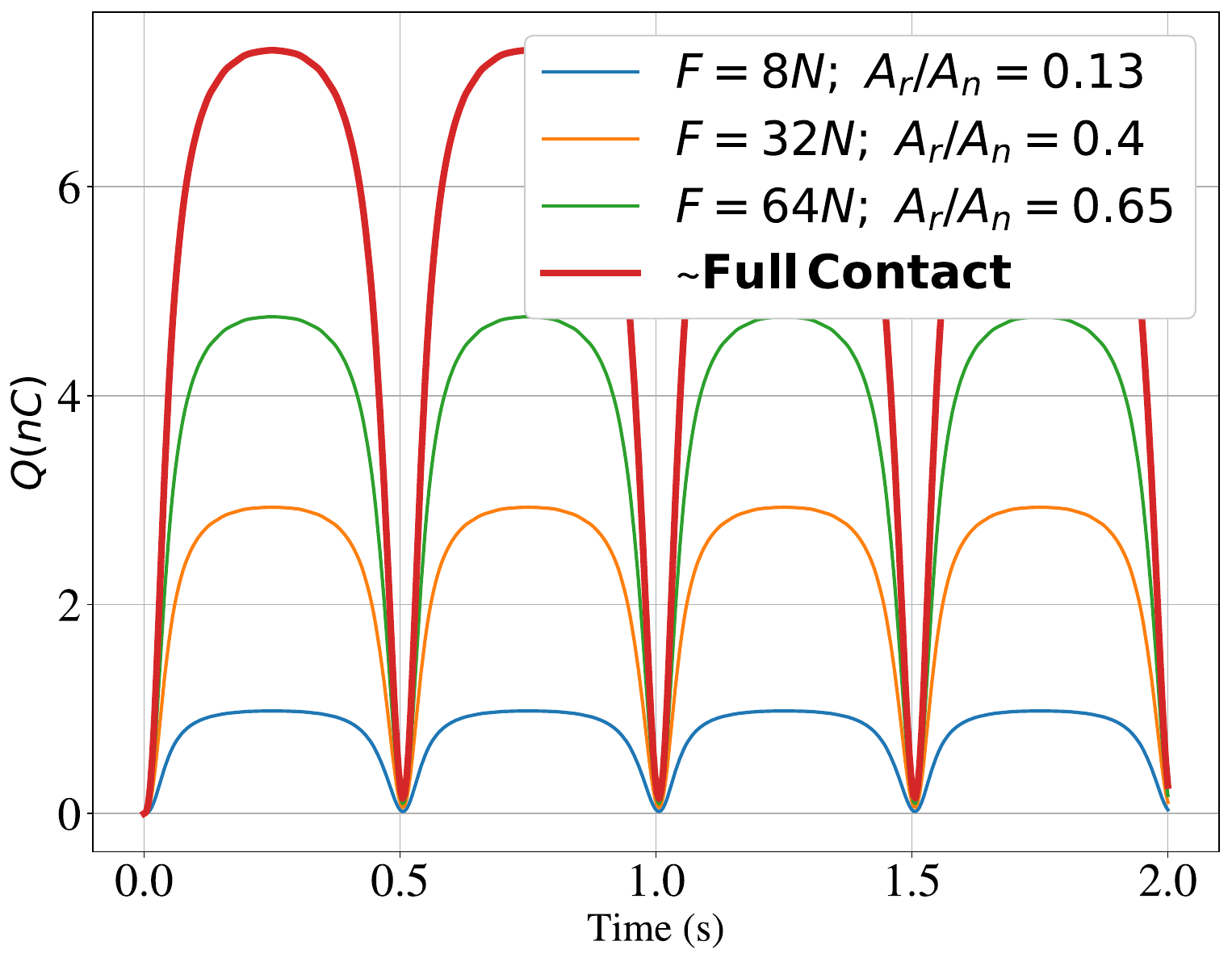}
        \caption*{(a)}
        \label{fig:Q_F}
    \end{minipage}
    \hfill
    \begin{minipage}{0.46\linewidth}
    \centering
    \includegraphics[width=\linewidth]{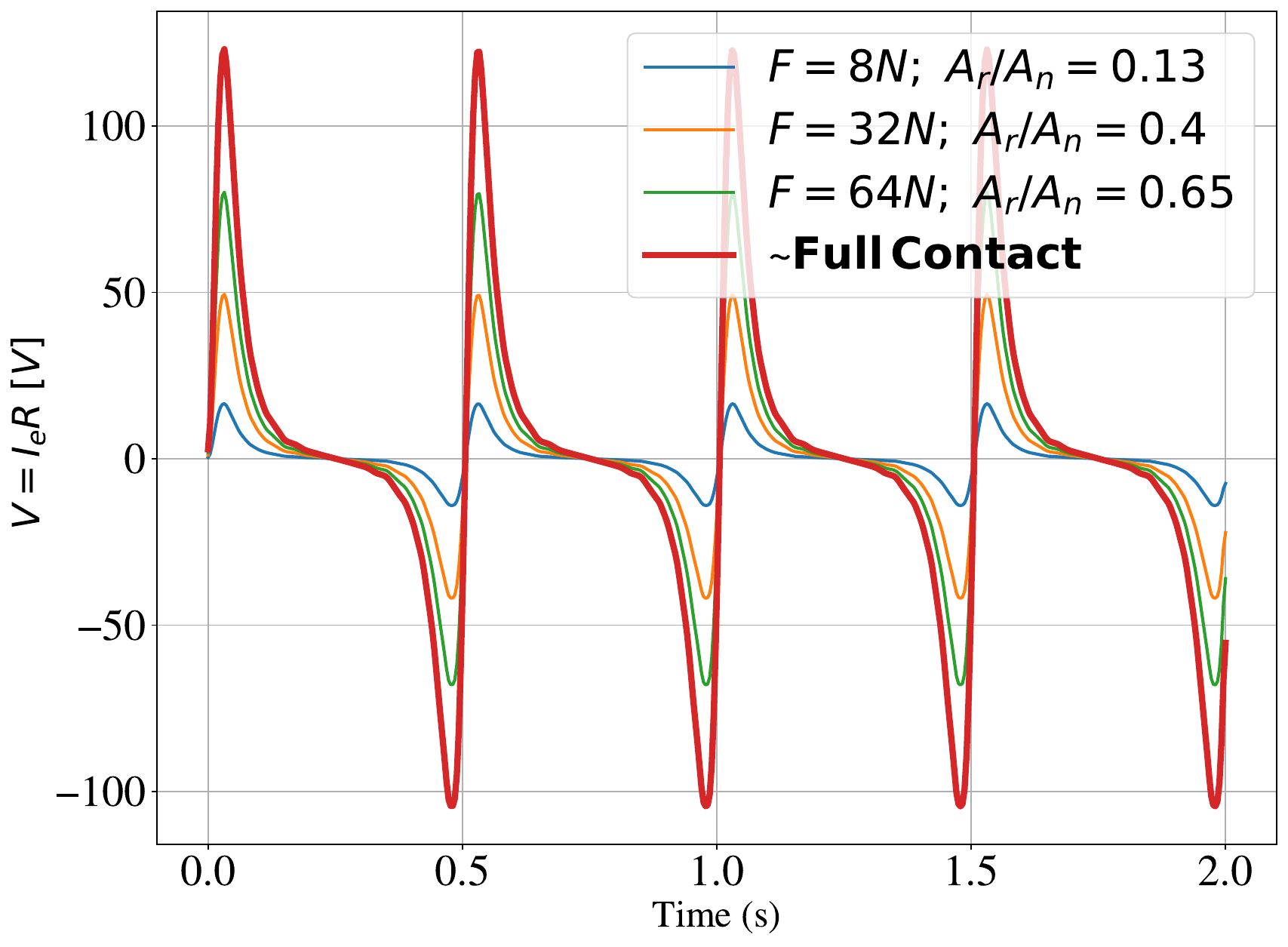}
    \caption*{(b)}
    \label{fig:V_F}
    \end{minipage}
  
    \caption{Illustrates the influence of excitation frequency on the electrical output characteristics of the TENG under varying contact conditions under a fixed resistive load of $100 M\Omega$ at $2$Hz for tapping frequency for $2$ seconds a)  Evolution of the transferred charge $Q(t)$ for different applied normal loads, illustrating the increase in charge transfer with increasing real contact area and (b) Corresponding output voltage $V(t)$.}
  \label{fig:IV_vs_F}
\end{figure}

The results shown in \textbf{Figure~\ref{fig:IV_vs_F}} are consistent with the assumptions of the ODE-based TENG circuit model, in which the time-dependent electrical response is governed by the evolution of the transferred charge $Q(t)$, the voltage, and the effective capacitance. The increase in the applied normal load leads to a larger real contact area ratio $A_r/A_n$, resulting in higher transferred charge and voltage amplitudes, in agreement with the model assumption that tribocharge generation scales with $A_r/A_n$.

\begin{figure}[H]
    \centering
        \begin{minipage}{0.47\linewidth}
        \centering
        \includegraphics[width=\linewidth]{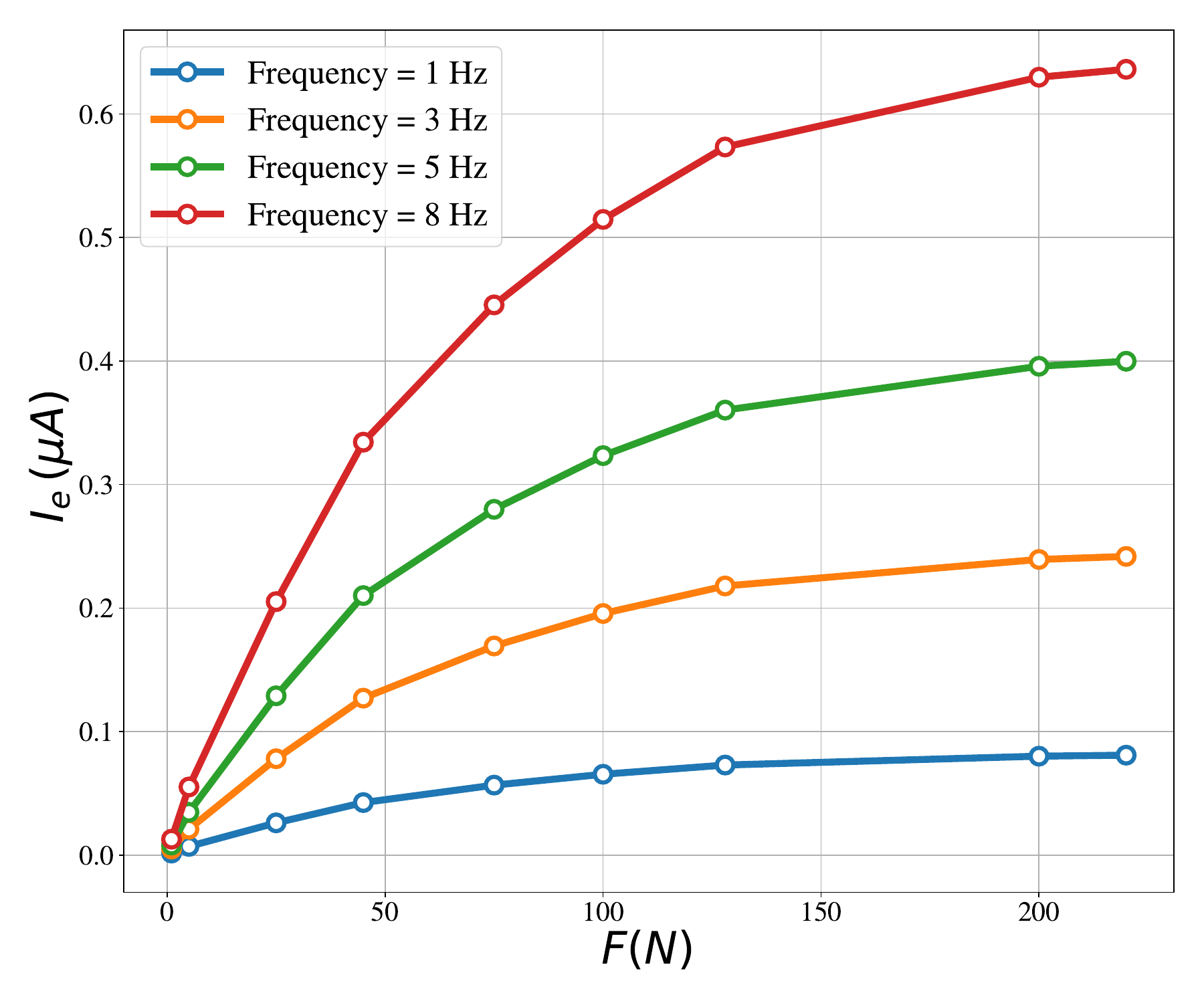}
        \caption*{(a)}
        \label{fig:Isc_F}
    \end{minipage}
    \hfill
    \begin{minipage}{0.48\linewidth}
    \centering
    \includegraphics[width=\linewidth]{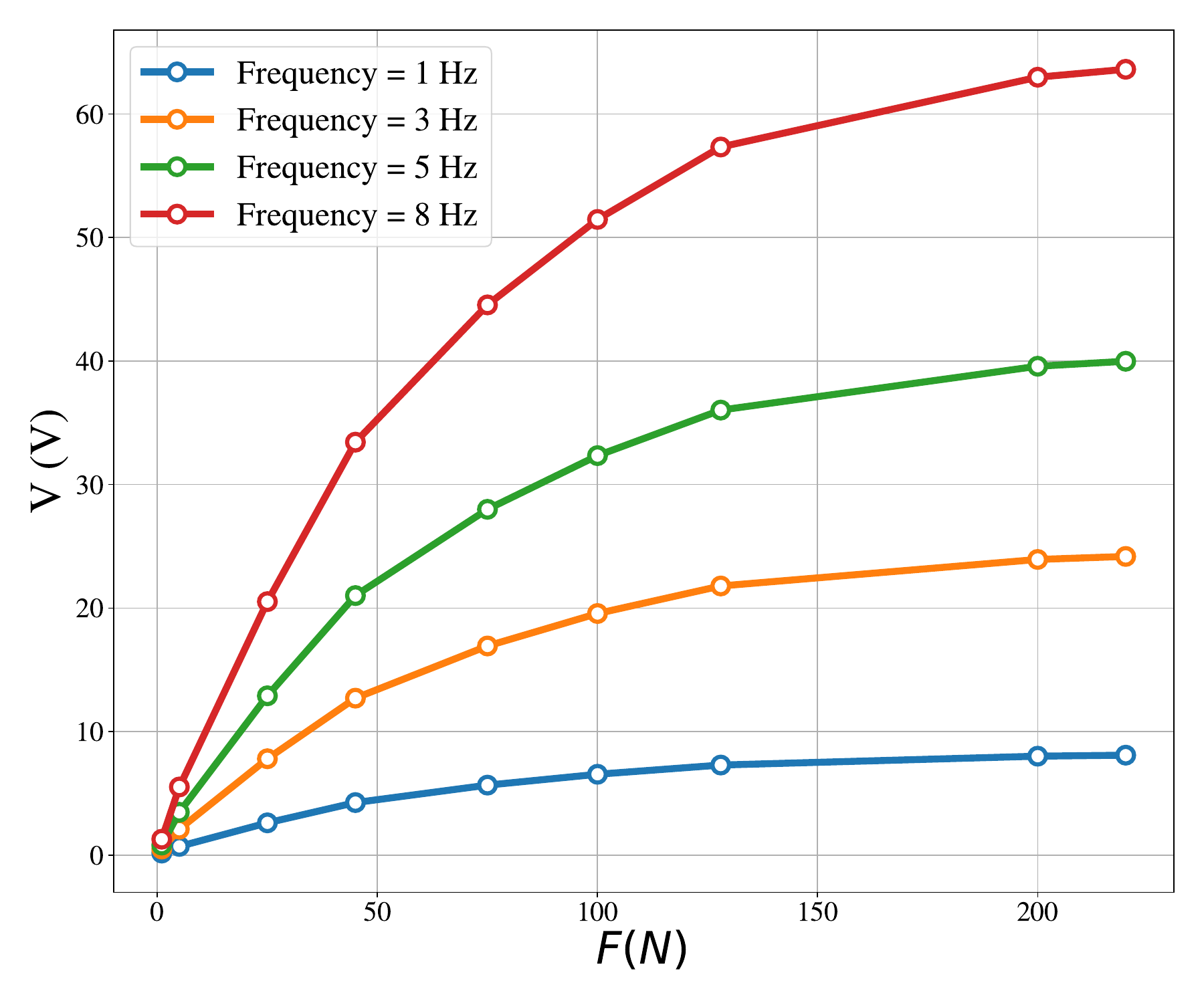}
    \caption*{(b)}
    \label{fig:V_F_f}
    \end{minipage}

    \caption{(a) Variation of peak-to-peak current $I_{e}$ with contact area ratio for different excitation frequencies, highlighting the combined influence of contact area and frequency on current output. (b) Computed voltage output as a function of applied normal force for different excitation frequencies under demonstrating increased voltage levels at higher forces and frequencies. (c) Variation of p2p current ($I_{e}$) with contact area ratio for different frequencies, showing that higher frequencies result in greater current output and stronger sensitivity to $A_{r}/A_{n}$ (d) computed voltage upon resistive loading, however saturates at high forces.}
  \label{fig:peak_current_volatge_Force}
\end{figure}

Furthermore, the non-linear increase (because of the relation between $F$ and  $A_r/A_n$) of $I_e$ with contact area ratio observed in Figure~\ref{fig:peak_current_volatge_Force}(a), where the current is obtained directly from $\mathrm{d}Q/\mathrm{d}t$. In addition, the current and voltage outputs enhanced at higher excitation frequencies, shown in \textbf{Figure~\ref{fig:peak_current_volatge_Force}}, arise from the faster periodical variation of $Q(t)$ and $C_T(t)$ within each contact-separation cycle, while the reduced rate of increase at higher frequencies reflects the finite charge transfer time implicitly captured by the circuit dynamics. Overall, these results validate the use of contact-area-scaled electrostatic inputs within the ODE framework for predicting frequency- and load-dependent TENG performance.

\section*{SI-5: Contact area analysis to compare load-dependent $V_{OC}$ in Xu et al.~\cite{XUVoc}}

\subsection*{Analytical contact area calculation}
Analytically, for statistical surface roughness, the fraction $A_r/A_n$, can be evaluated as a function of the nominal pressure $p = F/A_n$.  The BGT asymptotic model [SR3] provides a linear approximation at low loads to the Persson et al. model [SR4].

In the BGT asymptotic model, the contact area ratio is assumed to vary linearly with pressure as

\begin{equation}
\frac{A_r}{A_n} = k \, \frac{p}{E^{*} S_{dq}}; k=\sqrt{2\pi}.
\end{equation}

In Persson's contact model, the contact area ratio is expressed as

\begin{equation}
\frac{A_r}{A_n} = \mathrm{erf} \left( \frac{\sqrt{2}\, p}{E^{*} S_{dq}} \right),
\end{equation}

where $\mathrm{erf}(\cdot)$ is the error function and the effective composite modulus $E^*$ is given by,

\begin{equation}
E^{*} = \left( \frac{1-\nu_1^2}{E_1} + \frac{1-\nu_2^2}{E_2} \right)^{-1},
\end{equation}

where $E_1$ and $E_2$ are the Young's moduli, and $\nu_1$ and $\nu_2$ are the Poisson's ratios of the two contacting materials.
\subsection*{Surface height maps and contact area comparison }
\begin{figure}[H]
    \centering
     \begin{minipage}{0.49\linewidth}
        \centering
        \includegraphics[width=\linewidth]{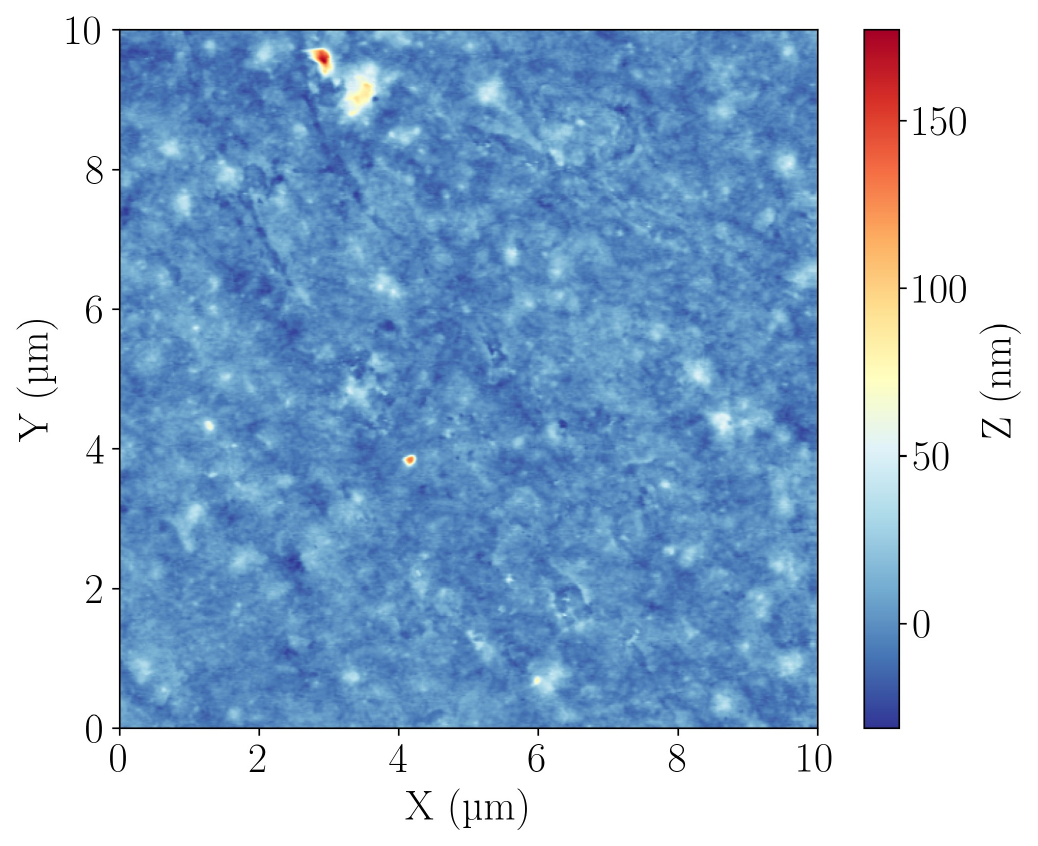}
        \caption*{(a)}
        \label{fig:sub1}
    \end{minipage}
    \hfill
    \begin{minipage}{0.49\linewidth}
        \centering
        \includegraphics[width=\linewidth]{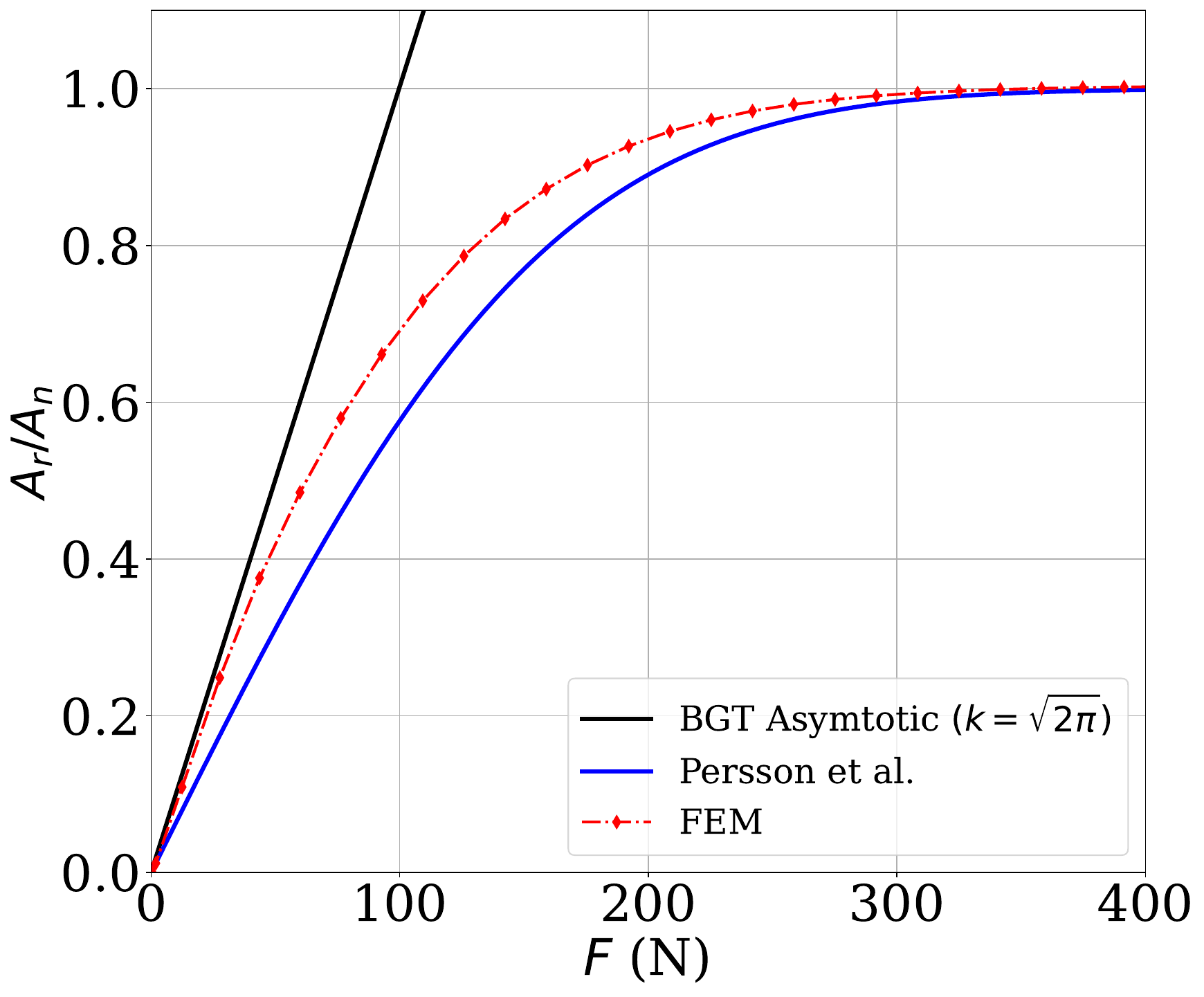}
        \caption*{(b)}
        \label{fig:patch2_result}
    \end{minipage}
    \caption{(a) PDMS surface height maps with $S_{dq}$= 0.22 based on the AFM measures, which have been reproduced with permission \cite{XUVoc}, (copyright: elsevier, 2020). 
  (b) Comparison of contact area investigation between FEM and analytical outcome while pressing a rigid $25\times25$ PET on a rough PDMS.}
  \label{fig:surface_yang}
\end{figure}

Surface roughness data in \textbf{Figure~\ref{fig:surface_yang}(a)}, has been taken from Xu et al.framework [SR5], where AFM measurements (Dimension Icon, Bruker, USA) were performed over PDMS and PET interfaces on $10 \times 10~\mu\text{m}^2$ areas with $512 \times 512$ resolution. The contact area analysis in \textbf{Figure~\ref{fig:surface_yang}(b)} and corresponding $V_{OC}$ were validated using these surface profiles to ensure consistency with the methodology.

\section*{Supplementary References}

[SR1] A. M. Loveless, A. L. Garner, Physics of Plasmas 2017, 24, 11. \\

[SR2] Y. Zi, C. Wu, W. Ding, Z. L. Wang, Advanced Functional Materials 2017, 27, 24 1700049.\\

[SR3] A. Bush, R. Gibson, T. Thomas, Wear 1975, 35, 1 87.\\

[SR4] B. Persson, F. Bucher, B. Chiaia, Physical review B 2002, 65, 18 184106.\\

[SR5] Y. Xu, G. Min, N. Gadegaard, R. Dahiya, D. M. Mulvihill, Nano Energy 2020, 76 105067.

\clearpage
\newpage




\end{document}